\documentclass{aa}  

\usepackage{graphicx}
\usepackage{txfonts}
\usepackage[colorlinks=true,linkcolor=blue,citecolor=blue,urlcolor=blue]{hyperref}
\usepackage{tikz}
\usepackage{import}
\newcommand{\paramalias}[2][\newcommand]{#1#2}

\begin{document} 
\paramalias{\parampllogLlogMdot}{2.4} 
\paramalias{\paramQtoSTslope}{-0.28} 
\paramalias{\paramQtoSTintercept}{50.5} 
\paramalias{\paraxallOstars}{3.0} 
\paramalias{\paraxerrallOstars}{0.1} 
\paramalias{\paraDallOstars}{9.7} 
\paramalias{\paraDerrallOstars}{0.5} 
\paramalias{\paraZall}{2.0} 
\paramalias{\paramall}{3.3} 
\paramalias{\paraalphaall}{3.1} 
\paramalias{\paraalphahigherL}{4.1} 
\paramalias{\paraalphaLslope}{-1.2} 
\paramalias{\paraalphaLintercept}{10.3}

%


 \title{Stellar wind properties of the nearly complete sample of O stars in the low metallicity young star cluster NGC~346 in the SMC galaxy\thanks{Based on observations with the NASA/ESA {\em Hubble Space Telescope}, which is operated by the Association of Universities for Research in Astronomy, Inc., under NASA contract NAS 5-2655. Also based on observations collected at the European Organisation for Astronomical Research in the Southern Hemisphere.}}


   \author{M. J. Rickard\inst{1,2}
        \and
          R. Hainich\inst{1}
          \and
          W.-R. Hamann\inst{1}
        \and
          L. M. Oskinova\inst{1}
          \and
          R. K. Prinja\inst{2}
          \and
          V. Ramachandran\inst{1,3}
          \and
          D. Pauli\inst{2}
          \and\\
          H. Todt\inst{1}
          \and
          A. C. C. Sander\inst{3,4}
          \and
          T. Shenar\inst{5}
          \and
          Y.-H. Chu\inst{6}
          \and
          J. S. Gallagher III\inst{7}}

   \institute{Institut f\"{u}r Physik und Astronomie, Universit\"{a}t Potsdam, Karl-Liebknecht-Str. 24/25, D-14476 Potsdam, Germany\\
              \email{matthew.rickard.18@ucl.ac.uk}
         \and
             Department of Physics and Astronomy, University College London, Gower Street, London WC1E 6BT, UK
        \and
            Zentrum für Astronomie der Universit\"{a}t Heidelberg, Astronomisches Rechen-Institut, M\"{o}nchhofstr. 12-14, 69120 Heidelberg
        \and
             Armagh Observatory and Planetarium, College Hill, Armagh, BT61 9DG, Northern Ireland
        \and
            Institute of Astrophysics, KU Leuven, Celestijnlaan 200D, 3001 Leuven, Belgium
        \and
            Institute of Astronomy and Astrophysics, Academia Sinica, No. 1, Sec. 4, Roosevelt Rd., Taipei 10617, Taiwan, R.O.C.
        \and
            Dept. Astronomy, University of Wisconsin, Madison, WI, USA}

   \date{Received 9 February 2022; accepted 6 July 2022}

 
  \abstract
   {Massive stars are among the main cosmic engines driving the evolution of star-forming galaxies. Their
   powerful ionising radiation and stellar winds inject a large amount of energy in the interstellar medium. Furthermore, mass-loss ($\dot{M}$) through radiatively driven winds plays a key role in the evolution of massive stars. 
   Even so, the wind mass-loss prescriptions used in stellar evolution models, population synthesis, and stellar feedback models often disagree with mass-loss rates empirically measured from the UV spectra of low metallicity massive stars.}
   {The most massive young star cluster in the low metallicity Small Magellanic Cloud galaxy is NGC~346. This cluster contains more than half of all O stars discovered in this galaxy so far. A similar age, metallicity ($Z$), and extinction, the O stars in the NGC~346 cluster are uniquely suited for a comparative study of stellar winds in O stars of different subtypes. We aim to use a sample of O stars within NGC~346 to study stellar winds at low metallicity.}
   {We mapped the central 1$\arcmin$ of NGC~346 with the long-slit UV observations performed by the Space Telescope Imaging Spectrograph (STIS) on board of the {\em Hubble Space Telescope} and complemented these new datasets with archival observations. Multi-epoch observations allowed for the detection of wind variability. The UV dataset was supplemented by optical spectroscopy and photometry. The resulting spectra were analysed using a  non-local thermal equilibrium model atmosphere code (PoWR) to determine wind parameters and ionising fluxes. }
  {The effective mapping technique allowed us to obtain a mosaic of almost the full extent of the cluster and resolve stars in its core. Among hundreds of extracted stellar spectra,  21 belong to O stars.  Nine of them are classified as O stars for the first time. We analyse, in detail, the UV spectra of 19 O stars (with a further two needing to be analysed in a later paper due to the complexity of the wind lines as a result of multiplicity). This more than triples the number of O stars in the core of NGC~346  with constrained wind properties.
  We show that the most commonly used theoretical mass-loss recipes for O stars over-predict mass-loss rates.  We find that the empirical scaling between mass-loss rates ($\dot{M}$) and luminosity ($L$), $\dot{M}\propto L^{\parampllogLlogMdot}$, is steeper than theoretically expected by the most commonly used recipes. In agreement with the most recent theoretical predictions, we find within $\dot{M}\propto Z^\alpha$ that $\alpha$ is dependent upon $L$.
    Only the most luminous stars dominate the ionisation feedback, while the weak stellar winds of O stars in NGC~346 and the lack of previous supernova explosions in this cluster restrict the kinetic energy input.}    
  {}

    \titlerunning{Stellar wind properties of O stars in NGC~346}
   \maketitle

\section{Introduction}
Massive stars ($M_\mathrm{initial} \ga 8 M_\sun$) are a critical component in the understanding of galaxy evolution. Their feedback sets the conditions  for star formation and determines the fates of star clusters. Massive stars, their winds, and their evolution are still poorly understood, especially at low metallicity ($Z$).  

\begin{figure}
  \center
  \begin{tikzpicture}
     \node[anchor=south west,inner sep=0] at (0,0) {\resizebox{\hsize}{!}{\includegraphics[angle=270, origin=c, trim={8cm 2cm 2cm 0cm},clip]{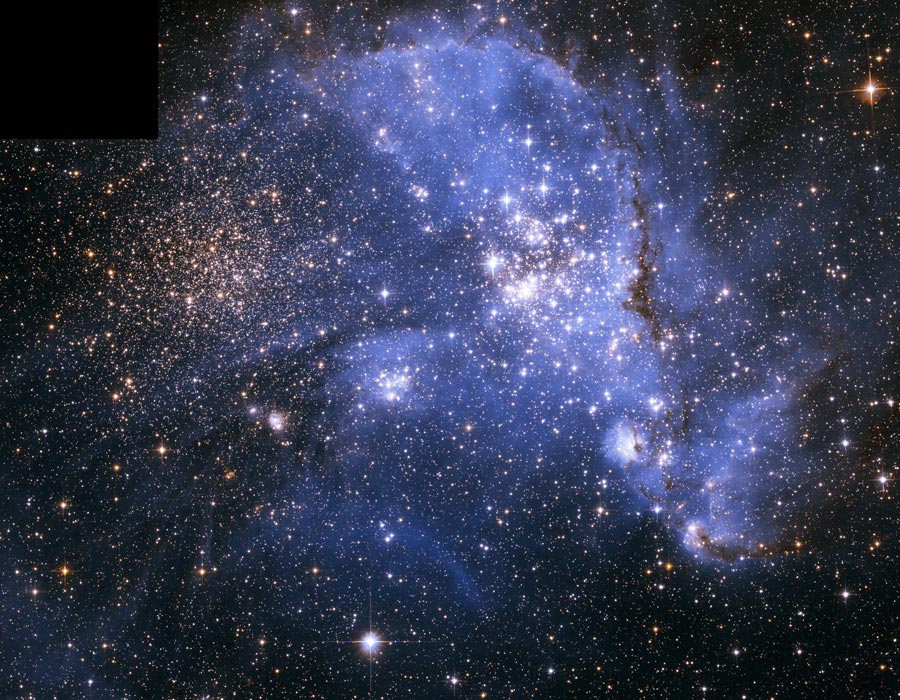}}};
     \draw[green,ultra thick] (3.9,3.0) rectangle (7.8,5.9);
    \end{tikzpicture}
    \resizebox{\hsize}{!}{\includegraphics{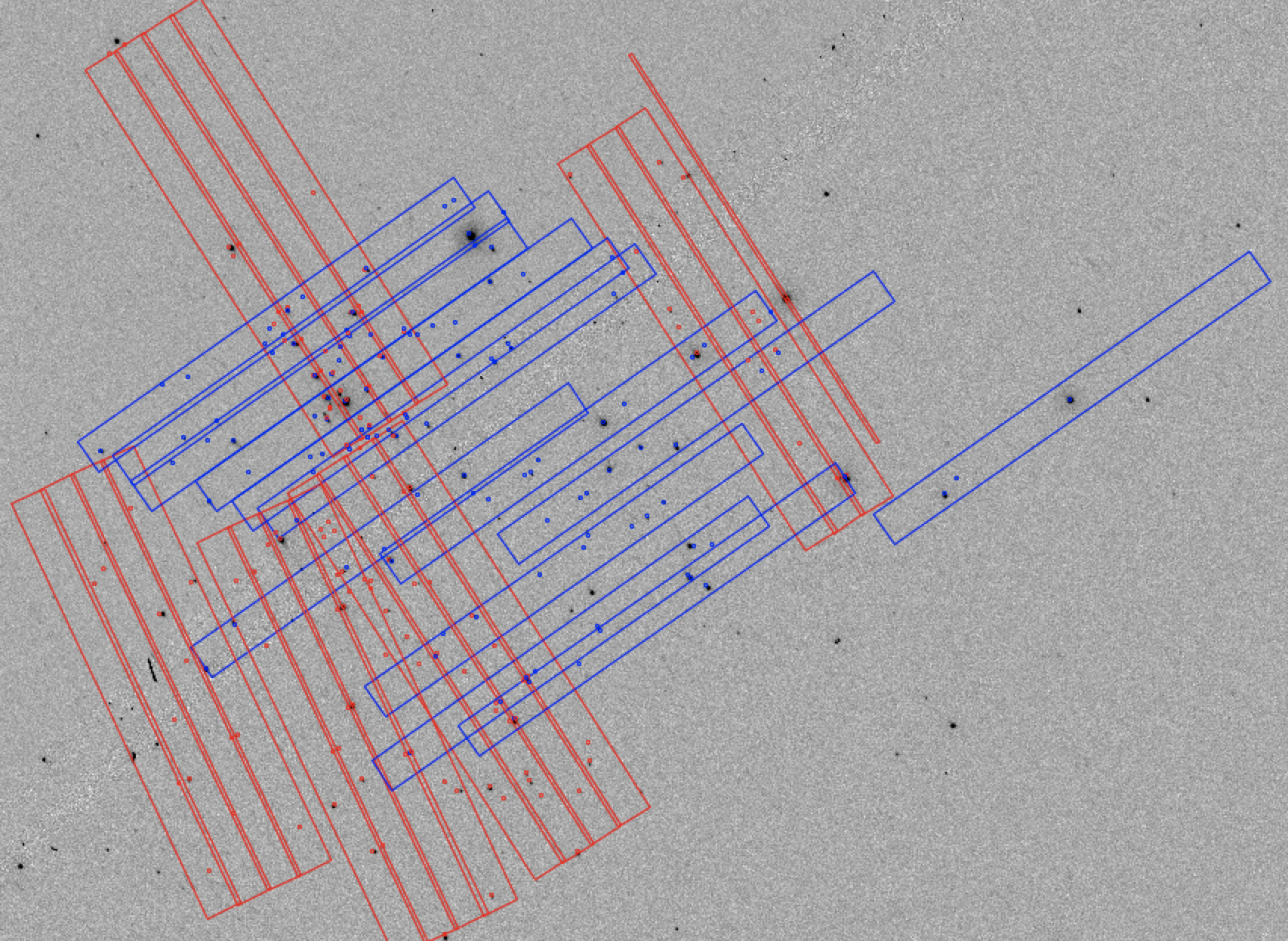}}
   \caption{NGC346 in the SMC. {\em Top panel:} Optical colour composite of HST
ACS image of the H\,{\sc ii} region N~66, covering 
a total area of $3.5'\times 3.5'$ ($\approx 62$\,pc$\times$ 62\,pc at 
d=61\,kpc) 
on the sky. North is up, and east is to the left. The cluster of bright stars 
in the centre  of the image is NGC~346. Credit: A.\,Nota (ESA/STScI).
   {\em Lower panel:} Zoom to the region indicated by the green box in the top panel. The rectangles show our spectroscopic mosaic covering the central part of NGC~346. The GO 15112 STIS G140L slit positions are in red, while GO 8629 slit positions are in blue, see text for details. Background F225W image from {\em  HST} archive (GO 12940).}
  \label{fig:NGC_346_Obs}
\end{figure}

The Small Magellanic Cloud (SMC) is a nearby dwarf galaxy at $d=61$\,kpc \citep[distance modulus, DM = 18.9 dex,][]{2005MNRAS.357..304H}. It has a metallicity $5{-}7$ times lower than the Galaxy \citep{1982ApJ...252..461D, 2000A&A...364..455L, 2007A&A...471..625T}. Thus the SMC offers an excellent laboratory to study  massive stars in environments that resemble the earlier cosmic epochs. The brightest and largest H\,\textsc{ii} region in the SMC is LHA 115-N~66 \citep{1956ApJS....2..315H}. Supernovae have not occurred yet in N~66 \citep{2003ApJ...586.1179D}, making it the closest unpolluted laboratory optimally suited for studies of massive stars at low $Z$. N~66 is ionised by the young compact cluster NGC~346 
(Fig.\,\ref{fig:NGC_346_Obs}), which contains about half of all O-type stars in the entire SMC \citep{1989AJ.....98.1305M}. NGC~346 is less than ${\sim}3$ Myr old and is amoung the youngest and the most massive low-Z star clusters in the Local Group.

\citet{Walborn1986} identified the major ionising source of the N~66 region as the O2 type star NGC~346 
W 3 \citep[SSN 9,][]{2007AJ....133...44S}. \citet{1989AJ.....98.1305M} used the  CCD UBV photometry and spectroscopy to investigate the stellar content of NGC~346 and found 33 O type stars, with 11 being of type O6.5 or earlier; their catalogued stars have the prefix MPG (e.g.\ the O2 star NGC~346 W\,3 is the same object as MPG~355). 
The field size in the \citet{1989AJ.....98.1305M} survey was $2\farcm5\times 4\farcm1$, that is, tracing nearly full extent of the N~66 H\,{\sc ii} region (see upper panel in Fig.\,\ref{fig:NGC_346_Obs}) but the core of NGC~346 remained unresolved prior to the study we present in this paper.
\citet{Gouliermis2008} presented evidence that massive star feedback induces on-going star formation in N~66. The star formation rate in N~66 complex is $\approx 4\times 10^{-3}\,M_\sun$\,yr$^{-1}$ \citep{Hony2015}.
\citet{Cignoni2011} found the star formation started in NGC~346 about 6\,Myr ago, continued at a high rate ($\geq 2\times 10^{-5}\, M_\sun$\,yr$^{-1}$\,pc$^{-2}$) for about 3\,Myr, and is now progressing at a lower rate.  They also noticed the deficiency of upper main sequence stars in NGC~346 and suggested that the youngest most massive stars did not yet have time to form.

\citet{Kudritzki1989} obtained optical spectra of four of the brightest stars in NGC~346, derived their high masses from spectroscopic analysis, and concluded that these four stars are responsible for half of the ionising flux in N~66. \citet{Walborn2000} present a UV and optical spectral atlas of 15 O stars in the SMC, among which 5 are in NGC 346. The UV spectra were measured by the Space Telescope Imaging Spectrograph (STIS) in high-resolution E140M echelle mode covering the wavelength interval $1150{-}1700\,\AA$. These data were exploited by \citet{2003ApJ...595.1182B} who produced detailed spectroscopic analysis of six O stars in NGC~346. \citet{2007AJ....133...44S} presented a catalogue of objects detected in the deep F555W ($\sim$V) and F814W ($\sim$I) {\em Hubble Space Telescope} ({\em HST}) ACS images of N~66 including NGC~346; their catalogue has prefix SSN, for example, the O2 star NGC~346 W\,3 has aliases MPG~355 and SSN~9. 

Up to now, the largest spectroscopic survey of NGC~346 massive stars was presented by \citet{2019AA...626A..50D}. Using the VLT-FLAMES spectroscopy in the optical they determined spectral types of 47 O-type stars in the wide area around NGC~346, and estimated their temperatures, gravities, and projected rotational velocities using \textsc{tlusty} model atmospheres \citep{1995ApJ...439..875H}. However, optical spectra do not allow reliable measurements of stellar winds at lower mass-loss rates where there is no H$\alpha$ emission. 

The stellar winds of massive stars are best studied with UV spectroscopy. P-Cygni resonance lines formed in stellar winds as well as numerous metal lines are in the UV range, making UV spectroscopy the best tool to measure  mass-loss rates,  wind velocities, and to probe  wind structure. Spectral analysis is completed by comparison of observed spectra to non-LTE models. These solve the radiative transfer equation in the co-moving frame whilst ensuring statistical equilibrium for rate equations \citep{1987ApJS...63..947H, 1987A&A...174..173H, 2003MNRAS.339..157H, 2015A&A...577A..13S}. 

The assumed mass-loss rates of massive stars are one of the key parameters defining their evolution. The most commonly employed mass-loss recipe for massive stars is the one of \citet{2000A&A...362..295V, 2001A&A...369..574V}, from here on $\dot{M}_{\rm Vink}$. These mass-loss prescriptions are implemented in stellar evolution models, population synthesis, and stellar feedback models and highly impact our understanding of the universe. However, increasing observational evidence question this theoretical  prediction for low~$Z$ stars and indicate that the non-supergiant O stars have in fact much smaller mass-loss rates \citep{2003ApJ...595.1182B, 2013A&A...555A...1B, Hainich2018, 2019A&A...625A.104R}. This highlights the need for larger samples of reliably derived low-Z massive stars wind parameters.

Mass-loss rate prescriptions as well as non-LTE stellar atmosphere models assume that stellar 
winds are homogeneous and spherically-symmetric on sufficiently large scales. However, the time 
series observations of Galactic massive stars consistently show the presence of discrete absorption components (DACs) moving towards the bluest extent of the absorption trough in UV wind profiles over the stellar rotation time-scale \citep{Kaper1996,Massa2015}. Typically, DACs migrate towards the blueward end of the absorption trough, while stationary narrow absorption components (NACs) are constant within the velocity space. The DACs are interpreted as 
a spectroscopic evidence for the large-scale structures in the winds of rotating stars, known as co-rotating interaction regions \citep[CIRs,][]{Mullan1984, Cranmer1996, Lobel2008}. 
While well established in Galactic O stars, very little is known about line profile variability of 
their SMC counterparts. In this paper, we investigate this question whenever we have muti-epoch UV observations, and discuss how wind line profile variability affects the measurements of stellar wind properties.

With the above key goal, the present study uses long slit UV spectroscopic observations to dissect the heart of the N~66 star forming region and measure for the first time the UV spectra of the nearly complete census  of O stars in the dense parts of NGC~346 cluster (see lower panel in Fig.\,\ref{fig:NGC_346_Obs}).  In this work we more than triple the number of the NGC~346 O stars spectroscopically analysed in the UV, and hence we draw robust conclusions on the wind properties of young O stars in general.

In Sect.\,\ref{sec:obs} we describe our new {\em  HST} STIS UV  observations, our search for complementary archival {\em  HST} observations, and introduce supplementary Multi Unit Spectroscopic Explorer (MUSE) Integrated Field Unit (IFU) observations. In Sect.\,\ref{sec:data_analysis} we describe the spectroscopic analyses using the synthetic spectra from models generated with the Potsdam Wolf-Rayet (PoWR) code to constrain mass-loss rates and other wind parameters and compare these to theoretical predictions and previous observations. Where multi-epochs exist, we compare observations of the wind profiles to look for signs of variability and large-scale wind structure.  In Sect.\,\ref{sec:results} we detail our results for each target and the relationships derived. The discussion is presented in Sect.\,\ref{sec:discussion}, while the conclusions are drawn in 
Sect,\,\ref{sec:concl}.

\begin{figure}
  \center
  \resizebox{\hsize}{!}{\includegraphics[trim={1.0cm 1.0cm 9.5cm 9.5cm},clip]{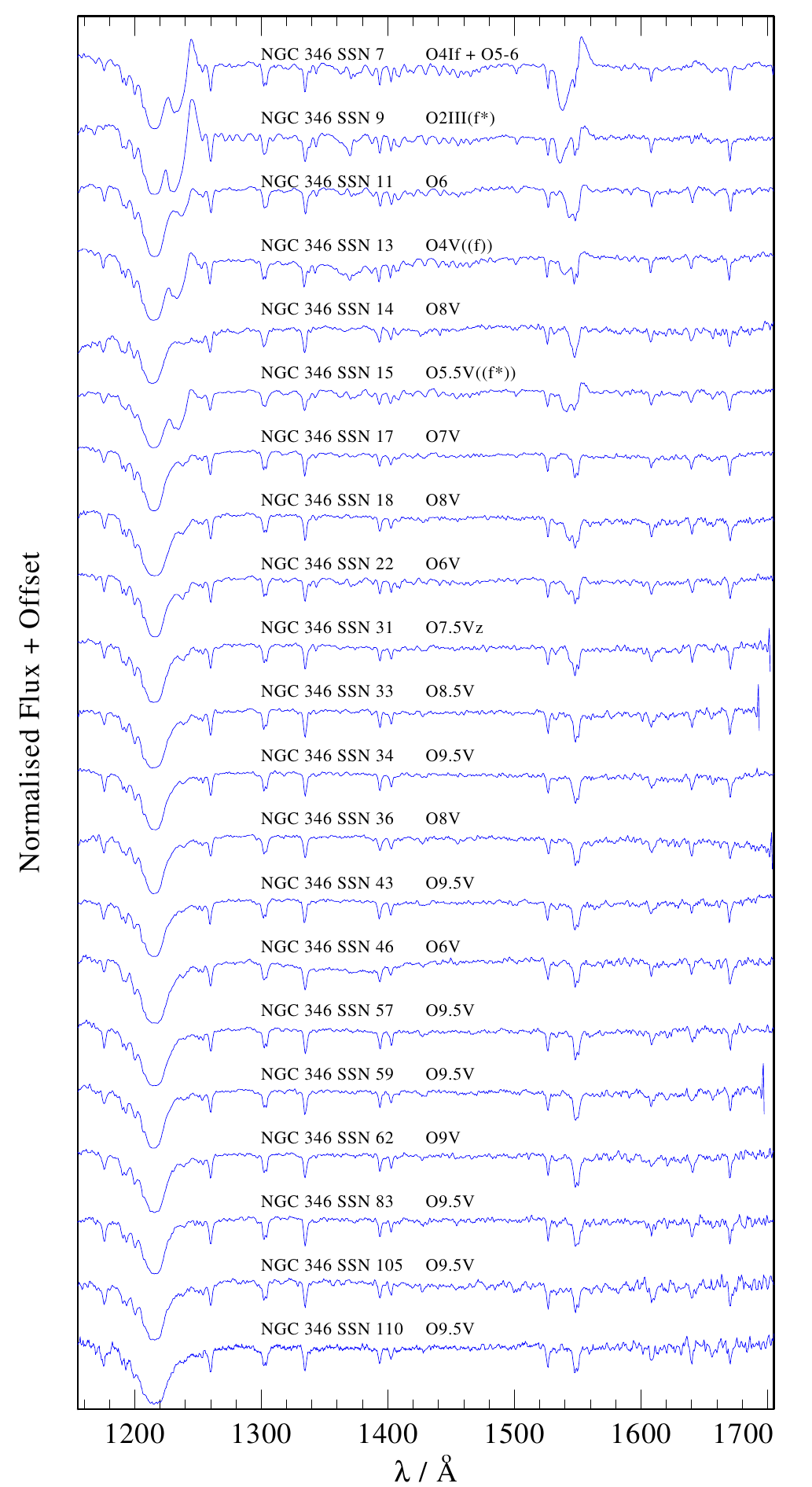}}
  \caption{Normalised UV spectra of all identified O stars extracted within the G140L observations of both GO 15112 and GO 8629.}
  \label{fig:UV_NORM}
\end{figure}

\section{Observations}
\label{sec:obs}

Our goal is to study stellar winds which are best observed in UV resonance lines. Therefore, the core of our observing samples are UV spectroscopy. These observations are complemented by optical spectra for each of our sample stars. Because the SSN catalogue overlaps completely with all MPG objects, throughout this paper we use the SSN numbers as star identifiers. A full list of the archival observations exploited for this work is contained within Appendix~\ref{sec:obs_diary}.

\subsection{Ultraviolet spectroscopy}
\label{UV_SPEC_OBS}

HST observations were made using the STIS with the G140L grism in the long slit grating mode. All slits are $25\arcsec \times 2\arcsec$ except for one where the luminosity of the brightest object required the use of a $25\arcsec \times 0.2\arcsec$ slit.  A total of 20 observations were made with General Observer programme 15112 (PI L. Oskinova). These are combined with twelve similar G140L observations from the {\em  HST} archive (GO 8629, PI F. Bruhweiler). These observations mosaic the central $1\arcmin$ of NGC~346.

Each of the 2D spectra are integrated across the wavelength range and for each 2D spectra a plot of Y pixel positions to flux is examined. A simple peak finding routine located all sources for spectral extraction. Background regions are identified from sections along the slit of ten pixels in length or greater with a continuous range of less than $0.5\,\sigma$ deviation from the median flux value from the whole length of the slit. For each potential star to extract, two background regions, one above and one below the target, are selected. For the next step, the 1D spectra are extracted from the 2D spectra using the \textsc{calstis} software package for {\em  HST} data processing \citep{STIS_DATA_HANDBOOK}. The \textsc{extract\_1d} function extracts a 1D spectrum based on a pixel position for the centre of the object as well as the pixel offset and widths of the two background regions.

Of the 247 spectra successfully extracted from both observation sets, including multiple spectra for some sources with overlapping slits, 61 OB stars have signal-to-noise (S/R) $>10$. In this work, we consider only O stars; B type stars will be considered in future work. These spectra have a resolution of $R~\sim~2400-2950$. The spectra for all O type stars extracted from the observations are shown in Fig.\,\ref{fig:UV_NORM}.

A search of the Mikulski Archive for Space Telescopes (MAST) was conducted for additional archival spectra. For four of the O star sources, SSN~9 (MPG~355), SSN~13 (MPG~324), SSN~15 (MPG~368), and SSN~18 (MPG~487), high resolution ($R=45\,800$) echelle spectra were identified from the {\em  HST} GO 7437 (PI: D. J. Lennon). These were previously analysed by \citet{2003ApJ...595.1182B, 2013A&A...555A...1B}. The star SSN~17 was observed with {\em  HST} GO 7667 (PI: R. Gilliland) in the G140M mode. These observations are a series of long slit high resolution ($R~\sim~11\,400-17\,400$) spectra, with incrementally increasing central wavelengths. To assemble the full spectrum, each extracted 1D spectrum is spliced using the IRAF \textsc{splice} function. In addition, a E140M spectrum of SSN~11 was obtained as part of GO 15837 (PI: L. Oskinova) and two E140M spectra of SSN~9 were obtained during GO 16098 (PI: J. Roman-Duval), as a part of the ULLYSES Director Discretionary Time (DDT) programme.

\subsection{Far ultraviolet spectroscopy}
\label{sec:FUSE_OBS}
Four of the brightest objects within our sample, SSN~7, SSN~9, SSN~11, and SSN~13, have been observed with the Far Ultraviolet Spectroscopic Explorer (FUSE). We accessed spectra from the FUSE Legacy in the Magellanic Clouds online archive \citep{2009PASP..121..634B}, covering a range in the Far Ultraviolet (FUV) of $900-1190\,\AA$. Lines within the FUV that are useful for wind diagnostics inculde the P\,\textsc{v}  $\lambda\,1118,\,1128$ doublet, which is a useful diagnostic of wind structure such as macroscopic clumping \citep{2007A&A...476.1331O}.

\subsection{Optical spectroscopy}
\label{sec:OPT_spectra_obs}

NGC~346 has been a target of multiple optical spectroscopic surveys \citep[e.g.][]{2006AA...456..623E}. We use the ESO VLT integral field spectrograph (MUSE) to observe the same central $1\arcmin$ of NGC~346 as covered by the STIS mosaic (Programme ID 098.D-0211(A), PI WR Hamann). The results of MUSE observations will be detailed in a forthcoming publication (M. Rickard, in prep). For this work, we combined three of our data cubes with the highest S/R and extracted from this the optical and near infrared spectra ($R \sim 1770{-}3590$, $4620{-}9265\,\AA$) of all the O stars in our sample.

Furthermore, we carefully search ESO VLT archives and whenever available, use existing archival spectra. Specifically for some of our objects, spectra are available from the Fibre Large Array Multi Element Spectrograph (FLAMES) from the GIRAFFE spectrograph ($R \sim 5000{-}2,300$, $3850{-}6690\AA$ depending on object, as detailed in Table~\ref{table:observation_record}). This data was presented by \citet{2019AA...626A..50D}. Therefore, for all our 19 analysed O stars, we have assembled UV and optical spectroscopy which allows for detailed multi-line spectroscopic analysis.

\begin{table*}
\tiny
\center
\caption{Stellar and Wind Parameters Roadmap. Where no wind profile can be fit at this resolution within the UV data, the assumed values are listed.}
\begin{tabular}{ c c c } 
\hline
\hline
Property & Source & Assumed Value  \\
 &  &  \\
\hline
Spectral Type & Literature (Table\,\ref{table:spt}), MUSE spectrum\textsuperscript{(1)} & \\
$T_\ast$ & \textsc{tlusty} fits of FLAMES spectrum\textsuperscript{(3)}, PoWR grid fit to MUSE spectrum\textsuperscript{(2)} & \\
$\log\,g$ & \textsc{tlusty} fits of FLAMES spectrum\textsuperscript{(3)}, PoWR grid fit to MUSE spectrum\textsuperscript{(2)} & \\
$\log L$ & SED fitting to UV spectrum\textsuperscript{(1)}& \\
$E(B{-}V)$ & SED fitting to UV spectrum\textsuperscript{(1)}& \\
$\varv \sin i$ & Literature\textsuperscript{(3)}, Fourier Transform fitting of He \textsc{i}\,$\lambda\,4921$ \ \& $\lambda\,6678$  with \textsc{iacob broad}\textsuperscript{(1)} & \\
$\log \dot{\mathrm{M}}$ & PoWR fitting to UV spectrum\textsuperscript{(1)} & \\
$\beta$ & PoWR fitting to UV spectrum\textsuperscript{(1)} & 0.8 \\
$D$ & PoWR fitting to UV spectrum\textsuperscript{(1)} & 10 \\
$\varv_\infty$ & PoWR fitting to UV spectrum\textsuperscript{(1)} & From $\varv_\textnormal{esc}$  \\
$\xi$ (Hydrostatic Equation)  & & $10\,\mathrm{km\,s}^{-1}$\\
$\xi(r)$ (Emergent Spectrum) & $\mathrm{max}(\xi_\mathrm{min},\,0.1\cdot\varv(r))$,\, $\xi_\mathrm{min}$ from Spectral Type / Luminosity Class (Table\,\ref{table:turb_velo}) & \\
\hline
\end{tabular}
\tablebib{(1) this work; (2) Rickard et al. (in prep); (3) \citet{2019AA...626A..50D}.}
\label{table:params_recipe}
\end{table*}

\section{Data analysis}
\label{sec:data_analysis}

The roadmap describing the methods we use for determining the various stellar properties of each O star in our sample stars is detailed in Table~\ref{table:params_recipe}. In this section we provide additional information for each step.

\begin{table}
\tiny
\center
\caption{Spectral types of O stars in this work, with references where drawn from published literature.}
\begin{tabular}{ c c c c } 
\hline \hline
SSN& MPG& Spectral Type      & Reference          \\
\hline
7   & 435 & O4\,If + O5-6 & \citet{2019AA...626A..50D}  \\
9   & 355 & O2\,III(f *)  & \citet{Walborn2000}  \\
11  & 342 & O6: V-III((f))      & \citet{2019AA...626A..50D}  \\
13  & 324 & O4V((f))    & \citet{Walborn2000}  \\
14  & 470 & O8\,V         & This work                  \\
15  & 368 & O5.5\,V((f+)) & \citet{1989AJ.....98.1305M}  \\
17  & 396 & O7\,V         & \citet{1989AJ.....98.1305M}  \\
18  & 487 & O8\,V         & \citet{2009ApJ...692..618M}  \\
22  & 476 & O6\,V         & This work                     \\
31  & 417 & O7.5\,Vz       & This work                     \\
33  & 467 & O8.5\,V       & \citet{2006AA...456..623E} \\
34  & 370 & O9.5\,V       & \citet{1989AJ.....98.1305M}  \\
36  & 495 & O8\,V         & \citet{1989AJ.....98.1305M}  \\
43  & 481 & O9.5\,V       & This work                     \\
46  & 500 & O6\,V         & \citet{1989AJ.....98.1305M}  \\
57  & 455 & O9.5\,V       & \citet{2010AA...517A..39H} \\
59  & 375 & O9.5\,V       & This work                     \\
62  & 468 & O9\,V         & \citet{1989AJ.....98.1305M}  \\
83  & 429 & O9.5\,V       & This work                     \\
105 & 486 & O9.5\,V         & This work                     \\
110 & 499 & O9.5\,V       & This work   \\
\hline     
\end{tabular}
\label{table:spt}
\end{table}

\begin{figure}
  \center
  \resizebox{\hsize}{!}{\includegraphics[trim={0.5cm 2.2cm 0.8cm 2.5cm},clip]{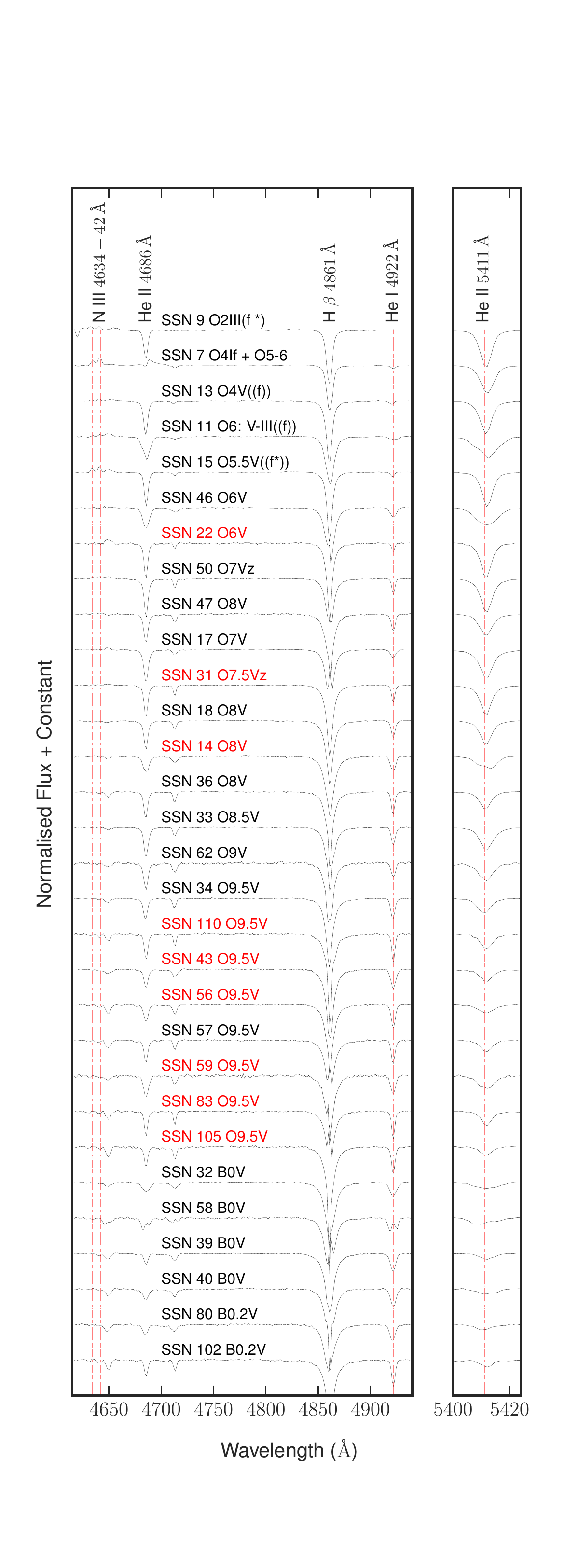}}
  \caption{Normalised MUSE spectra of all candidate OV-B0\,V types stars in our sample. Black labels denote spectral types from literature. Red and bold labels denote spectral type assigned from this work.}
  \label{fig:SPT_MUSE}
\end{figure}

\subsection{Spectral typing}
\label{sec:sp_typing}

Commonly used spectral typing classification schemes are based on optical spectra, typically using lines blueward of $4600\,\AA$ \citep{1971ApJ...170..325C, 1990PASP..102..379W, 2011ApJS..193...24S}. This is not possible for our MUSE spectra due to their wavelength range which covers $\lambda\,4620-9350\,\AA$. As all of our objects have MUSE spectra, we follow the methods described by \citet{2020A&A...634A..51B} and \citet{1999AJ....117.2485K} to determine spectral type based on lines available within MUSE.

\citet{2020A&A...634A..51B} assign a spectral type for O types based on the equivalent widths ($W_\lambda$) of He\,\textsc{i}\,$\lambda\,5411$ and He\,\textsc{i}\,$\lambda\,6678$  In this work, we are interested in O type stars and the border between O9.5\,V and B0\,V spectral type. According to \citet{2020A&A...634A..51B}, the $W_\lambda$ He\,\textsc{ii}\,$\lambda\,5412$ will be below $\sim0.1 \AA$ for spectral types later than B0.5. Therefore we exclude all such objects.

\begin{figure}
  \center
  \resizebox{\hsize}{!}{\includegraphics[trim={0.7cm 21.013cm 10.3cm 2.19cm},clip]{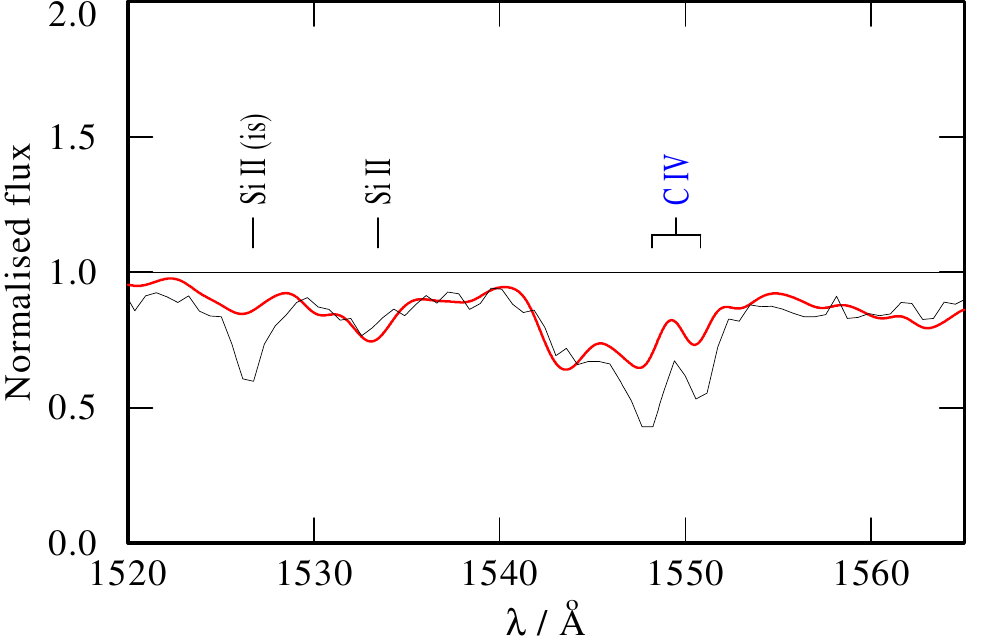}}
  \resizebox{\hsize}{!}{\includegraphics[trim={10.7cm 21.013cm 0.3cm 2.15cm},clip]{SSN_0031_p1_noISM.pdf}}
  \caption{C\,\textsc{iv}\,$\lambda\lambda\,1548.2,\,1550.8$ wind profile in the UV spectrum of SSN~31 (O7.5\,Vz). Synthetic spectrum in red, observations in black. {\em Top panel}: synthetic spectrum with wind parameters $\varv_\infty=1000 \,\mathrm{km\,s}^{-1}$, $\beta=0.8$, D=10, and $\log\,\dot{M} =-8.9$. See details in Sect.\,\ref{sec:indv_stars}. Without ISM line models applied. {\em Bottom panel}: the same synthetic but with the addition of ISM lines C\,\textsc{iv} $\lambda\lambda\,1548.2,\,1550.8$ doublet, and Si\,\textsc{ii}\,$\lambda\,1526.7$ with the ISM properties detailed in Sect.\,\ref{sec:UV_WAVE_CAL}.}
  \label{fig:CIV_wISM}
\end{figure}

\citet{1999AJ....117.2485K} developed a spectral type classification for heavily reddened O stars using the ratio of $W_\lambda$ (He\,\textsc{i}\,$\lambda\,4922 / He\,\textsc{ii}$\,$\lambda\,5411$). We take advantage of this by ordering the MUSE spectra of our candidate O\,V to B0\,V stars according to the ratio of $W_\lambda$ He\,\textsc{i}\,$\lambda\,4922$ to $W_\lambda$ He\,\textsc{ii}\,$\lambda\,5411$(Fig.\,\ref{fig:SPT_MUSE}).  We use the spectral types determined by \citet{2019AA...626A..50D} and others based on FLAMES spectra, combined with this spectral sequence to assign spectral types to stars in our sample. Finally, 21 stars within the STIS mosaic are identified as having O spectral type (Table\,\ref{table:spt}).

All newly identified O type stars have strong He\,\textsc{ii}\,$\lambda\,4686$ absorption, and so are assigned the dwarf luminosity class. The ratio of $W_\lambda$ He\,\textsc{ii}\,$\lambda\,4686$ / $W_\lambda$ He\,\textsc{i}\,$\lambda\,4471$ \ $> 1.1$ identifies O stars with a Vz luminosity class \citep{2016AJ....152...31A}. Amongst the O type stars in our sample with archival FLAMES spectra which covers the He\,\textsc{i}\,$\lambda\,4471$ and He\,\textsc{ii}\,$\lambda\,4542$ lines, only SSN~31 has the ratio $> 1.1$ and hence we assign it the Vz luminosity class.

\subsection{Interstellar medium (ISM) line modelling and spectra wavelength correction for slit position}
\label{sec:UV_WAVE_CAL}
High resolution {\em  HST} UV echelle E140M spectra show multiple lines too narrow to be of stellar origin and are instead attributed to the intervening ISM. These lines include Si\,\textsc{ii}\,$\lambda\,1260.4$, O\,\textsc{i} $\lambda\lambda\,1302.1,\,1304.9$ doublet, and C\,\textsc{ii} $\lambda\lambda\,1334.5,\,1335.7$ doublet, as well as Si\,\textsc{ii} $\lambda\ 1526.7$ and C\,\textsc{iv} $\lambda\lambda\,1548.2,\,1550.8$ doublet, which contaminate low resolution observations of the C\,\textsc{iv}\,$\lambda\lambda\,1548.2,\,1550.8$ wind lines and the photospheric lines of the same ions.

For the G140L spectra, the position of the source across the $2\arcsec$ width of the slit can shift the wavelength of the extracted spectra by $\sim10\,\AA$, therefore the known positions of ISM lines can be used to improve the wavelength calibration of the extracted spectra (Fig.~\ref{fig:CIV_wISM}). The Si\,\textsc{ii}\,$\lambda\,1260.4$, O\,\textsc{i} $\lambda\lambda\,1302.1,\,1304.9$ doublet and C\,\textsc{ii} $\lambda\lambda\,1334.5,\,1335.7$ doublet ISM lines are modelled following the formalism from \citet{1989A&AS...79..359G} to simulate absorption based on a column density from the extinction, $E(B{-}V)$, with a Lorentzian absorption profile assuming the natural line broadening dominates and using the square root part of the curve of growth. Two components of ISM absorption are simulated, from Galactic ISM and the SMC ISM. All our sources are within $1\arcmin$ of each other and hence we use a uniform Galactic extinction for the Galactic line of sight, which we adopt as $E(B{-}V)=0.07$\,mag. As a result, the model Spectral Energy Distribution (SED, see Sect.\,\ref{sec:lum_ebv_fit}) is modified by both a galactic extinction which is constant for all targets and an SMC contribution which is varied for each target.

To simplify the ISM model, we assume a standard gas to dust ratio for the SMC.  To account for the low SMC metallicity we scale the SMC hydrogen column density from the Galactic hydrogen column density $N_H=3.8\times 10^{21} \times E(B{-}V)$ \citep{1989A&AS...79..359G} by a factor of 7. Column densities of the relevant components of the ISM are then calculated from relative solar abundances scaled in the same way. A value of 165 $\,\mathrm{km\,s^{-1}}$ is used for radial velocity (RV) shift of the SMC ISM lines as this is found to best match the relative wavelength difference between the galactic and SMC components. This value is similar to that of stars within the SMC \citep{2003ApJ...595.1182B} . A {\em Voigt function} is applied to describe the line shape. A cool ISM temperature of $10\,\mathrm{K}$ is selected for the Galactic ISM line contribution, while a temperature of $16\,000\,\mathrm{K}$ is adopted for the SMC component to this line of sight, based on radio continuum studies \citep{2006MNRAS.367.1379R}. This leaves only the turbulent velocity as a free parameter to adjust to match the profile strength. The Galactic contribution is kept constant with a value of $10\,\mathrm{km\,s^{-1}}$. In the SMC ISM,  turbulent velocity is adjusted for each star based on the shape of the observed Si\,\textsc{ii}, O\,\textsc{i}, and C\,\textsc{ii} ISM lines, with values within the range of $10-30 \,\mathrm{km\,s^{-1}}$ (e.g Fig.\,\ref{fig:CIV_wISM}).

As a next step, a wavelength shift is applied to each G140L spectrum to match the positions of the observed ISM lines with those modelled in our synthetic spectrum. Due to the instrument resolution and instrumental broadening, the precision of our wavelength calibration method is of the order of $\pm 0.1 \, \AA$.

\subsection{Stellar atmosphere models}
\label{sec:sam}

In this work, we use the state-of-the-art PoWR code, which is a non-LTE model atmosphere code for hot stars with expanding atmospheres that can be applied to a range of metallicities \citep{2014A&A...565A..27H, 2015A&A...581A..21H, 2011MNRAS.416.1456O, 2015ApJ...809..135S}. PoWR iteratively solves the radiative transfer equations in the co-moving frame while balancing the rate equations for each ion level to ensure statistical equilibrium. The model produces a synthetic spectrum based on a detailed model that accounts for mass-loss, line blanketing, and wind clumping. Full details of PoWR are described within \citet{2002A&A...387..244G, 2004A&A...427..697H, 2015A&A...577A..13S, 2015A&A...581A..21H}.
Detailed atomic data for a range of ion levels is used along with a `superlevel' approach to account for `iron blanketing' \citep{2002A&A...387..244G}. PoWR includes the handling of microclumping, dense regions within the wind smaller than the mean free photon path length.

In the subsonic part of the atmosphere, the velocity field is defined such that a hydrostatic density stratification is approached \citep{2015A&A...577A..13S}. In the supersonic part of the atmosphere, the velocity field is defined by a so-called $\beta$-law \citep{1975ApJ...195..157C}, given by; 

\begin{equation}
  \label{eq:beta_law}
    \varv(r) \simeq \varv_\infty \left(1-\frac{r_0}{r}\right)^\beta
\end{equation}

\noindent where $r_0 \sim R_\ast$.

PoWR relates temperature to radius and luminosity via the Stefan–Boltzmann law ($L=4 \pi R_\ast^2 \sigma_\mathrm{SB}T_\ast^{-4}$). $T_\ast$ is defined within the PoWR code as the temperature at the Rosseland continuum optical depth. The effective temperature is typically defined at $\tau_\mathrm{eff}=2/3$. However, as the winds of O stars are optically thin, the difference between $T_\ast$ and $T_\mathrm{eff}$ is negligible. Group elements are incorporated in PoWR models with the atomic data for H, He, C, N, O, Si, Mg, S, and P, with Fe group elements included with a `superlevel' approach \citep{2002A&A...387..244G}.



Clumping within the PoWR code is specified in the form of a density contrast $D$, which denotes the inverse of the filling factor $f_\text{V}$:

\begin{equation}
  \label{eq:clump_def}
    D=\frac{\rho_\text{clump}}{\langle\rho\rangle}=f_\text{V}^{-1},
\end{equation}

\noindent where the medium between the clumps is assumed to be void. As microclumping increases the density, this affects the wind recombination lines. Far ultraviolet lines are also sensitive to microclumping. However, in all the stars considered in the present paper with FUSE observations, the lines such as P\,\textsc{v} have no identifiable wind profile.

\subsection{Temperature, surface gravity, and abundances}

The temperature ($T_\ast$) and surface gravity ($\log\,g$) are determined by fitting the optical (MUSE) spectra to a published grid of PoWR models of SMC O stars \citep{2019A&A...621A..85H}. These grids use the SMC abundances from \citet{2007A&A...466..277H} and \citet{2007A&A...471..625T}, with additional elements scaled from solar abundances\footnote{\url{http://astro.uni-tuebingen.de/~rauch/TMAP/TMAP_solar_abundances.html}} by $1/7$ for $Z_{\mathrm{SMC}} \,/\,Z_\sun$.  Table\,\ref{table:Abundances_Ions} shows the abundances employed in the PoWR models grids. These grids have a resolution of $1000\,\mathrm{K}$ in $\mathrm{T}_\ast$ and $0.2 \ \mathrm{dex}$ in $\log\,g$. The results of this process are included in Sec.~\ref{sec:teff_logg_grid_fits}. The grids are publicly available and can be downloaded from the PoWR website\footnote{\url{http://www.astro.physik.uni-potsdam.de/~wrh/PoWR/PoWRgrid1.php}}. Full details of the MUSE observations and data analysis will be presented in a forthcoming paper (Rickard et al. in prep).

\begin{table}
\footnotesize
\center
\caption{Abundances and Ions used within the PoWR model SMC grid.}
\begin{tabular}{ c c c } 
\hline
\hline
\centering
Element & Abundance & Ions  \\
 & $\mathrm{X}/\/\mathrm{X}_\sun$ &   \\
\hline
H & 1.00 & \textsc{i}, \textsc{ii}\\
He & 1.04 & \textsc{i}, \textsc{ii}, \textsc{iii}\\
C & 0.09& \textsc{i}, \textsc{ii}, \textsc{iii}, \textsc{iv}, \textsc{v}\\
N & 0.05 &\textsc{i}, \textsc{ii}, \textsc{iii}, \textsc{iv}, \textsc{v}, \textsc{vi}\\
O & 0.20 & \textsc{i}, \textsc{ii}, \textsc{iii}, \textsc{iv}, \textsc{v}, \textsc{vi}, \textsc{vi}, \textsc{vii}\\
Mg & 0.14 & \textsc{i}, \textsc{ii}, \textsc{iii}, \textsc{iv}, \textsc{v}\\
Si & 0.19& \textsc{i}, \textsc{ii}, \textsc{iii}, \textsc{iv}, \textsc{v}\\
P & 0.14& \textsc{iv}, \textsc{v}, \textsc{vi}\\
S & 0.14& \textsc{iii}, \textsc{iv}, \textsc{v}, \textsc{vi}\\
Fe & 0.29&\\
\hline     
\end{tabular}
\label{table:Abundances_Ions}
\end{table}

\begin{figure*}
  \center
  \includegraphics[width=17cm, trim={0.0cm 20.52cm 0cm 2.2cm},clip]{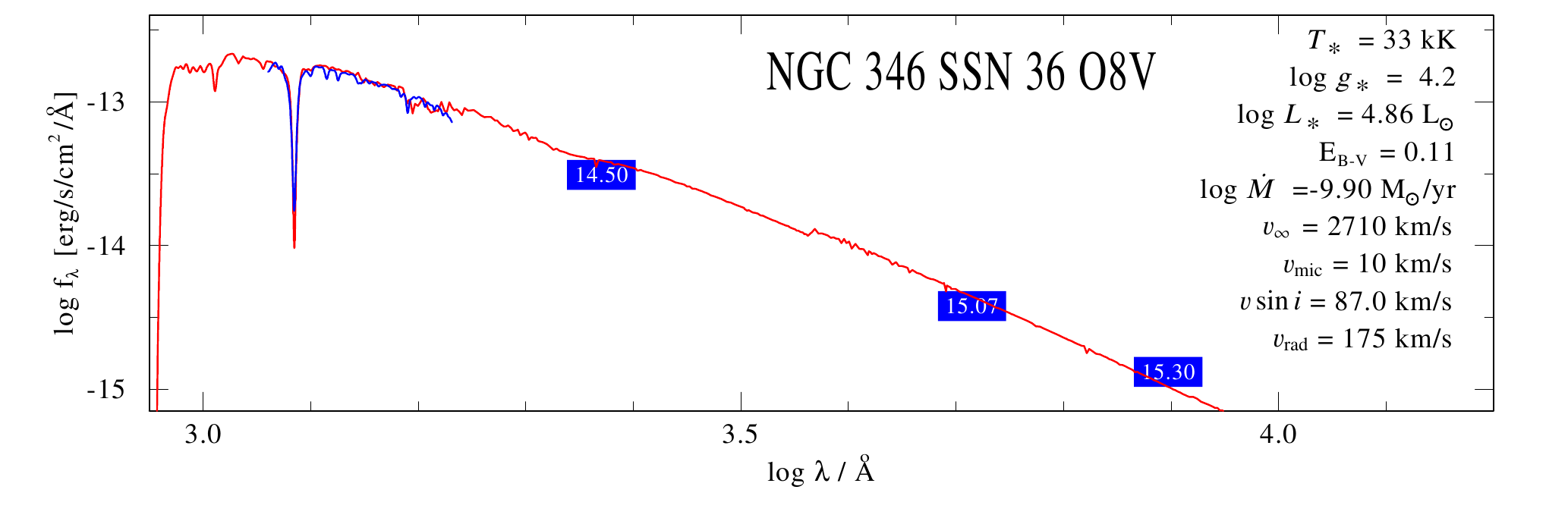}
  \caption{Example SED illustrating the method to estimate $L$ and $E(B{-}V)$. Spectrum and photometry (HST F225W, F555W, and F814W) of SSN~36 are shown by blue line and boxes correspondingly. The best model, modified for extinction (Sect.\,\ref{sec:lum_ebv_fit}), is in red. Model parameters are listed in the insert.}
  \label{fig:SED_example}
\end{figure*}

In the majority of the optical spectra, there are few metal lines available to individually tailor abundances. Therefore, the grid abundances as detailed in Table~\ref{table:Abundances_Ions}, are carried forward into the individual models for each O star. The exception to this is the CNO abundances for the hottest stars in our sample. In these cases, CNO abundances are adjusted per star based upon UV and optical metal lines, as detailed in Sect.~\ref{sec:indv_stars}.

\subsection{Luminosity}
\label{sec:lum_ebv_fit}
The bolometric luminosity of each star is determined using its flux calibrated UV spectrum. This is compared to a synthetic SED modified for extinction ( Sect.~\ref{sec:UV_WAVE_CAL}). The synthetic spectra are modified using the \citet{1979MNRAS.187..785S} law for the Galactic components and the \citet{1985A&A...149..330B} extinction law while for the SMC component using the PoWR code plotting package which has a form of these laws integrated across the whole SED. The combined Galactic and SMC value of $E(B{-}V)$ is selected to match both the slope of the extinction modified synthetic SED to the observations and the width of the $\mathrm{Ly}\alpha$ absorption compared to a synthetic $\mathrm{Ly}\alpha$ profile generated following the formalisation in \citet{1989A&AS...79..359G}, see example Fig.\,\ref{fig:SED_example}. With a model luminosity and extinction, the PoWR code generates a model flux continuum which is then used to normalise the observed spectra.

The synthetic SED is shifted to match the observed UV spectra and photometry from {\em  HST} bands F555W and F814W \citep{2007AJ....133...44S}, and F225W DAOPhot extraction from the Hubble Legacy Archives\footnote{\url{https://hla.stsci.edu}}. $\log L$ can be found to $\pm 0.01$ dex and $E(B{-}V)$ to $\pm 0.01$\,mag, but the degeneracy between the two means that an error of $\pm 0.01$\,mag in $E(B{-}V)$ translates to a total error of $\pm 0.05$ dex in $\log L$.

\subsection{Projected rotational velocities}

For half of our sample stars the projected rotational velocities, $\varv\sin i$, are taken from \citet{2019AA...626A..50D}, calculated from \textsc{tlusty} fits to the FLAMES spectra. For the remaining objects, the photospheric He\,\textsc{ii}\,$\lambda\,4921$ and $\lambda\,6678$ lines within our MUSE spectra are examined with the \textsc{iacob-broad} tool \citep{2014A&A...562A.135S}. This uses a combination of a Fourier Transform (FT) and a goodness of fit method to measure $\varv\sin i$. The synthetic model spectra are then convolved with a Gaussian with a full width half maximum equal to the measured $\varv \sin i$.

\subsection{Terminal wind velocities}

Terminal velocities ($\varv_\infty$) are measured from the blue edge of the black trough of a saturated wind profile. Only one O star in our sample has a wind profile that is fully saturated, aving a saturated N\,\textsc{v} $\lambda\lambda\,1238.8,\,1242.8$ doublet. In an unsaturated wind profile, $\varv_\infty$ may be beyond the measurable blueward shift in velocity space of the profile. In this case, the blue edge of the absorption trough gives a lower limit to $\varv_\infty$. For stars where no wind profiles are seen in the UV spectra, we adopt a terminal velocity calculated from the escape velocity of the star following \citet{1995ApJ...455..269L}, scaled to SMC metallicity  \citep{1992ApJ...401..596L}.

\subsection{Microturbulence}

In stellar atmosphere models, the microturbulent velocity parameter $\xi(r)$ accounts for small scale turbulent motions. Within PoWR, this parameter is handled separately for the  hydrostatic part of the atmosphere and for the expanding wind. For the hydrostatic part, a turbulent pressure term is incorporated in the hydrostatic equation, adding to the gas pressure term, in line with the arguments by \citet{1991ApJ...377L..33H}. By varying this value, we see that the value of $\xi$ for the hydrostatic part of the wind has little impact on the synthetic spectrum and so adopt a constant $\xi=10\,\mathrm{km\,s}^{-1}$. 

For the emergent spectrum of the stellar wind, we specify $\xi(r)=\mathrm{max}(\xi_\mathrm{min},\,0.1\cdot\varv(r))$ \citep{2015ApJ...809..135S, 2015A&A...577A..13S}, where $\xi_\mathrm{min}$ is dependent upon spectral type.  A value of $10 \ \,\mathrm{km\,s^{-1}}$ is commonly invoked for O stars, but earlier spectral types can be best matched by models with values as high as $25 \ \,\mathrm{km\,s^{-1}}$. For the lines that are affected, a general trend of higher turbulent velocities for earlier spectral types and higher luminosity classes is determined \citep{2003ApJ...595.1182B}.  We follow these results and set $\xi_\mathrm{min}$ as detailed in Table\,\ref{table:turb_velo}.

\subsection{Wind parameters}
\label{sec:wind_param_fitting}

For each O star within our sample, a tailored model is run with the fundamental stellar parameters of $T_\ast$, $\log g$, $L$, and abundances so far established. The velocity law exponent ($\beta$; Eq.~\ref{eq:beta_law}), mass-loss rate ($\dot{M}\, [M_\sun\mathrm{yr}^{-1}]$), and clumping factor ($D$; Eqs.~\ref{eq:clump_def}~and~\ref{eq:Hillier}) are constrained by the fitting the diagnostic resonance P-Cygni wind lines of the N\,\textsc{v} $\lambda\lambda\,1238.8,\,1242.8$ doublet,  the O\,\textsc{v}\,$\lambda\,1371.0$ line, and the C\,\textsc{iv} $\lambda\lambda\,1548.2,\,1550.8$ doublet. The mass-loss rate is measured by the overall strength of the wind profile. The selected $\beta$-law affects the shape of the blue-ward absorption.

For our tailored models, we adopt clumping distribution using the equation:

\begin{equation}
  \label{eq:Hillier}
    f_\text{V}=f_{\text{V},\infty} +(1 - f_{\text{V},\infty})\,\exp(\varv/\varv_\text{cl}),
\end{equation}

\noindent with clumping starting at $\varv_\text{cl}=30\,\mathrm{km\,s^{-1}}$ \citep{1998ApJ...496..407H}. Models with a higher $D$ have lower population of N\,\textsc{v} within the clumps due to recombination and so the N\,\textsc{v}\,$\lambda\lambda\,1238.8,\,1242.8$ wind profile becomes less pronounced. The effect is smaller for the O\,\textsc{v}\,$\lambda\,1371.0$ and C\,\textsc{iv}\,$\lambda\lambda\,1548.2,\,1550.8$ wind profiles, and so the simultaneous adjustment of $\dot{M}$ and $D$ is required whilst considering all three UV wind lines. The clumping factor, $D$, is limited to possible values of 1 (for a smooth wind), 5, 10, 20, 50, or 100 to limit the parameter space.

Eight stars in our sample (Table~\ref{table:results_Ostars}) have detectable wind profile in either the N\,\textsc{v}\,$\lambda\lambda\,1238.8,\,1242.8$ or C\,\textsc{iv}\,$\lambda\lambda\,1548.2,\,1550.8$ resonance lines. For these stars, the wind parameters are fitted simultaneously.  Four among them have only slight blueward absorption in the C\,\textsc{iv}\,$\lambda\lambda\,1548.2,\,1550.8$ profile. For the low resolution spectra, the ISM C\,\textsc{iv}\,$\lambda\lambda\,1548.2,\,1550.8$ has the effect of enhancing the core of this wind line, see Fig.~\ref{fig:SSN_14_comparison}. The line contamination by the ISM means that very slight blue-ward asymmetry ($\la 300\,\mathrm{km\,s^{-1}}$) in the C\,\textsc{iv}\,$\lambda\lambda\,1548.2,\,1550.8$ wind line cannot be determined from the G140L spectra.

The N\,\textsc{v}\,$\lambda\lambda\,1238.8,\,1242.8$ line is sensitive to X-ray radiation field in stellar wind. However, the X-ray fluxes of individual stars 
in our sample are not unknown \citep{2002ApJ...580..225N}. The single stars in the SMC seem to be intrinsically faint in X-rays, as shown in sensitive X-ray observations  
did not detect the O3V-type star SK\,183 in the SMC \citep{2013ApJ...765...73O}. 
Since X-ray fluxes of our sample stars  are not know, to avoid introduction of additional free parameters describing X-ray properties, we exclude X-rays from our models.

For stars without observed UV wind profiles, only an upper limit for $\dot{\mathrm{M}}$ can be determined. In this case, $\beta$ and $D$ are the standard values for O stars (0.8 and 10 respectively, \citeauthor{2019A&A...621A..85H} \citeyear{2019A&A...621A..85H} and references therein).

\begin{table}
\footnotesize
\center
\caption{Minimum turbulent velocities ($\xi_\mathrm{min} \ [\mathrm{km\,s^{-1}}]$) adopted according to spectral type \citep{2003ApJ...595.1182B}.}
\begin{tabular}{ l c c } 
\hline
\hline
\centering Spectral Type & \multicolumn{2}{c}{Luminosity Class} \\
  & III-IV & V     \\
\hline
O2...$<$ O4 & 25 &- \\
O4...$<$ O6 &-& 15 \\
O6...$<$ O8 &-& 10  \\
$\geq$ O8 &-& 5  \\
\hline     
\end{tabular}
\label{table:turb_velo}
\end{table}

\begin{table*}
\footnotesize
\center
\caption{Model Parameters derived for O stars within sample with a UV spectrum. SSN~9-SSN~31 are derived from model fits. SSN~33-SSN~110 mass-loss rates are upper limits.}
\begin{tabular}{ c c c c c c c c c c c c c} 
\hline
\hline
\rule[0mm]{0mm}{3.0mm}
SSN & MPG & $\varv \sin i$ & $T_\ast$ & $\log g$ & $\log L$ & $E(B{-}V)$ & $\xi_\mathrm{min}$ & $\log \dot{M}$ &  $\varv_\infty$  & $D$ & $\beta$ &Comment\\
 &  & $(\mathrm{km\,s^{-1}})$  &(kK) & (cgs) & ($\mathrm{L}_\sun$) & (mag) &$(\mathrm{km\,s^{-1}})$ & ($\mathrm{M}_\sun \mathrm{yr}^{-1}$) & $(\mathrm{km\,s^{-1}})$ & & \\
\hline
9 & 355  & 130\textsuperscript{(2)} &  51.7& 4.0 & 6.12 &  0.13 & 35 &  $-6.65$ & $2800$ & 200 & 0.8 &MSB\\
13 & 324  & $\geq 113$\textsuperscript{(2)} & 42.0 & 3.8 & 5.60 & 0.11 & 15\textsuperscript{(1)} & $-7.4$ & $\geq2300$ &  20 & 1.0 &SB1\textsuperscript{(2)}, MSB\\
14 & 470 & 145  & 37.0 & 4.4 & 5.37 &0.16& 5  & $-9.5$ & $ \geq 600$ & 20 & 1.0 & MSB\\
15 & 368  & 58\textsuperscript{(2)}  & 39.0 & 4.0 & 5.45  &0.12 & 15\textsuperscript{(1)} &  $-7.6$ & $\geq 2100$ & 20 & 1.0 &SB1\textsuperscript{(2)}, MSB\\
17 & 396 & 196 & 37.0 & 4.0 & 5.30  &0.11 & 10 & $-8.7$ & $\geq 1000$ & 1 &0.8&MSB \\
18 & 487 & 131  & 37.0 & 4.4 & 5.25 & 0.15 & 5 & $-8.45$ / $\leq -9.1$ \textsuperscript{*} & $\geq 1500$ & 5 & 1.0 &\\
22 & 476 & 100 & 38.0 & 4.0 & 5.32  &0.13 & 10 & $-8.2$ & $\geq 1600$ & 5 &0.8 \\
31 & 417  & 98\textsuperscript{(2)} & 37.0 & 4.4 & 5.09  &0.12 & 10 & $-8.9$ & $\geq 1000$ &  10 & 0.8&MSB\\
33 & 467 & 87 & 34.0 & 4.2 & 4.97 &0.14 & 5 &  $\leq -9.5$  & $3210$\textsuperscript{$\dagger$} & 10 & 0.8&SB1\textsuperscript{(2)}, MSB \\
34 & 370  & 158\textsuperscript{(2)}  & 33.0 & 4.2 & 4.97  & 0.13 & 5 &  $\leq -9.8$ & $2820$\textsuperscript{$\dagger$} & 10& 0.8&SB1\textsuperscript{(2)}, MSB \\
36 & 495 & 87\textsuperscript{(2)} & 33.0 & 4.2 & 4.86 & 0.11 & 5 & $\leq -9.9$ & $2710$\textsuperscript{$\dagger$} & 10 & 0.8 &SB1\textsuperscript{(2)}, MSB \\
43 & 481 & 119 & 32.3 & 4.4 & 4.82  & 0.12 & 5 & $\leq -10.0$ & $3460$\textsuperscript{$\dagger$} & 10 &0.8\\
46 & 500 & 316\textsuperscript{(2)} & 37.0 & 4.4 & 4.91 & 0.15 & 10 & $\leq -9.1$ & $3140$\textsuperscript{$\dagger$} & 10&0.8& MSB\\
57 & 455 & 120\textsuperscript{(2)}  & 33.0 & 4.2 & 4.89 & 0.14 & 5 &  $\leq -9.9$ & $2850$\textsuperscript{$\dagger$} & 10 &0.8\\
59 & 375 & 115 & 32.5 & 4.2 & 4.66 & 0.11 &5 & $\leq -9.8$   & $2460$\textsuperscript{$\dagger$} & 10 & 0.8\\
62 & 468 & 113 & 35.0 & 4.2 & 4.74 &  0.12 &5 & $\leq -9.8$ & $2370$\textsuperscript{$\dagger$} & 10 & 0.8\\
83 & 429 & 112 & 32.0 & 4.2 & 4.53 &  0.12 & 5 & $\leq -10.0$ & $2210$\textsuperscript{$\dagger$} & 10 & 0.8\\
105 & 486 & 120 & 32.0 & 4.2 & 4.40 &  0.12 &5 & $\leq -10.0$ & $2140$\textsuperscript{$\dagger$} & 10 & 0.8\\
110 & 499 & 100 & 33.0 & 4.2 & 4.58 &  0.17 & 5 &  $\leq -10.0$ & $2430$\textsuperscript{$\dagger$} & 10 & 0.8 \\
\hline
\end{tabular}
\tablebib{(1) \citet{2003ApJ...595.1182B}; (2) \citet{2019AA...626A..50D}.}
\tablefoot{
\tablefoottext{$\dagger$}{$\varv_\infty$ is scaled from $\varv_\mathrm{esc}$.} \tablefoottext{*}{Higher value from G140L low resolution spectrum extraction. Lower value from E140M spectrum, see Sect.\,\ref{sec:variability}. MUSE Spectral Binary (MSB) are candidate spectroscopic binaries based on observed RV shifts in MUSE observations (Rickard et al. in prep). SB1, SB2, and SB3 are spectral binaires with one set of photospheric lines observed to be variable, while SB2 and SB3 have sets of lines with 2 and 3 different RV variations respectively.}
}
\label{table:results_Ostars}
\end{table*}

\subsection{Spectroscopic binaries}

\citet{2019AA...626A..50D} have identified spectroscopic binaries in NGC~346 by subtracting two FLAMES spectra of different epochs from each other. This method can identify spectral binaries with RV variation $\ga 10 \,\mathrm{km\,s^{-1}}$ between two observations. They also identify a number of SB2 and SB3 binaries within NGC~346. Our multi-epoch MUSE observations allow further identification of spectroscopic binaries via RV variations between epochs (Rickard et al. in prep.) with MSB used as a label (MUSE Spectroscopic Binary candidate) for this case, see Table~\ref{table:results_Ostars}. These results are preliminary, but have a good correlation with the spectroscopic binaries from \citet{2019AA...626A..50D} which is encouraging.

In the absence of spectral lines that can be attributed to a secondary, we consider a single star fit to be appropriate, as the secondary is going to be a considerably lower luminosity star with little contribution to the flux. On the other hand, when spectral lines are clearly seen from a second or even third component, such as in the SSN~7 and SSN~11 spectra, we do not fit a single star model. These objects will be considered in a forthcoming publication.

\section{Results}
\label{sec:results}

In this section, we detail the results for each O star within our sample and present the general trends and relationships which emerge from our analysis. The full model, observed SED, and UV spectra for each object are included in Appendix~\ref{sec:appedix_fits}.

\subsection{Individual stars}
\label{sec:indv_stars}
We present the model parameters of the synthetic spectra that match spectroscopic observations. For four stars (SSN~14, SSN~17, SSN~22, and SSN~31) we present the first measured wind parameters from UV observations. We also establish for the first time upper limits for wind parameters for evelven objects.\\

\subsubsection{SSN~7 (MPG~435)}

\citet{1989AJ.....98.1305M} assigned the spectral type of O5.5If, assuming that the object is a single star. However, RV variations indicate that the star is a binary with a period of $24.2\mathrm{d}$ \citep{1986PASP...98.1133N, 2004NewAR..48..727N}. \citet{2019AA...626A..50D} confirm it as an SB1 binary. We also observe rapidly shifting lines within our MUSE observations (Rickard et al. in prep).  As described by \citet{2019AA...626A..50D}, the N\,\textsc{iv}\,$\lambda\,4058$ emission line matches the RV of the primary therefore requiring the primary spectral subtype earlier than O5, with He\,\textsc{i}\,$\lambda\,4471$ absorption and weak He\,\textsc{ii}\,$\lambda\,4686$ emission indicating a spectral subtype O4\,If. The secondary has a spectral type of O5-6 from He\,\textsc{i}\,$\lambda\,4471$ absorption being weaker than the He\,\textsc{ii}\,$\lambda\,4542$ absorption. With spectral lines and components attributable to the RV of the individual components, we identify SSN~7 as a potential object for which a spectral fit can be made with a binary composite model. Such a fit will be shown in a future paper.\\

\begin{figure}
  \center
  \resizebox{0.95\hsize}{!}{\includegraphics[trim={0.7cm 21.0133cm 10.3cm 2.19cm},clip]{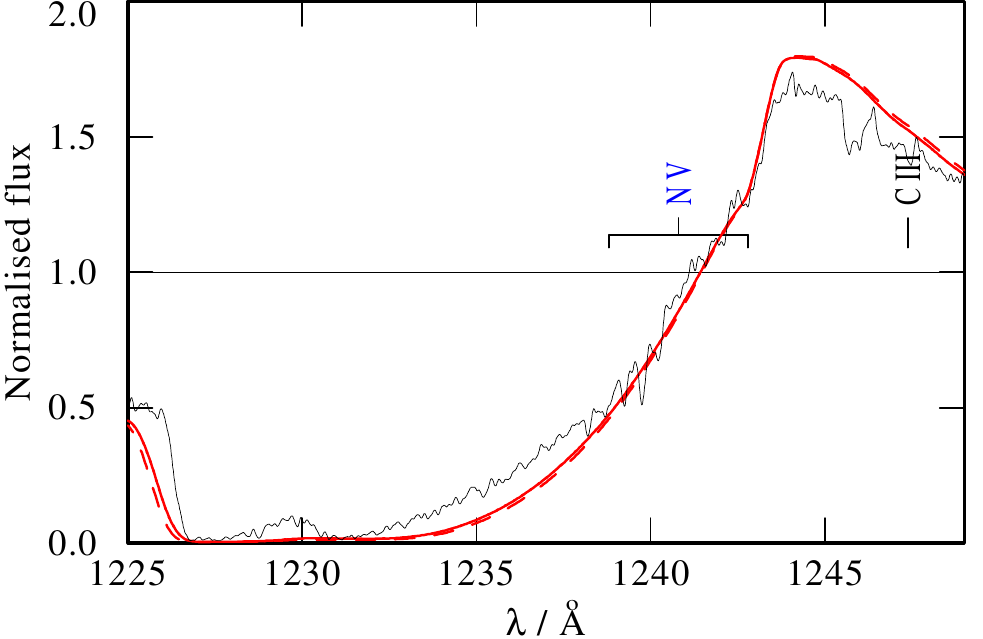}}
    \resizebox{0.95\hsize}{!}{\includegraphics[trim={10.7cm 14.44cm 0.3cm 8.66cm},clip]{SSN_0009_p3.pdf}}
 \resizebox{0.95\hsize}{!}{\includegraphics[trim={10.7cm 21.013cm 0.3cm 2.19cm},clip]{SSN_0009_p3.pdf}}
 \resizebox{0.95\hsize}{!}{\includegraphics[trim={0.7cm 14.44cm 10.3cm 8.66cm},clip]{SSN_0009_p3.pdf}}
   \caption{Spectral observations of wind lines of SSN~9, normalised and compared to various models. The observed spectra of SSN~9 is shown by black line. Synthetic spectra with $\log\,\dot{M}=-6.65$ and $D=200$ shown by solid red line. The dashed red line shows the spectra with $\log\,\dot{M}=-6.5$ and $D=10$.  All shown models have $\varv_\infty=2800 \,\mathrm{km\,s^{-1}}$.}
  \label{fig:SSN_9_fitting}
\end{figure}

\begin{figure}
  \center
  \resizebox{\hsize}{!}{\includegraphics[ trim={10.7cm 21.013cm 0.3cm 2cm},clip]{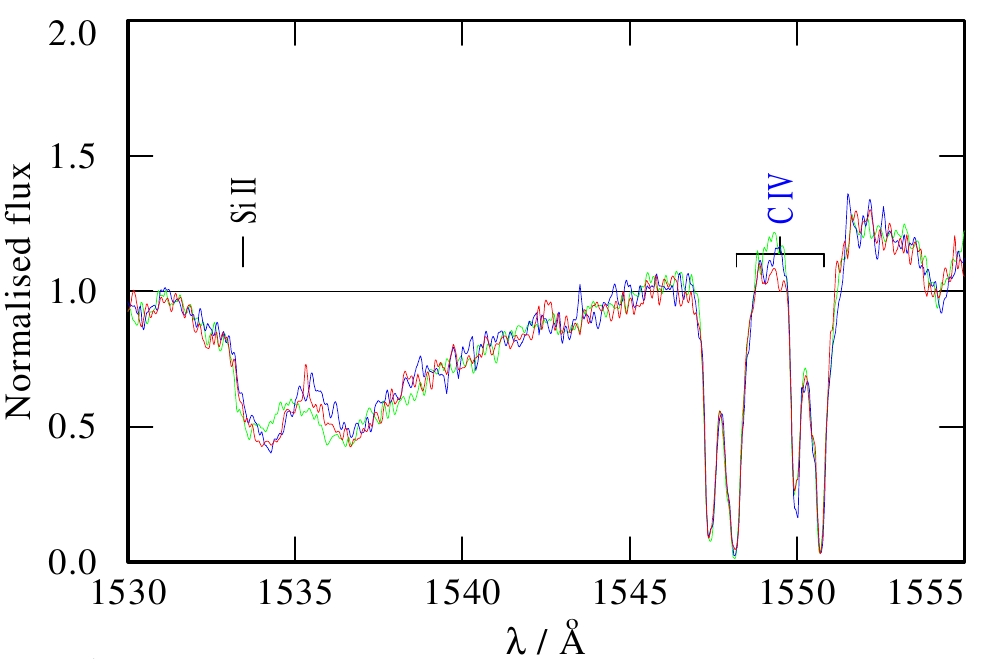}}
  \caption{Multi-epoch observations of SSN~9 C\,\textsc{iv}\,$\lambda\lambda\,1548.2,\,1550.8$ UV wind profile. Green line: observation taken on 22-11-2000.  Blue and red lines: observations taken on 01-09-2020 and separated by $\sim 1$ hr, with the blue-line spectrum observed prior to the red-line spectrum. Each of these exposures is $\sim 40$ min. A feature can be seen moving blueward on the timescale of hours.}
  \label{fig:SSN_9_multiepoh}
\end{figure}

\subsubsection{SSN~9 (MPG~355)}

SSN 9 has the spectral type of O2\,III(f*) from analysis by \citet{Walborn2000} and projected rotational velocity $\varv \sin i=130\mathrm{km\,s}^{-1} $ from \citet{2019AA...626A..50D}. The E140M UV spectrum we use has previously been analysed by \citet{2003ApJ...595.1182B, 2013A&A...555A...1B} who determined $T_\ast=51.7\,\mathrm{kK}$ and $\log{g}=4.00$. This temperature is beyond the  grid range, so we adopt the stellar parameters from \citet{2013A&A...555A...1B} and find that synthetic spectra produced to be a good match to the optical He\,\textsc{i} and He\,\textsc{ii} lines.

\citet{2013A&A...555A...1B} adopt $\mathrm{N} / \mathrm{N}_\sun=1.9$. We find that N\,\textsc{iv} $\lambda\lambda\,7105,\,7111,\,7123$ triplet emission, not available to \citet{2013A&A...555A...1B}, is lower in the MUSE observations and is best matched by $\mathrm{N} / \mathrm{N}_\sun=1.5$.

With an abundance of $\mathrm{O} / \mathrm{O}_\sun=0.2$ as found by \citet{2013A&A...555A...1B}, we find that the emission line O\,\textsc{iv} $\lambda 4654$ is over-estimated. We instead implement an abundance of $\mathrm{O} / \mathrm{O}_\sun=0.09$ to explain this line better. However, this then leaves the O\,\textsc{iv}\,$\lambda\lambda\,1337\,1343\,1344$ absorption too weak. This line is heavily affected by the microturbulence.  \citet{2013A&A...555A...1B} experimented with values of $\xi_\mathrm{min} \geq 25 \mathrm{km\,s}^{-1}$ but we are unable to fit their value with other lines. With the lower O abundance we adopt due to the observed O\,\textsc{iv} $\lambda\lambda\,7105,\,7111,\,7123$ triplet emission we find $\xi_\mathrm{min}=35 \mathrm{km\,s}^{-1}$ to match O\,\textsc{iv}\,$\lambda\lambda\,1337\,1343\,1344$.

Despite the lower O abundance, we are not able to match the O\,\textsc{v} $\lambda 1371$ wind line even with $D=100$. \citet{2013A&A...555A...1B} reproduce this line  with their higher O abundance and the wind parameters $\log\,\dot{M} =-6.74$, $D =50$,  and  $\varv_\infty=2800 \,\mathrm{km\,s^{-1}}$. We find even with $\log\,\dot{M} =-6.5$ and $D =100$ that, while we match the C\,\textsc{iv}\,$\lambda\lambda\,1548.2,\,1550.8$ wind profile, we are still unable to match the O\,\textsc{v} $\lambda 1371$ (Fig.\,\ref{fig:SSN_9_fitting}). By invoking a high density contrast ($D=200$) we find a match to all UV wind lines and H$\alpha$ with  $\log\,\dot{M} =-6.65$. We were unable to match the slope of the 
blue wing of the N\,\textsc{v} $\lambda\lambda\,1238.8,\,1242.8$ doublet in any model. We speculate that a model which includes X-rays might better describe the shape of this line.

While there is evidence of a shift in the He\,\textsc{ii} lines in the MUSE observations (Rickard et al. in prep), it was not noted by \citet{2019AA...626A..50D} as a spectral binary and there are no spectral lines that could be attributed to a secondary. As a result, the wind lines are entirely attributed to the primary.

At the bluest extent of the C\,\textsc{iv}\,$\lambda\lambda\,1548.2,\,1550.8$ wind profile, a NAC is present. Due to the high $\varv_\infty$ of this wind, the NAC is close to the Si\,\textsc{ii} $\lambda\,1533$ line but is of a greater strength and centered redward. SSN~9 has been observed with E140M grating as a part of the ULLYSES programme, enabling high resolution multiple epoch spectral comparisons  (Fig.\,\ref{fig:SSN_9_multiepoh}). This NAC feature can be seen to be shifting blueward on the timescale of hours.\\  

\subsubsection{SSN~11 (MPG~342)}


The He\,\textsc{i} $\lambda\,4471$ and He\,\textsc{ii} $\lambda\,4542$ lines in the FLAMES spectra of SSN~11 show at least two distinct spectral line features. \citet{2012ApJ...748...96M} mentioned a third component, confirmed by \citet{2019AA...626A..50D} who show three components in both the He\,\textsc{ii} $\lambda\,4542$ and He\,\textsc{ii} $\lambda\,4685.8$ lines within the AAT-UCLES spectra they present. With 11 MUSE observations over $\sim 10$ days, we see three components in He\,\textsc{i} and He\,\textsc{ii} lines shifting with large ($\ge 250\, \mathrm{km\,s}^{-1}$) RV changes over time. This is evident in lines such as the He\,\textsc{i} $\lambda\,5875.6$ and He\,\textsc{i} $\lambda\,7065.3$ lines.

Due to SSN~11 being an SB3, the $T_\ast$ and $\log g$ fit to the SMC grid cannot be relied on. The fit temperature we find is not consistent with the  N\,\textsc{iii} $\lambda\lambda\,4634,\,4641$ doublet emission. We do not attempt a single star fit for SSN~11 as we cannot reasonably believe our $T_\ast$ and $\log g$ is representative of the primary. We will present a three body solution to SSN~11 in a future paper (M. Rickard, in prep). \\

\begin{figure}
  \center
  \resizebox{0.95\hsize}{!}{\includegraphics[trim={0.7cm 21.013cm 10.3cm 2.19cm},clip]{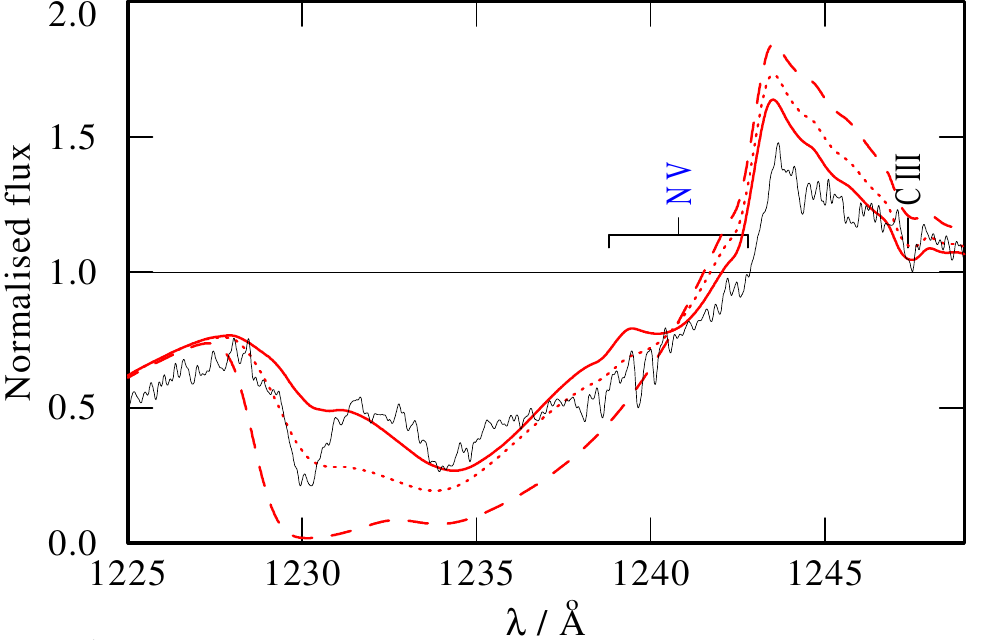}}
 \resizebox{0.95\hsize}{!}{\includegraphics[trim={10.7cm 14.40cm 0.3cm 8.66cm},clip]{SSN_0013_p3.pdf}}
  \resizebox{0.95\hsize}{!}{\includegraphics[trim={10.7cm 21.013cm 0.3cm 2.19cm},clip]{SSN_0013_p3.pdf}}
\resizebox{0.95\hsize}{!}{\includegraphics[trim={0.7cm 14.40cm 10.3cm 8.66cm},clip]{SSN_0013_p3.pdf}}
  \caption{Spectral observations of wind lines of SSN~13, normalised and compared to various models. The observed spectra of SSN~13 is shown by black line. Synthetic spectra with $\log\,\dot{M}=-7.4$ and $D=20$ shown by solid red line. The dotted red line shows model spectra with $\log\,\dot{M}=-7.2$ and $D =10$ while the dashed red line shows the spectra with $\log\,\dot{M}=-6.7$ and $D=1$.  All shown models have $\varv_\infty=2300 \,\mathrm{km\,s^{-1}}$.}
  \label{fig:SSN_13_windcomparison}
\end{figure}

\subsubsection{SSN~13 (MPG~324)}

This star has a spectral type of O4\,V((f)) from \citet{Walborn2000} and we use the rotational velocity $\varv \sin i=113\mathrm{km\,s}^{-1} $ from \citet{2019AA...626A..50D}. We determine $T_\ast=41\,\mathrm{kK}$ and $\log\,g=3.8$ from our MUSE observations. \citet{2003ApJ...595.1182B} find a higher temperature due to the balance of C\,\textsc{iii}\,$\lambda\,1176$ to  C\,\textsc{iv}\,$\lambda\,1169$. PoWR models also favour slightly higher temperatures when considering the carbon UV lines, so we adopt $T_\ast=42\,\mathrm{kK}$.

\citet{2003ApJ...595.1182B} show a model fit to the E140M spectra that we have used to aid our analysis. They require an N abundance of N/$\mathrm{N}_\sun=0.2$. We find agreement with PoWR, with this abundance being the best fit for both the N\,\textsc{iii} $\lambda\lambda\,1183,\,1185$ doublet, the N\,\textsc{iv}\,$\lambda\,1718$ absorption lines, and the N\,\textsc{iii} $\lambda\lambda\,4634,\,4641$ doublet emission and the  N\,\textsc{iii} triplet $\lambda\lambda\,7105,\,7111,\,7123$ emission lines in MUSE, which were not available for the \citet{2003ApJ...595.1182B} analysis.

The wind parameters from \citet{2003ApJ...595.1182B} are $\log\,\dot{M}=-7.00$, $\beta=1.0$, $D=10$, and $\varv_\infty=2300\,\mathrm{km\,s^{-1}}$. Our PoWR model fit to the observed wind profiles gives $\log\,\dot{M} =-7.4$, $\beta=1.0$, and $\varv_\infty=2300\,\mathrm{km\,s^{-1}}$, with a higher clumping factor of $D=20$. The small difference in mass-loss rate can be attributed to the difference of $0.2 \,\mathrm{dex}$ in $\log\,g$ and $500\,\mathrm{K}$ in $T_\ast$ between our stellar parameters and those used by \citet{2003ApJ...595.1182B}.

We assume that the deep absorption feature seen at the blue extent of the N\,\textsc{v}\,$\lambda\lambda\,1238.8,\,1242.8$ absorption trough is a NAC. Thus when adjusting mass-loss rate, we select model that does not produce as deep a profile as would be required to match the depth of the NAC (Fig.\,\ref{fig:SSN_13_windcomparison}).\\

\begin{figure}
  \center
  \resizebox{\hsize}{!}{\includegraphics[trim={4cm 3cm 4cm 0cm},clip]{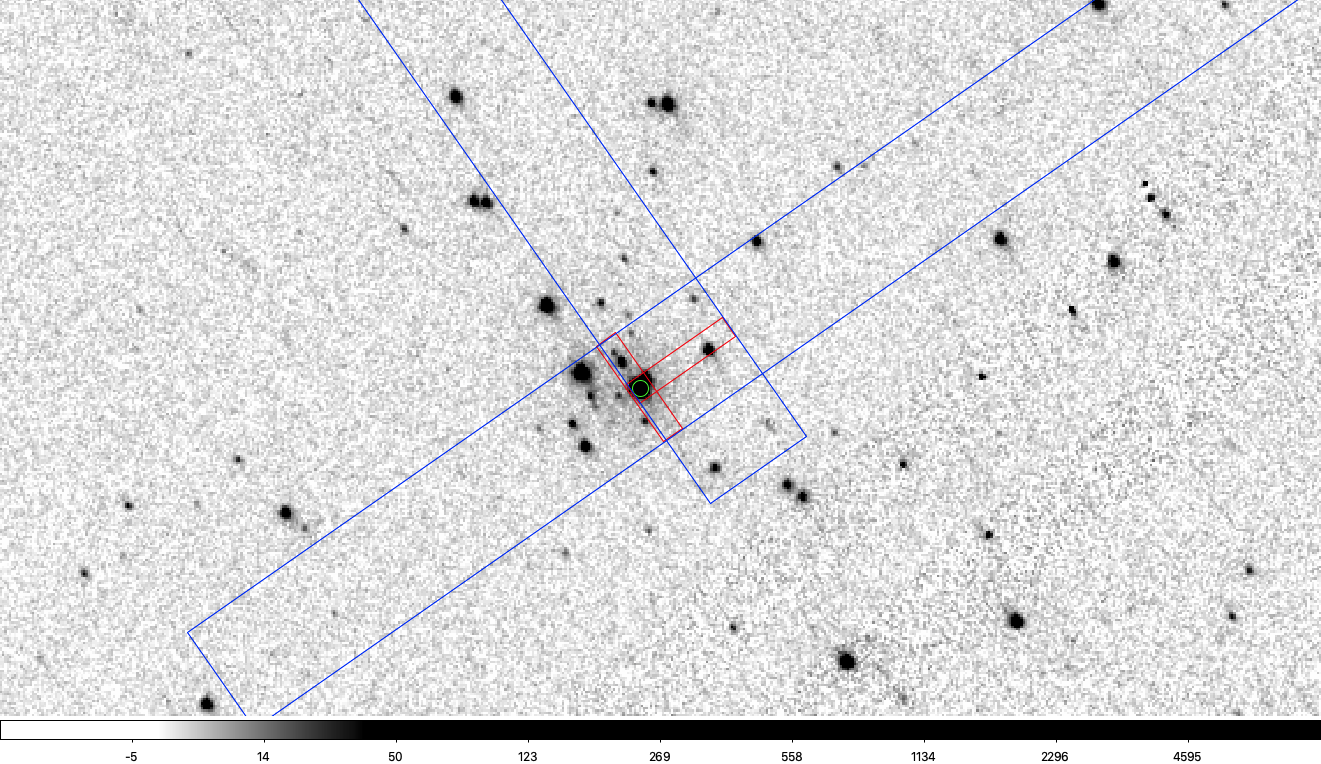}}
   \caption{Area on the sky around SSN~14. Blue boxes show the approximate positions of the two G140L slits with $2\arcsec$ widths, while red boxes show the position along the slit from where the spectra of SSN~14 were extracted. Background is the F225W {\em  HST} image, flux is log scaled.}

  \label{fig:SSN_14_overlap_extraction}
\end{figure}

\begin{figure}
  \center
  \resizebox{\hsize}{!}{\includegraphics[ trim={10.7cm 21.013cm 0.3cm 2.19cm},clip]{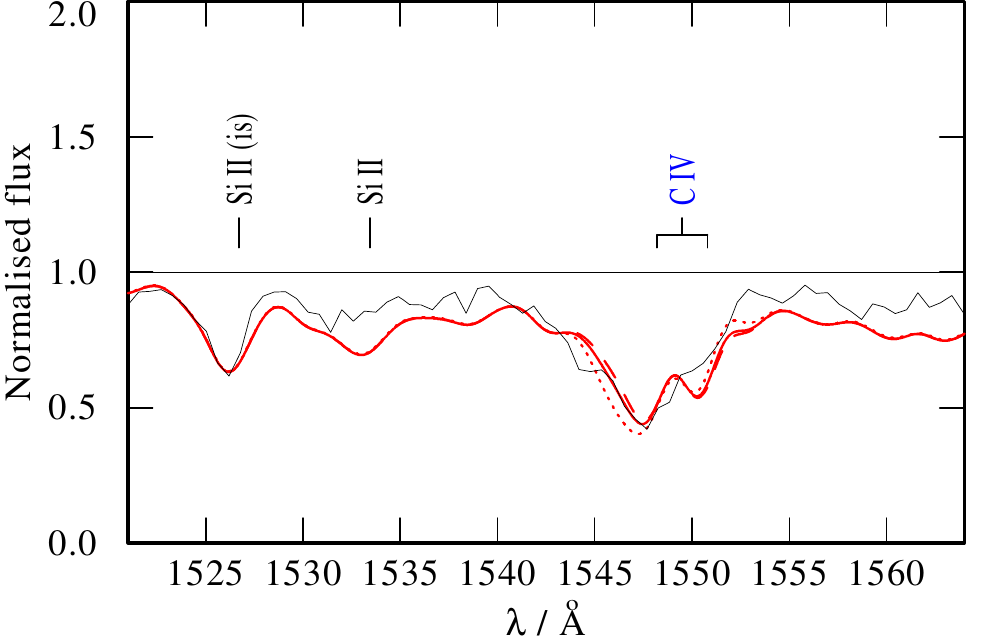}}
  \caption{Spectral observations of wind lines of SSN~14, normalised and compared to various models. The observed spectra of SSN~14 is shown by black line. Synthetic spectra with $\log\,\dot{M}=-9.5$ shown by solid red line. The dotted red line shows model spectra with $\log\,\dot{M}=-9.3$ while the dashed red line shows the spectra with $\log\,\dot{M}=-9.6$.  All shown models have $\varv_\infty=600 \,\mathrm{km\,s^{-1}}$ and $D=20$.}  
  \label{fig:SSN_14_comparison}
\end{figure}

\subsubsection{SSN~14 (MPG~470)}

This source is in a very crowded field (Fig.\,\ref{fig:SSN_14_overlap_extraction}). We believe that SSN~14 is MPG~470 as both represent the brightest UV sources in their respective surveys within the tight group, as seen in Fig.\,\ref{fig:SSN_14_overlap_extraction}, though \citet{2007AJ....133...44S} do not make this connection. MPG~470 has previously been given a classification of O8.5III \citep{1989AJ.....98.1305M}, but this classification is not supported by the wings of the observed MUSE He\,\textsc{ii}\,$\lambda\,4860$ line which fit to a 
high $\log\,g =  4.4$.

SSN~14 was observed twice in G140L long slit mode (Fig.\,\ref{fig:SSN_14_overlap_extraction}). The SEDs from these observations have different slopes, though they both agree at the blue end of the UV. This would suggest that SSN~14 is the main contributor in UV. From the long slit sloping SE to NW, we extract two additional UV bright objects on the axis, such that these two objects are contaminants in the extraction of SSN~14 in the NE to SW long slit. We subtract these two flux calibrated spectra from the spectra of SSN~14. As these containments are shifted in the dispersion direction, they are shifted in to their observed frames based on the morphology of the ISM lines. The resultant observed SED of SSN~14 give the luminosity as $L / L_\sun=5.37 \pm 0.05$. We assign the spectral type of O8\,V in this work. The fit to grid models yields $T_\ast=33\,\mathrm{kK}$ and $\log\,g=4.4$. 


This star is a known eclipsing binary with a period 9.7\,d \citep{2016AcA....66..421P}. However both the UV spectra taken in 2001 and in 2018 have nearly identical UV wind features. In the absence of a time series of spectra we model it as a single star. Both spectra show no wind profiles in any UV lines with the exception of C\,\textsc{iv}\,$\lambda\lambda\,1548.2,\,1550.8$ which shows a very slight blueward asymmetry up to $600\,\mathrm{km\,s^{-1}}$ present in both spectra.

There is no emission in neither the  N\,\textsc{iii} $\lambda\lambda\,4634,\,4641$ doublet, nor in the  N\,\textsc{iii}\,$\lambda\lambda\,7105,\,7111,\,7123$ lines, suggesting that there is no surface N enrichment.  For the UV spectra extracted from the long slits, contamination of other sources along the width of the slit means that the narrow UV metals lines are hard to attribute solely to SSN~14. A bright object also within the width of the slit will contribute to the spectrum observed but will be shifted in the wavelength direction. We attribute the broadening of the various N and C photospheric UV lines to this. As such, we do not use UV lines for abundance fitting and in the absence of any evidence for N enrichment in the MUSE spectra, we adopt the grid abundances. 

The slight blueward asymmetry in the C\,\textsc{iv}\,$\lambda\lambda\,1548.2,\,1550.8$ wind profile can be matched with the synthetic spectrum with  $\log\,\dot{M}= -9.5$ and $\varv_\infty=600\,\mathrm{km\,s^{-1}}$. A density contrast $D=10$ produces slight N\,\textsc{v}\,$\lambda\lambda\,1238.8,\,1242.8$ blueward absorption, not present in the observations, so a slightly increased density of  $D=20$ is invoked 
(Fig.\,\ref{fig:SSN_14_comparison}). This slight blueward asymmetry could be attributed to another object within the slit. However, due to the depth of the feature, it would require the contaminating object to be a significant contributor to the UV flux. No such sources are seen in the {\sc HST} F225W image.\\

\begin{figure}
  \center
  \resizebox{0.95\hsize}{!}{\includegraphics[trim={0.7cm 21.013cm 10.3cm 2.19cm},clip]{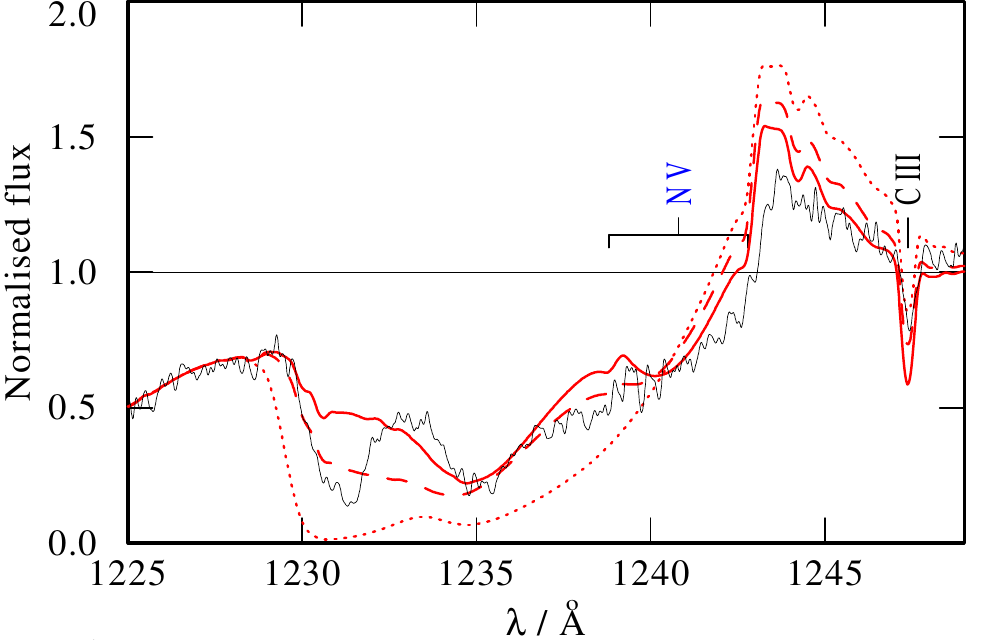}}
  \resizebox{0.95\hsize}{!}{\includegraphics[trim={10.7cm 14.50cm 0.3cm 8.66cm},clip]{SSN_0015_p3.pdf}}
  \resizebox{0.95\hsize}{!}{\includegraphics[trim={10.7cm 21.013cm 0.3cm 2.19cm},clip]{SSN_0015_p3.pdf}}
  \resizebox{0.95\hsize}{!}{\includegraphics[trim={0.7cm 14.50cm 10.3cm 8.66cm},clip]{SSN_0015_p3.pdf}}
  \caption{Spectral observations of wind lines of SSN~15, normalised and compared to various models. The observed spectra of SSN~15 is shown by the black line. Synthetic spectra with $\log\,\dot{M}=-7.6$ and $D=20$ are shown by solid red line. The dotted red line shows model spectra with $\log\,\dot{M}=-7.0$ and $D =1$ while the dashed red line shows the spectra with $\log\,\dot{M}=-7.5$ and $D=10$.  All shown models have $\varv_\infty=2100 \,\mathrm{km\,s^{-1}}$.}
  \label{fig:SSN_15_4comparison}
\end{figure}

\subsubsection{SSN~15 (MPG~368)}

\citet{2003ApJ...595.1182B} assign this star the spectral type of O4–5 V((f*)). Rotational broadening, $\varv \sin i=58\,\mathrm{km\,s}^{-1}$ is from \citet{2019AA...626A..50D}. From our MUSE observations, we find $T_\ast=40\,\mathrm{kK}$ and $\log\,g=4.0$. This temperature is similar to that found by  \citet{2003ApJ...595.1182B}, however they find a lower  $\log\,g=3.75$. To reproduce the minimal emission in  N\,\textsc{iii}\,$\lambda\lambda\,7105,\,7111,\,7123$ seen in the MUSE spectra, we reduce the temperature to $T_\ast=39\,\mathrm{kK}$ and hence avoid this line emission produced in all hotter models.

We adopt an abundance of N/$\mathrm{N}_\sun=0.6$, in agreement with \citet{2003ApJ...595.1182B}. This produces the best fit for N\,\textsc{iii}\,$\lambda\lambda\,1183,\,1185$ , N\,\textsc{iv}\,$\lambda\,1189$, and  N\,\textsc{iv}\,$\lambda\,1718$ in the UV, in addition to showing emission in N\,\textsc{iii}\,$\lambda\lambda\,4634,\,4641$. All other lines are best fit with the abundances used in the grid models, as detailed in Table\,\ref{table:Abundances_Ions}.

We find the best matching wind parameters of $\log\,\dot{M}=-7.6$, $\beta=1.0$, $D=20$, and $\varv_\infty=2100 \,\mathrm{km\,s^{-1}}$ (Fig.\,\ref{fig:SSN_15_4comparison}). This mass-loss rate is lower than $\log\,\dot{M}=-7.24$, reported by \citet{2003ApJ...595.1182B}. Similar to \citet{2003ApJ...595.1182B}, for measuring wind parameters we exclude the NAC 
within the N\,\textsc{v} and C\,\textsc{iv} absorption troughs.\\

\begin{figure}
  \center
  \resizebox{\hsize}{!}{\includegraphics[trim={0.0cm 17.765cm 10.0cm 0cm},clip]{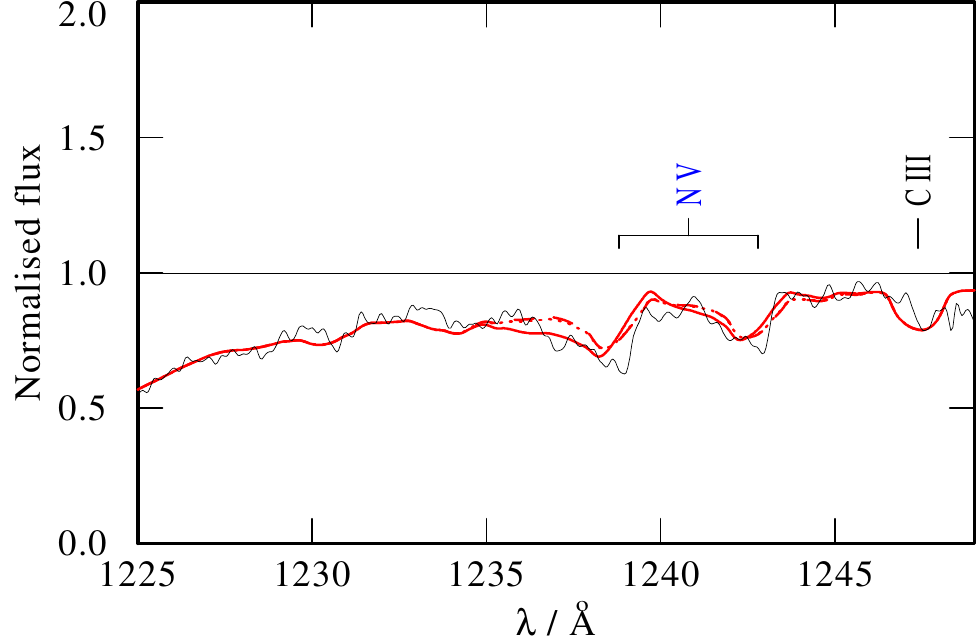}}
  \resizebox{\hsize}{!}{\includegraphics[trim={10.0cm 17.765cm 0.0cm 0cm},clip]{SSN_0017_p3.pdf}}
  \resizebox{\hsize}{!}{\includegraphics[trim={0.0cm 11.2cm 10.0cm 6.5cm},clip]{SSN_0017_p3.pdf}}
  \caption{Normalised spectral observations of wind lines of SSN~17, normalised and compared to various models. The observed spectra of SSN~17 is shown by the black line. Synthetic spectra with $\log\,\dot{M}=-8.7$ and $D=1$ shown by the solid red line. The dotted red line shows model spectra with $\log\,\dot{M}=-9.0$ and $D =5$ while the dashed red line shows the spectra with $\log\,\dot{M}=-9.0$ and $D=10$.  All shown models have $\varv_\infty=1000 \,\mathrm{km\,s^{-1}}$.}
  \label{fig:SSN_17_3comparison}
\end{figure}

\subsubsection{SSN~17 (MPG~396)}

The spectral type of O7\,V has been assigned by \citet{1989AJ.....98.1305M}. From our MUSE observations $\log\,g=4.0$ and $T_\ast=37\,\mathrm{kK}$. \citet{2019AA...626A..50D} gives $\varv \sin i=196 \,\mathrm{km\,s}^{-1}$ based on the He\,\textsc{i}\,$\lambda\,4471$ line. There is no evidence for CNO enrichment in the MUSE observations so we adopt the grid abundances.

The C\,\textsc{iv}\,$\lambda\lambda\,1548.2,\,1550.8$ wind profile is not seen in the UV spectra. A slight blueward asymmetry is evident in the N\,\textsc{v}\,$\lambda\lambda\,1238.8,\,1242.8$ profile. This is confirmed by the comparison to a model with lower mass-loss rate ($\log \dot{M}=9.0$) which featured no synthetic blueward asymmetry (Fig.\,\ref{fig:SSN_17_3comparison}). A model with $D=10$ could not replicate this lack of C\,\textsc{iv}\,$\lambda\lambda\,1548.2,\,1550.8$ wind profile along with the slight observed N\,\textsc{v}\,$\lambda\lambda\,1238.8,\,1242.8$ wind profile, so instead $D=1$ (homogeneous winds) and $D=5$ models were tried. The best fit wind parameters were therefore $D=1$, $\log\,\dot{M} =-8.7$, and $\varv_\infty=1000 \,\mathrm{km\,s^{-1}}$.\\

\begin{figure}
  \center
  \resizebox{\hsize}{!}{\includegraphics[ trim={0.7cm 13.8cm 10.3cm 2.19cm},clip]{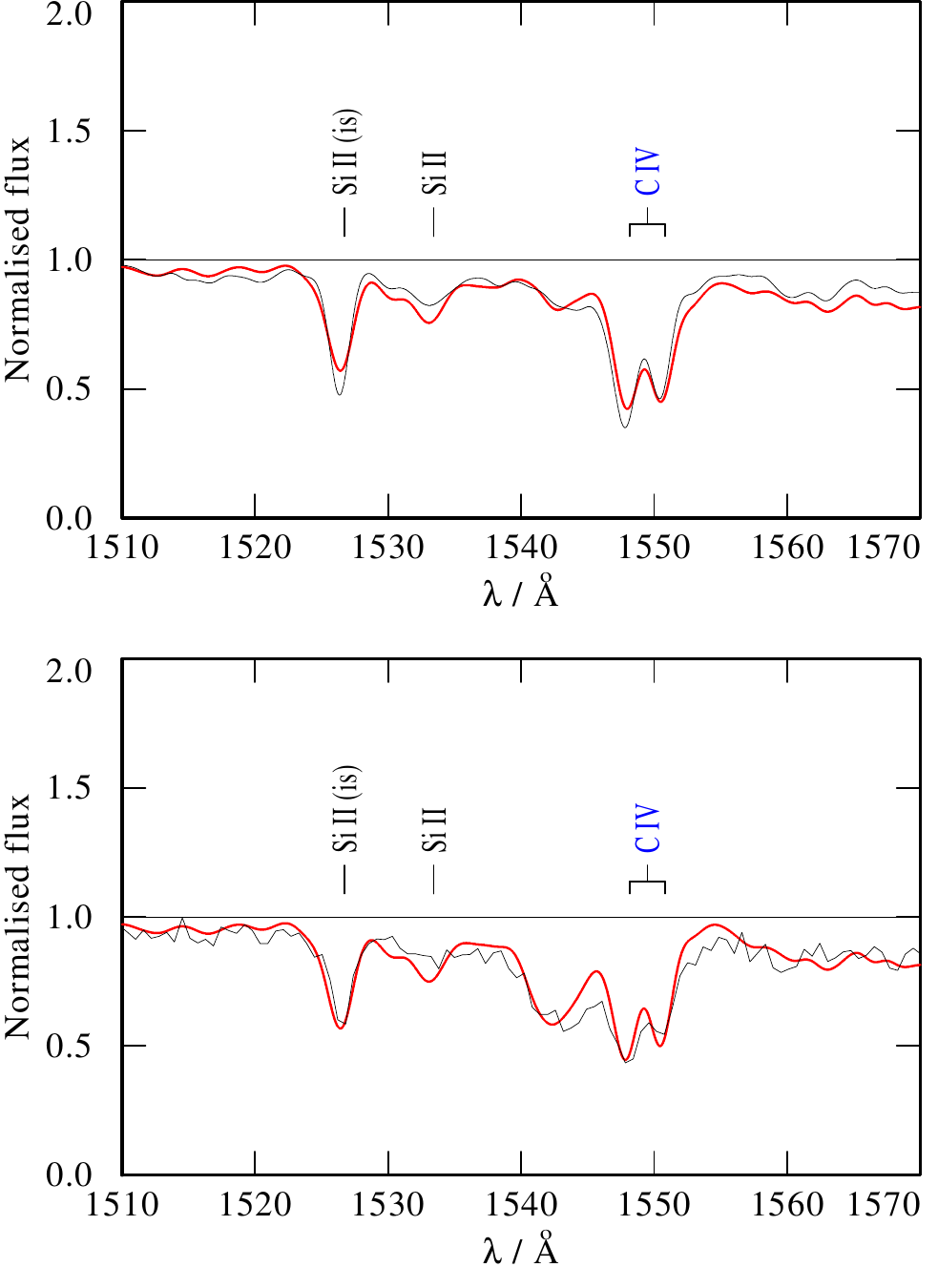}}
   \caption{Two seperate normalised spectral observations of the C\,\textsc{iv}\,$\lambda\lambda\,1548,\,1551$ wind lines of SSN~18, normalised and compared to various models. The observations are shown by the black lines. 
   Red lines are models. {\em Top panel}: high resolution E140M observations of SSN~18 (obtained on 28-09-2000) compared to a model with  $\log\,\dot{M}=-9.1$, convolved to the same resolution as the model in the lower panel. {\em Lower panel}: low resolution G140L spectrum of the same target (obtained on 30-12-2018) compared to a model with $\log\,\dot{M}=-8.45$.}
  \label{fig:SSN_18_LOW_HIGH_MOD_COMP}
\end{figure}

\begin{figure}
  \center
  \resizebox{\hsize}{!}{\includegraphics{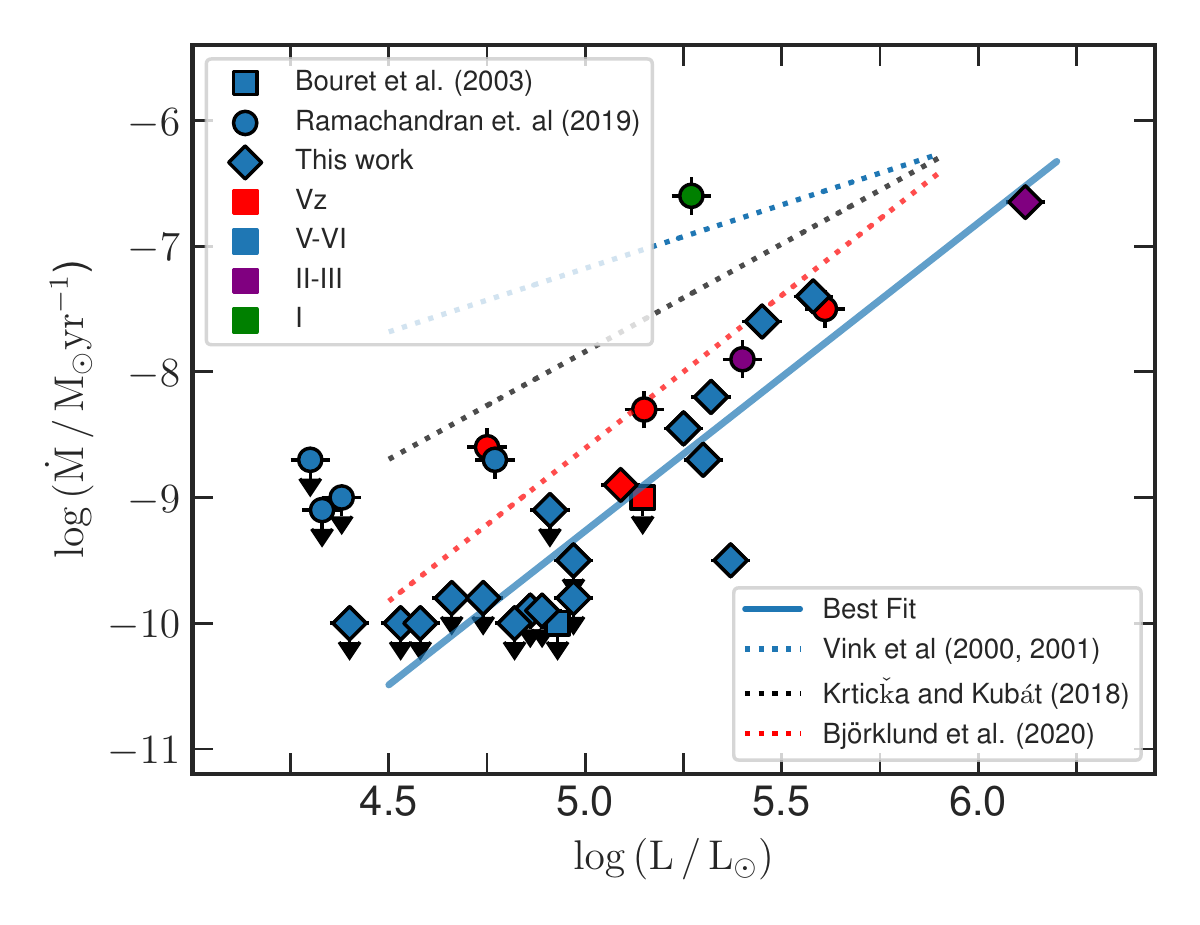}}
   \caption{Mass-loss rate as a function of stellar luminosity for the analysed O stars in NGC~346. Luminosity classes are distinguished by different colours as given in the legend. A power-law fit to the empirical results is represented by the blue solid line (see Sec.~\ref{sec:Mdot_L_relation} for details). For comparison, we include stars with $\dot{M}$ from UV observations from \citet{2019A&A...625A.104R} (circles), two additional objects within NGC~346 with $\dot{M}$ from \citet{2003ApJ...595.1182B} (squares) and the theoretical SMC predictions from \cite{2000A&A...362..295V, 2001A&A...369..574V} (dashed blue line), \citet{Krticka2018} (black dashed line) and \citet{2021A&A...648A..36B} (dashed red line).} 
  \label{fig:logMdot_logL}
\end{figure}

\begin{figure}
  \center
  \resizebox{\hsize}{!}{\includegraphics{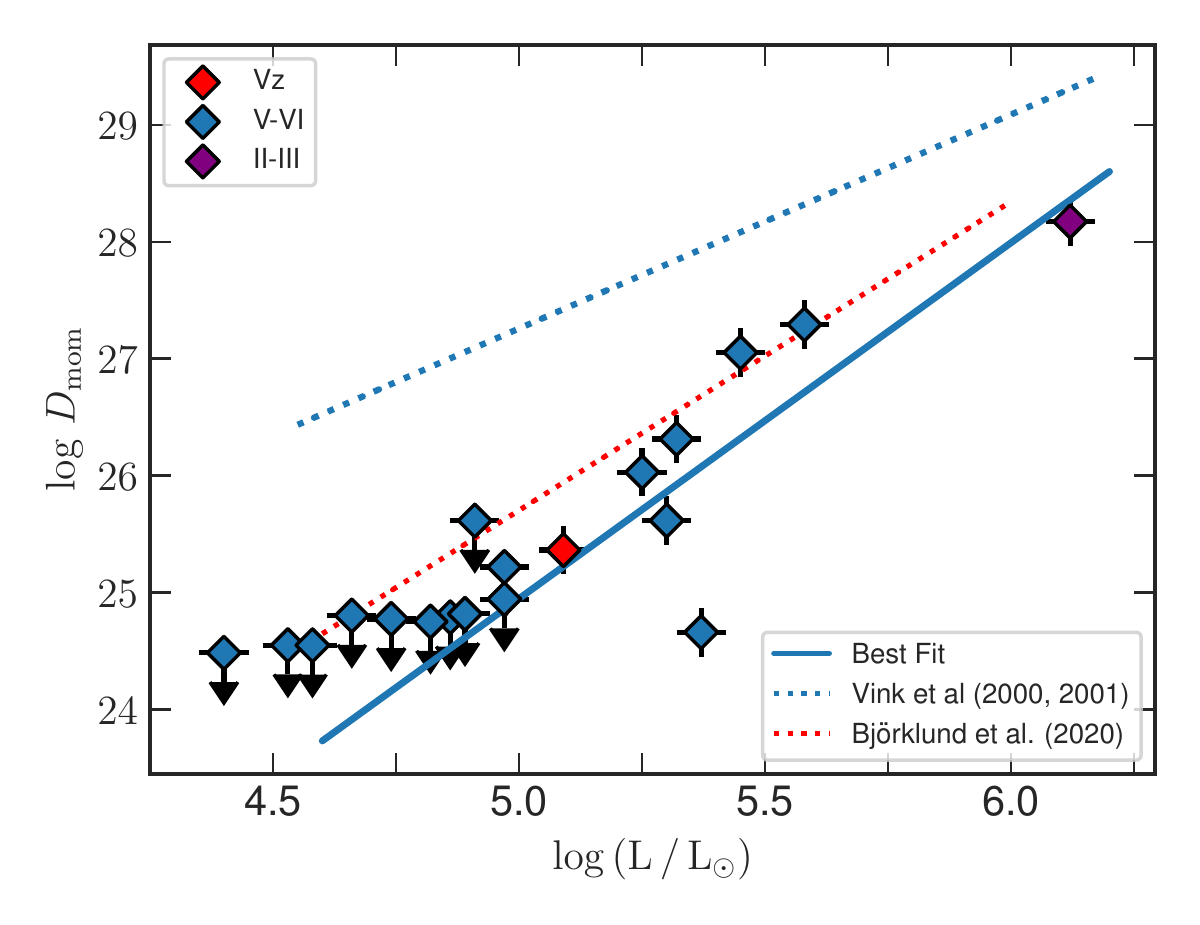}}
   \caption{Modified wind momentum $D_\mathrm{mom}$ 
   as a function of luminosity. Legend as Fig.\,\ref{fig:logMdot_logL}. The WLR to the empirical results is represented by the blue solid line. For comparison, theoretical WLR predictions from \citet{2000A&A...362..295V, 2001A&A...369..574V} are included for the SMC (dashed blue line) and the theoretical predictions from \citet{2021A&A...648A..36B} with comparable metallicity (dashed red line).}  
  \label{fig:logL_to_mod_wind_momentum}
\end{figure}

\subsubsection{SSN~18 (MPG~487)}

The spectral type O8\,V is from \citet{2003ApJ...595.1182B} who analysed the echelle E140M  high resolution UV spectrum.  \citet{2019AA...626A..50D} provide $\varv \sin i=131\,\mathrm{km\,s}^{-1}$ from He\,\textsc{ii}\,$\lambda\,4921$. In the E140M spectrum, the  
wind profile in the N\,\textsc{v}\,$\lambda\lambda\,1238.8,\,1242.8$ line is only marginally seen while 
there is no wind profile in C\,\textsc{iv}\,$\lambda\lambda\,1548.2,\,1550.8$.  \citet{2003ApJ...595.1182B} model the UV spectra with a homogeneous wind ($D=1$) and derive $\log\,\dot{M} =-8.52$ with an upper limit on $\varv_\infty \leq 1500\,\mathrm{km\,s^{-1}}$.

We find that in our observations with G140L, the spectrum shows blueward absorption in the C\,\textsc{iv}\,$\lambda\lambda\,1548.2,\,1550.8$ wind line (Fig.~\ref{fig:SSN_18_LOW_HIGH_MOD_COMP}). Our G140L spectrum gives the same luminosity as the E140M spectra, showing there is no contamination in our G140L observation. This is also shown by an inspection of the F225W {\em  HST} image which reveals no sources along the extraction of the G140L spectrum that could cause this feature through contamination.

By fitting the C\,\textsc{iv}\,$\lambda\lambda\,1548.2,\,1550.8$ wind profile in the G140L spectra we find $\log\,\dot{M} =-8.45$, $\beta=0.8$, $\varv_\infty=1500\,\mathrm{km\,s^{-1}}$. Density contrast $D=5$ is needed to match the N\,\textsc{v}\,$\lambda\lambda\,1238.8,\,1242.8$ wind lines simultaneously with the C\,\textsc{iv}\,$\lambda\lambda\,1548.2,\,1550.8$ line. Keeping all other wind parameters constant and only varying the mass-loss rate, we adjust the model to produce synthetic spectra to fit the lack of C\,\textsc{iv}\,$\lambda\lambda\,1548.2,\,1550.8$ wind profile in E140M spectra with $\log\,\dot{M}=-9.1$, that is, a factor of $\sim 4.5$ lower mass-loss rate. This variability is discussed further in Sect.\,\ref{sec:large_scale_wind_profile_variation}.\\

\subsubsection{SSN~22 (MPG~476)}

Spectral type of O6\,V is determined in this work. Our MUSE observations give $\log\,g=4.0$ and $T_\ast=38\,\mathrm{kK}$. \citet{2019AA...626A..50D} derive $\varv \sin i=100\,\mathrm{km\,s}^{-1}$ from He\,\textsc{ii}\,$\lambda\,4921$ line. Similar to SSN~14, this target has a very weak blueshifted absorption in C\,\textsc{iv}\,$\lambda\lambda\,1548.2,\,1550.8$ extending to $1600\,\mathrm{km\,s^{-1}}$. There is a very slight N\,\textsc{v}\,$\lambda\lambda\,1238.8,\,1242.8$ wind profile but a model with a smooth wind ($D=1$) gives a synthetic profile that is too strong in N\,\textsc{v}\,$\lambda\lambda\,1238.8,\,1242.8$ while the assumed clumped wind with $D=10$ does not produce a strong enough synthetic profile. A density contrast of $D=5$ along with the other wind parameters of $\log\,\dot{M} =-8.2$, $\beta=1.0$, $\varv_\infty=1600\,\mathrm{km\,s^{-1}}$ is required to match the UV and H$\alpha$  observations.\\

\subsubsection{SSN~31 (MPG~417)}

SSN 31 has a spectral type O7.5\,Vz which is determined in this work. We adopt $\varv \sin i=98\,\mathrm{km\,s}^{-1}$ from \citet{2019AA...626A..50D}. As with other mid-O type stars in our sample, no wind profiles are seen in the UV spectra with the exception of C\,\textsc{iv}\,$\lambda\lambda\,1548.2,\,1550.8$ which shows a very slight blueward asymmetry, in this case up to $1000\,\mathrm{km\,s^{-1}}$. The observed spectrum is reproduced by a model with $\log\,\dot{M}=-8.9$, $\beta=1.0$, and $\varv_\infty=1000\,\mathrm{km\,s^{-1}}$. There is a marginally discernible N\,\textsc{v}\,$\lambda\lambda\,1238.8,\,1242.8$ wind profile, 
and a density contrast of $D=10$ fits both  N\,\textsc{v}\,$\lambda\lambda\,1238.8,\,1242.8$ and the H$\alpha$ absorption well. \\

\subsection{Scaling between mass-loss late and luminosity}
\label{sec:Mdot_L_relation}
Using spectral analyses we determined the stellar wind parameters for eighteen stars. In order to investigate how mass-loss rates depend on luminosity, we perform a power-law fit restricted to
the stars with luminosity class V and Vz.

Stars within the sample with spectral types later than O7-8 (${\log{L/L_\sun} \lesssim 5.0}$) do not show clear evidence of stellar winds, with the exception of a blueward absorption asymmetry in the C\,\textsc{iv}\,$\lambda\lambda\,1548.2,\,1550.8$. Therefore, we were only able do derive an upper limit for their mass-loss rate and they were excluded from the fit. Additionally, we excluded SSN~18 due to its strong wind variability (see Sect.~\ref{sec:large_scale_wind_profile_variation}). The resulting  fit  yields $\dot{M} \propto L^{\parampllogLlogMdot}$ and is shown in Fig.\,\ref{fig:logMdot_logL}. 

In Fig.\,\ref{fig:logMdot_logL} theoretically predicted mass-loss rates of various authors are shown for comparison. It is evident that the recipe of \citet{2000A&A...362..295V, 2001A&A...369..574V}, which is most commonly used in stellar evolutionary models, overestimates the empirically found mass-loss rates. The more recent mass-loss prescription of \citet{Krticka2018} yield somewhat lower mass-loss rates but are still overestimating our results. The predictions of \citet{2021A&A...648A..36B} are closest to our findings, even though they overestimate the mass-loss rates of the least luminous stars in our sample.

\begin{figure}
  \center
  \includegraphics[width=\hsize]{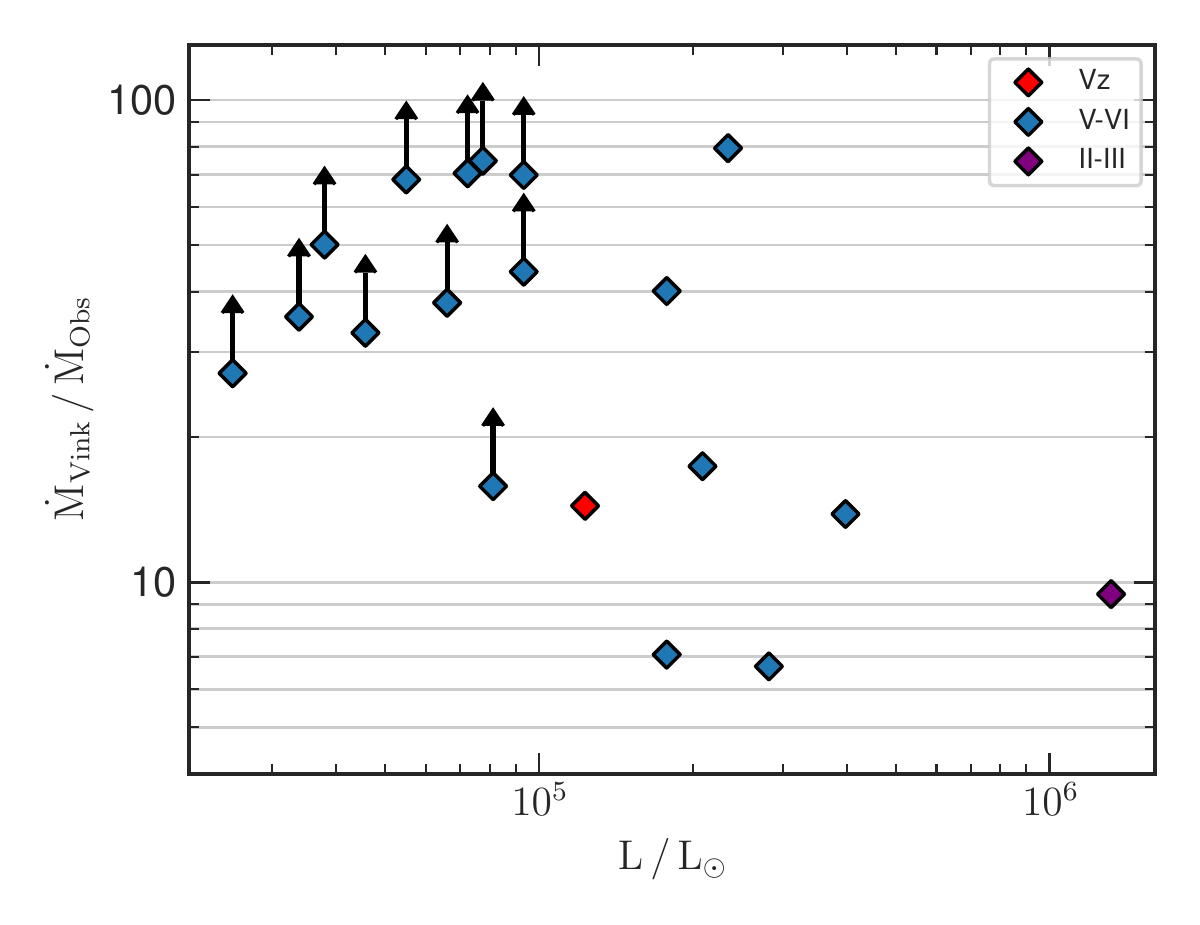}
   \caption{Ratio of theoretical $\dot{M}$  \citep{2000A&A...362..295V, 2001A&A...369..574V} to measured $\dot{M}$ as a function of luminosity. The $\varv_\infty$ for $\dot{M}_\mathrm{vink}$ is calculated from $\varv_\mathrm{esc}$ to ensure consistency.}
  \label{fig:logL_to_Mdot_ratio}
\end{figure}

In order to make our results easier to compare to theoretical prescriptions, we employed a modified wind momentum diagram (see Fig.\,\ref{fig:logL_to_mod_wind_momentum}). The modified wind momentum is defined as ${D_\mathrm{mom}=\dot{M} \varv_\infty R^{\,0.5}}$ and the wind-momentum luminosity relation (WLR) at a given metallicity can be expressed as a power-law of the form.

\begin{equation}
  \label{eq:WLR}
\log\,(D_\mathrm{mom} / (\mathrm{g} \ \mathrm{cm} \  \mathrm{s}^{-2} \ R_\sun^{\,0.5}))=x \ \log\,(L \ / \ L_\sun) + D_\circ,
\end{equation}
where x is the slope of our WRL and $D_\circ$ is the effective number of lines contributing to the wind acceleration and can be seen as a metallicity dependent offset. According to the theoretical predictions of \citet{2000A&A...362..295V, 2001A&A...369..574V} the Galactic, LMC, and SMC WLR all should have a slope of $x=1.83$  and offsets of $D_\circ=18.68$, $18.43$, and $18.11$, respectively. Observational studies of stars in the Galaxy \citep{2004A&A...415..349R} and the LMC \citep{2007A&A...465.1003M,2018A&A...615A..40R} agree with these theoretical WLRs.

Shown in Fig.\,\ref{fig:logL_to_mod_wind_momentum} and discussed in detail in Sect.\,\ref{sec:discussion_metallicity_dependence}, we find for our sample of O stars in the SMC follow a somewhat steeper WLR than the \cite{2000A&A...362..295V, 2001A&A...369..574V} gradient of $x = 1.83$;

\begin{equation}
  \label{eq:log_dmon_bright}
        \log\,(D_\mathrm{mom} / (\mathrm{g} \ \mathrm{cm} \  \mathrm{s}^{-2} \ {R}_\sun^{\,0.5}))=(\paraxallOstars\pm\paraxerrallOstars) \ \log\,(L \ / \ L_\sun) + (\paraDallOstars\pm\paraDerrallOstars).
\end{equation}

Recent theoretical predictions of \citet{2021A&A...648A..36B} are in much better agreement with our empirically found mass-loss rates than those of \citet{2000A&A...362..295V, 2001A&A...369..574V}. However we find that these predictions have a higher  $D_{\rm mom}$ than we measure. By finding a steeper relation at lower $Z$, we support the \citet{2021A&A...648A..36B} addition change from $\dot{M}\propto Z^\alpha$ to $\dot{M}\propto Z^{\alpha(l)}$ such that the exponent $\alpha$ is proportional to $L$. Following this idea, we find 

\begin{equation}
  \label{eq:alpha_L}
    \alpha(L) = \paraalphaLslope \log(L / L_\sun) + \paraalphaLintercept.
\end{equation}

\begin{figure*}
  \center
  \includegraphics[width=4cm, trim={6.5cm 20.8cm 10cm 3cm}]{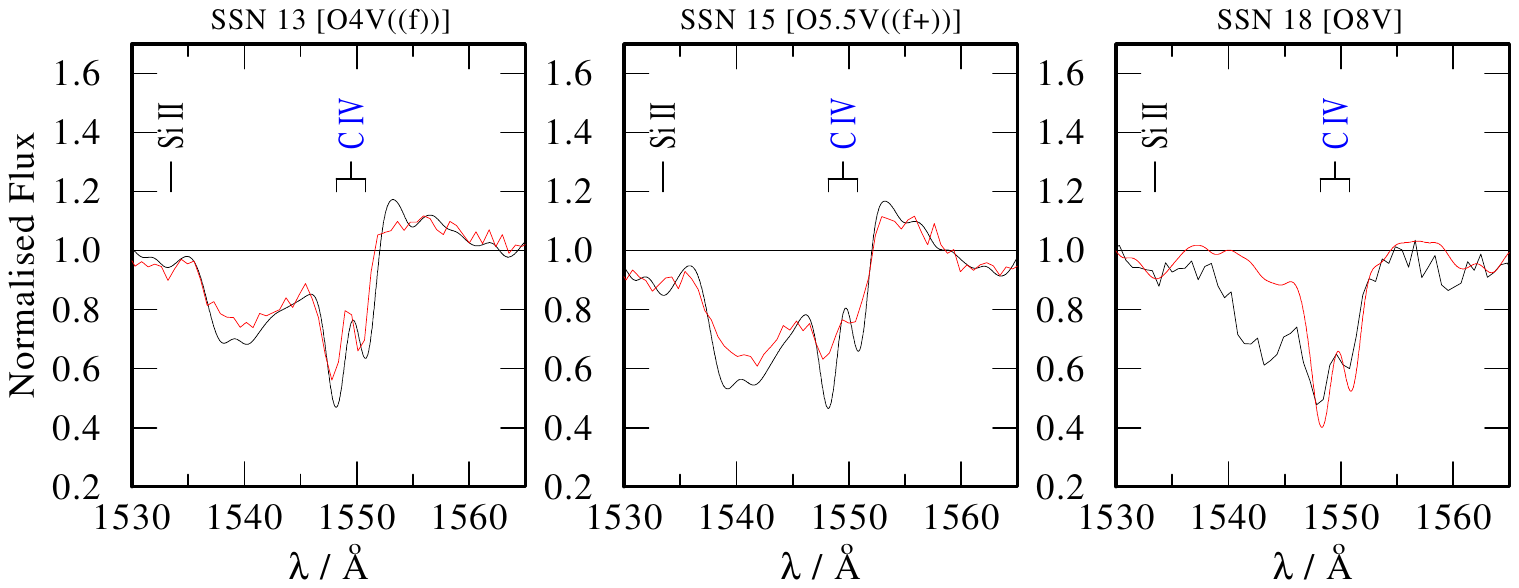}
   \caption{Normalised spectra of C\,\textsc{iv}\,$\lambda\lambda\,1548.2,\,1550.8$ wind lines in  SSN~13, SSN~15, and SSN~18 (from left to right) taken at two different epochs shown by black and red lines. All spectra are convolved with G140L resolution (see text for details). Variability in the wind lines is clearly seen.}
  \label{fig:NGC_346_VARIABILITY}
\end{figure*}

\subsection{Wind line variability}
\label{sec:variability}

Multi-epoch observations are available for six objects. Three stars, SSN~14 (O8\,V), SSN~17 (O7\,V), and SSN~57 (O9.5\,V), are late O types and have no significant wind lines. Three stars, SSN~13 (O6\,V), SSN~15 (O6\,V), and SSN~18 (O8\,V), display line variability (Fig.\,\ref{fig:NGC_346_VARIABILITY}). The strongest line variability can be seen in SSN~18, which has one of the latest spectral types showing wind lines.

Although our sample is small, we tentatively conclude that all stars with wind features in their spectra display spectral variability. This is  further discussed in the Sec.~\ref{sec:large_scale_wind_profile_variation}.

\subsection{Ionising and mechanic energy feedback}
\begin{figure}
  \center
  \includegraphics[width=\hsize]{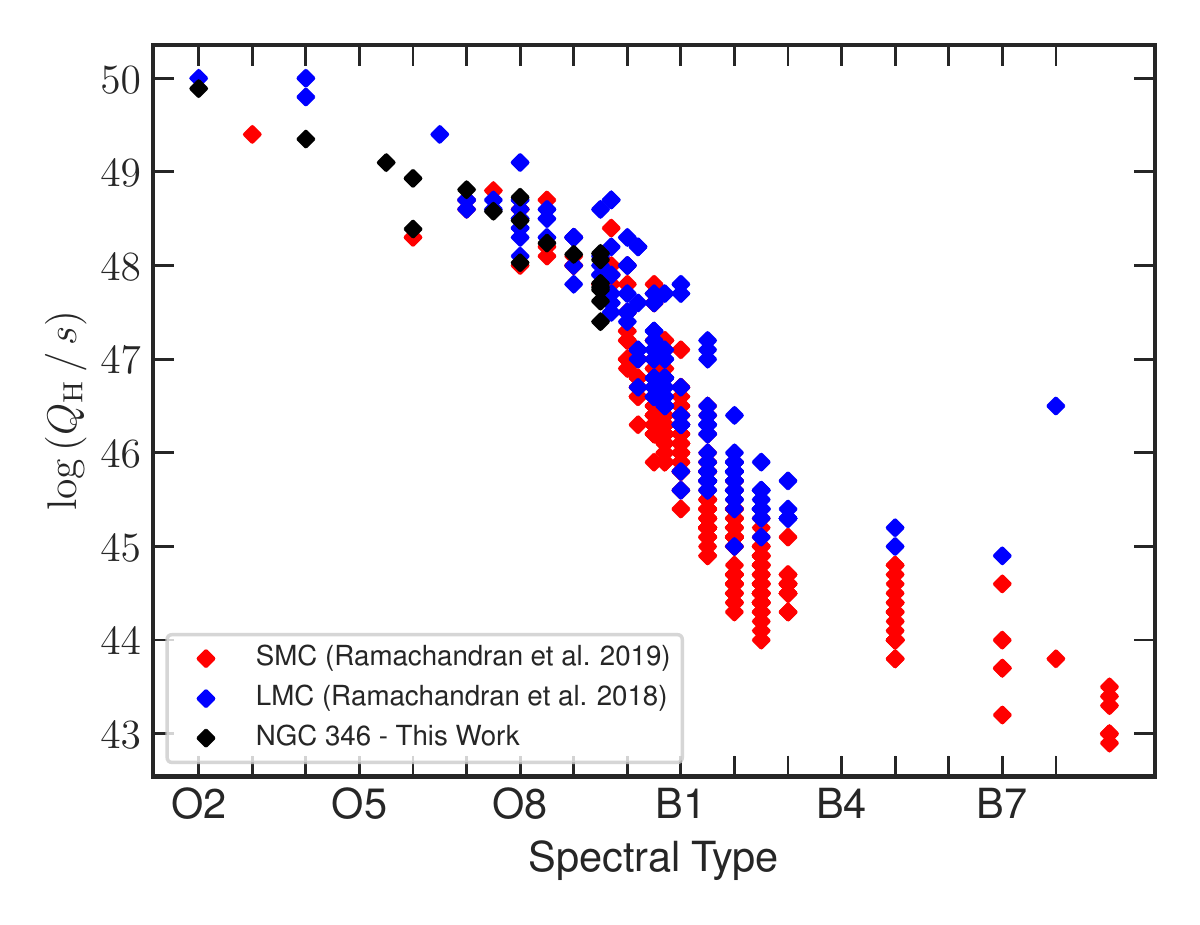}
   \caption{Ionising photon rate as a function of stellar spectral type. The NGC~346 stars are shown by black diamonds. For comparison, the SMC stars analysed by \citet{2019A&A...625A.104R} are included (red triangles) as well as LMC stars from \citet{2018A&A...615A..40R} (blue circles).}
  \label{fig:logQ_to_spt}
\end{figure}

The hydrogen ionising photon rates ($Q_\mathrm{H}$) of each star are obtained from its best fitting PoWR model and are listed in Table\,\ref{table:ionising_photons}. Fig.~\ref{fig:logQ_to_spt} shows $\log{Q_\mathrm{H}}$ of our sample of O stars against spectral type. Previous studies have identified three slopes in this relation; one for O types, another for early B types and a third for types later than B2. In our sample, the relation between spectral type and ionising photon rate is given by ${\log{Q_\mathrm{H}}= \paramQtoSTintercept  \paramQtoSTslope \times \mathrm{ST}}$, where ST refers to the subtype  (e.g.~$4=\mathrm{O}4$). This relation is in line with the result by \citet{2019A&A...625A.104R}. A full table of the ionisation photon rates of the models used for our objects is presented in Appendix~\ref{sec:Ionising_photons_table}.

\begin{figure*}
  \center
  \includegraphics[width=8.5cm]{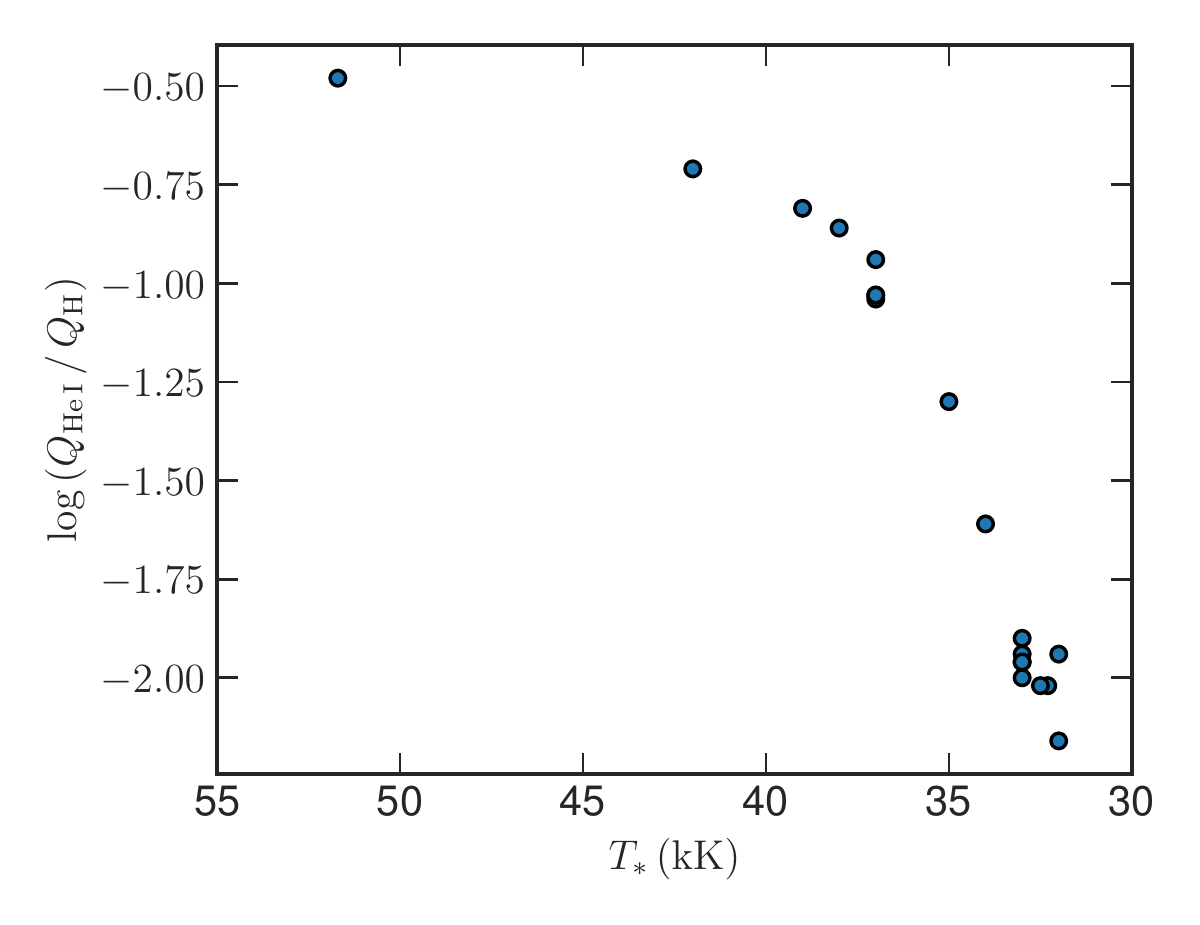}
  \includegraphics[width=8.5cm]{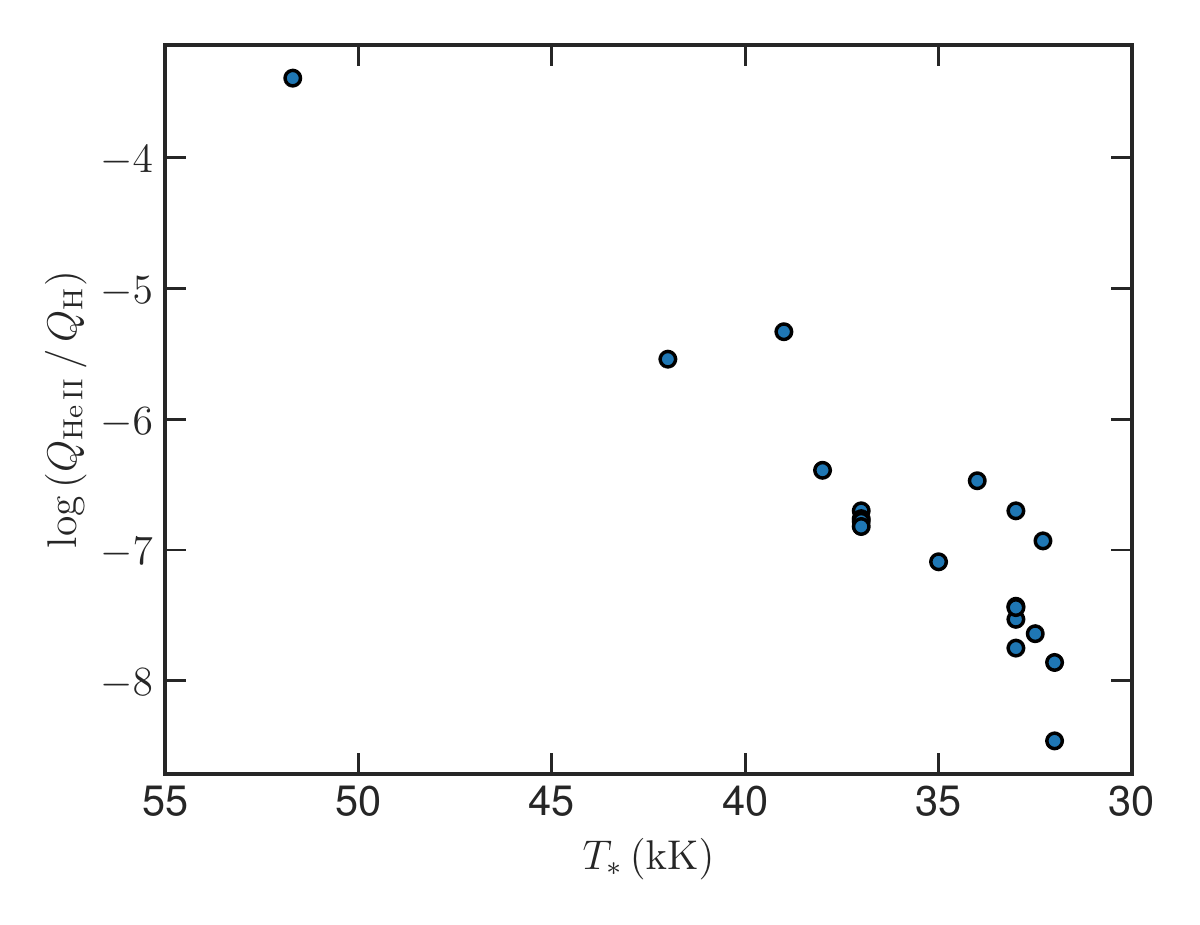}  
   \caption{Ratio of He\,{\sc i} to H ({\em left panel}) and He\,{\sc ii} to H ({\em right panel}) ionising photons of individual stars in our sample as functions of $T_\ast$.}
  \label{fig:logQHE_logQH}
\end{figure*}

From our models we find that, the feedback through ionising radiation in NGC~346 is dominated by SSN~9, which accounts for approximately $49\%$ of the H ionising flux, $71\%$ of the He\,{\sc i} ionising flux and $99\%$ of the He\,{\sc ii}, with no other object in our sample offering any significant contribution to this. This ionising flux is six times greater than the contribution of the next most ionising object, SSN~13.
We note that the binary system SSN~7, which was not part of this study, likely contains two early type stars which might be significant contributors of additional ionising flux.
Nevertheless, we believe that SSN~9 dominates the ionising flux production which is supported by the morphology of the N~66 H\,{\sc ii} region seen in the {\em  HST} images \citep[see top panel in Fig.\,\ref{fig:NGC_346_Obs} and][]{2022arXiv220505147G}. The prominent, crescent structure tracing the photo-dissociation region has SSN~9 in the centre.

\section{Discussion}
\label{sec:discussion}

\subsection{Impact of model assumptions}
\label{sec:discussion_assumptions}
We make multiple assumptions within the fitting process to save computational time and for consistency when analysing different stars from the same cluster. The stellar parameters are determined from a grid with the coarseness of $1000\,\mathrm{kK}$ in $T_\ast$ and $0.2\,\mathrm{dex}$ in $\log{g}$. In addition, we assume turbulent velocity dependence upon spectral type, as well as constant Galactic extinction and abundances for all our sample stars. We vary abundances only when there is clear spectroscopic evidence of CNO enrichment. In this work we neglect the influence of intrinsic X-ray emission from stellar winds, which likely affect the N\,{\sc v} wind lines.

In this work, the stellar wind inhomogeneity is treated using filling factor or  `microcluming' approximation (see Sect.\,\ref{sec:sam}). On the other hand, the strengths of resonance UV lines, and, 
correspondingly, derived $\dot{M}$ are affected by  the porosity (or `macroclumping' effects) 
\citep{2007A&A...476.1331O}. Using their 2D models,  \citet{Sund2010} examined in detail previous suggestions that the strongly non-monotonic wind velocity developing in 1D hydrodynamic simulations could lead  to a reduction of the resonance line strength that is insensitive to spatial scales (this effect is sometimes dubbed `vorocity').  However, the full 3D radiative transfer models revealed that vorocity has only moderate effect on the UV resonance lines \citep{Surlan2012}. 

The clumping factors derived for some of our sample stars are high, implying that the porosity
effects likely play a role in the resonance line formation. We computed a series of models which include porosity effects in both density and velocity, resulting in the increase of the empirically derived mass-loss rates. These will be a subject of our forthcoming publications.

Despite the assumptions made, the difference in the key parameters used for our results, $\log\,(L_ / L_\sun)$ and $\log\,\dot{M} $ between our results and the results in the literature are minor, as shown in Table\,\ref{Table:bouret_comp}. The only significant difference is our model for SSN~13 demonstrates the difference that can result from a discrepancy in $\log\,g$ as small as $0.2$. For SSN~9, while we find a different N and O abundance and a significantly different $D$, but the final mass-loss rate we find is very similar to that found by \citet{2013A&A...555A...1B}.

\begin{table}
\footnotesize
\center
\caption{Comparison of parameters for the stars analysed by \citet{2013A&A...555A...1B} 
using the {\sc cmfgen} code \citep{1998ApJ...496..407H} and for the stars in our sample analysed using the  PoWR code (see text on the discussion of model assumptions) }
\begin{tabular}{c c c c c} 
\hline\hline
\rule[0mm]{0mm}{3.0mm}
 &  \multicolumn{2}{c}{$\log{(L_ / L_\odot)}$} 
 &  \multicolumn{2}{c}{$\log({\dot{M} / \mathrm{M\, yr}^{-1})}$} 
 \\ \hline
SSN & 
{\sc cmfgen}   & PoWR & {\sc cmfgen}   & PoWR \\ \hline
 9	& $6.04$ & $6.12$ & $ -6.74$ & $-6.65$  \\
13	& $5.51$ & $5.58$ & $-6.85$ & $-7.4$  \\
15	& $5.38$ & $5.45$ & $-7.43$ & $-7.6$   \\ \hline
\end{tabular}
\label{Table:bouret_comp}
\end{table}

\begin{figure*}
  \center
  \includegraphics[width=17cm]{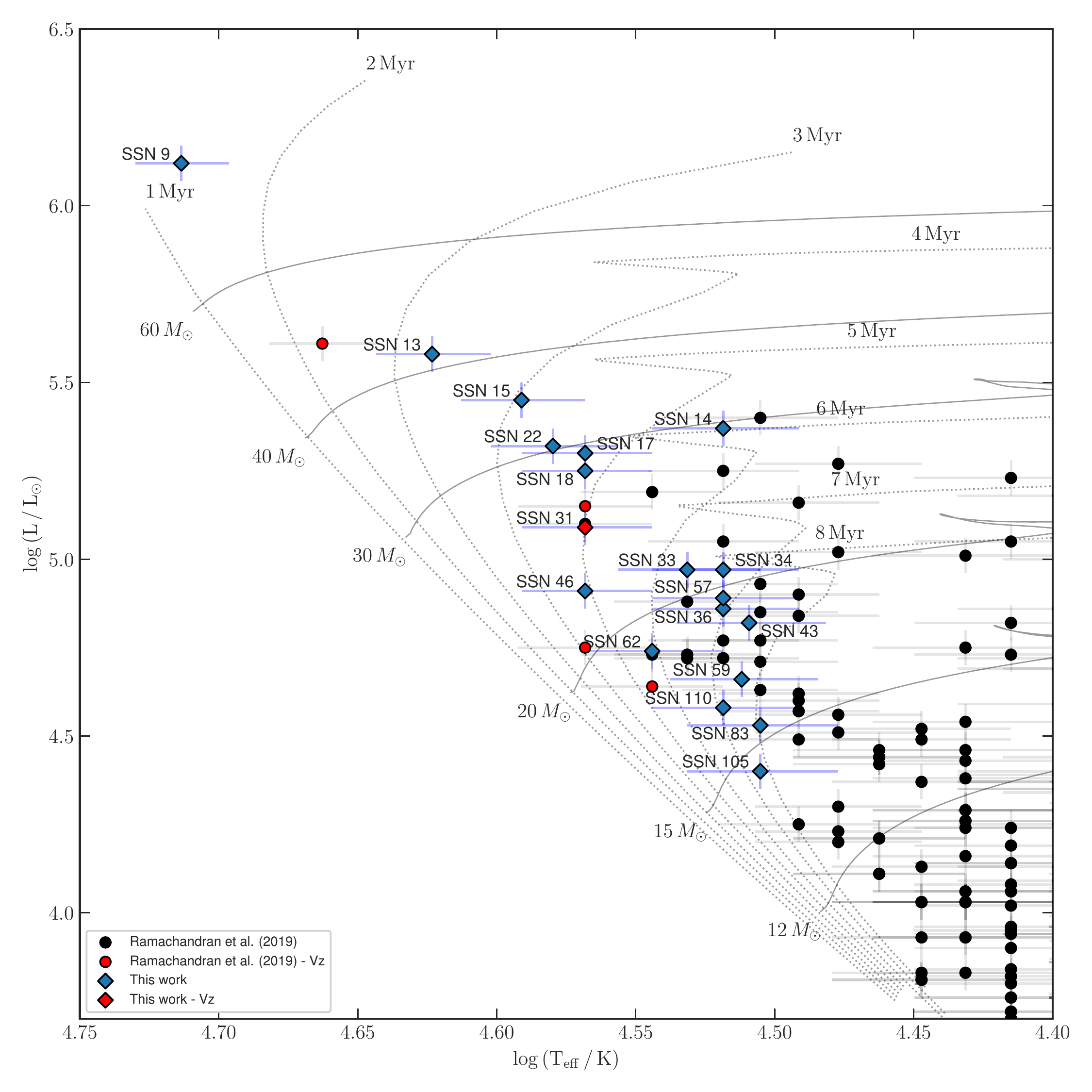}
   \caption{Hertzsprung–Russell diagram for massive stars in the SMC. The objects of NGC~346 studied in this work are shown as diamonds. For completeness, SMC stars from \citet{2019A&A...625A.104R} are illustrated as circles. Objects with luminosity class  Vz  are indicated by red colour. Solid lines are evolutionary tracks at SMC metallicity \citep{2011A&A...530A.115B}, dotted lines are isochrones with $Z=0.002$ \citep{2013A&A...558A.103G}. Isochrones and evolutionary tracks are for non-rotating stars.}
  \label{fig:HR}
\end{figure*}

\subsection{Metallicity dependency and differences between derived mass-loss rate and theoretical predictions}
\label{sec:discussion_metallicity_dependence}

We empirically derive a steeper WLR than predicted theoretically. The previous empirical study by \citet{2006A&A...456.1131M} found rough agreement with the theoretical predictions of \citet{2000A&A...362..295V, 2001A&A...369..574V} but their WLR was based on H$\alpha$ line fitted by smooth wind models in case of dwarf stars and overestimated $\dot{M}$ for these objects.  \citet{2004A&A...420.1087M, 2005A&A...441..735M} find a breakdown of the WLR below $\log \, (L / L_\sun) \sim 5.4$; we do not replicate this result (see Fig.\,\ref{fig:logL_to_mod_wind_momentum}).

\citet{2006A&A...456.1131M} showed that the \citet{2000A&A...362..295V, 2001A&A...369..574V} theoretical $\dot{M}$ predictions do match stars with ${\log{L/L_\sun} \ga 5.8}$ (${\log{\dot{M}} \ga -7.0}$). We only have one star, SSN~9, with this high luminosity. Using the theoretical mass-loss recipe of \citet{2000A&A...362..295V, 2001A&A...369..574V} the star should have a mass-loss rate of  ${\log{(\dot{M}_\mathrm{Vink}/(M_\odot\,\mathrm{yr^{-1}}))}=-5.68}$, which is a factor of ten higher compared to our value. Our sample does not contain enough objects of sufficiently high luminosity to validated the \citet{2000A&A...362..295V, 2001A&A...369..574V} relation in this range. Thus we are unable to explore if this discrepancy is also linked to the wind-metallicity relation as it is described above.

The main reason of our steeper WLR result is the accurate determination of the stellar wind properties of a large sample of low metallicity stars with luminosities below $\log{L/L_\sun}\lesssim 5.8$. From the WLR at different metallicities (Galactic, LMC, and SMC), the wind-metallicity relationship ($\dot{M} \propto Z^\alpha$) is derived. We followed the idea of \citet{2021A&A...648A..36B} and found that $\alpha\propto L^{\paraalphaLslope}$ (see  Eq.~\ref{eq:alpha_L}). This is a much stronger than $\alpha\propto L^{-0.32}$ predicted by \citet{2021A&A...648A..36B} and this is reflected in our steeper WLR.

\subsection{UV wind profile variation and large scale wind structure }
\label{sec:large_scale_wind_profile_variation}

Similar their Galactic counterparts, the winds of low~$Z$ O stars are variable \citep{Massa2000}. 
The multi-epoch observations available for some of our sample stars allow us search for  the variability in the wind lines and discuss its nature. 
SSN~18 (O8\,V) shows the strongest wind line variability for any star in sample. Comparing the G140L spectrum (lower panel, Fig.\,\ref{fig:SSN_18_LOW_HIGH_MOD_COMP}) to the E140M spectrum (upper panel, Fig.\,\ref{fig:SSN_18_LOW_HIGH_MOD_COMP}) one can see that the  C\,\textsc{iv}\,$\lambda\lambda\,1548.2,\,1550.8$ blueward asymmetry is no longer detected. \citet{2003ApJ...595.1182B}, who previously studied this object, only had access to the E140M spectrum without the C\,\textsc{iv}\,$\lambda\lambda\,1548.2,\,1550.8$ line and so report an upper limit to $\dot{M}$ and $\varv_\infty$ based on the N\,\textsc{v}\,$\lambda\lambda\,1238.8,\,1242.8$ asymmetry.

The differences between the mass-loss rates derived of these two observations is a factor of $\sim4.5$.
We speculate that the observed variability may be an extreme form of co-rotating interaction regions (CIRs), with no observable wind in the unstructured part of the wind and a C\,\textsc{iv}\,$\lambda\lambda\,1548.2,\,1550.8$ wind profile only visible during the part of the rotation phase where the structured part of the wind is in our line of sight. Neither synthetic wind profile can be described as the upper limit on $\dot{M}$, as we have no information about the phase of the variability  and its links with rotation. A time series of observations are hence required to reliably measure the average $\dot{M}$.

Overall, we find a high occurrence of variability in the C\,\textsc{iv}\,$\lambda\lambda\,1548.2,\,1550.8$ wind profile for the majority of the NGC~346 O6\,V-O8\,V stars in our sample (see Fig.\,\ref{fig:NGC_346_VARIABILITY}). 
Multi-epoch observations of SSN~9 suggest for the first time the presence of a structure moving blueward through the absorption trough of the wind line profile in low~$Z$ O star (see Fig,\,\ref{fig:SSN_9_multiepoh}). More extensive time coverage is needed to identify whether these features are DACs repeating on the stellar rotation timescale and hence could be explained as originating in CIRs, by analogy with the Galactic objects \citep{Massa2015}.

Beside SSN~9, we find strong evidence of a DAC and we show the presence of NACs in the wind lines of SSN~13 and 15. We may be also observing a NAC in the C\,\textsc{iv}\,$\lambda\lambda\,1548.2,\,1550.8$ absorption trough in SSN~11. Among all objects with available high resolution E140M or G140M spectra, only SSN~17 and 18 do not show  DACs or NACs.  However, SSN~17 has only very weak wind with low velocity ($1000 \, \,\mathrm{km\,s^{-1}}$) while SSN~18 shows the largest variation in its wind profiles at different epochs (see Sect.\,\ref{sec:variability}). 

To summarise, whenever the spectral resolution is sufficient
and the wind lines are  broad, we do observe the presence of narrow discrete absorption features, commonly explained by the large-scale structures in stellar winds. 
If this property is found to be common across all massive stars at low metallicity, then spectroscopic analysis with the non-LTE codes that assume spherically-symmetric expansion, stationary, and allow only for a small scale inhomogeneities (clumping), are likely to underestimate $\dot{M}$ from stars with dense, co-rotating structures within the wind.

\subsection{Empirical Hertzsprung–Russell diagram}

We construct an empirical upper Hertzprung-Russell diagram of the SMC where we include our new result on O stars in the NGC~346 cluster (Fig.\,\ref{fig:HR}). NGC~346 is commonly given the age $\sim 3\,\mathrm{Myr}$, while recent survey analysis by \citet{2019AA...626A..50D}  suggest that NGC~346 is undergoing continuous star formation. The absence of O stars close to the Zero Age Main Sequence (ZAMS) is prominent. Stars with luminosity class Vz are present yet they appear to the right of the $3\,\mathrm{Myr}$ and later isochrones. The newly classified SSN~31 O7.5\,Vz appears on the $5\,\mathrm{Myr}$ isochrone.

The absence of O stars close to the ZAMS in the Galaxy has previously been noted \citep{1982ApJ...263..777G, 1992A&A...261..209H, 2007ASPC..367...67H, 2004A&A...415..349R, 2014A&A...570L..13C, 2018A&A...613A..65H, 2020A&A...638A.157H}. This absence has also been noticed for the Magellanic Clouds \citep{1995ApJ...438..188M, 2014A&A...564A..39S} and other galaxies in the Local Group \citep{1993AJ....105..980M, 1995AJ....110.2715M}. This may be a result of lower accretion during star formation ($\log{\dot{M}_{\rm accretion}} \sim -5$) for objects more massive than $25\,M_\sun$. This results in the birthline being to the right of the ZAMS \citep{2020A&A...638A.157H}. An alternative explanation is that fast rotating massive stars evolve on a different track due to mixing \citep{1987A&A...178..159M}. However, the threshold rotation velocity for this is 
far above $\varv\sin{i}$ for any of the massive stars in our sample as well as in \citet{2019A&A...625A.104R}.

In addition to SSN~9 and SSN~15, whose luminosity and temperature has been determined by \citet{2003ApJ...595.1182B}, we derive the stellar parameters for SSN~13 and SSN~14, adding to the upper part of the HRD. If we consider the commonly adopted age of NGC346 ($\sim3\,\mathrm{Myr}$), all of our objects are to the right of the $3\,\mathrm{Myr}$ isochrome except for SSN~9, (O2\,III(f*)).

\section{Conclusions}
\label{sec:concl}
In this work we systematically and consistently analysed the UV spectra complemented by the optical spectra for the nearly complete sample of O-type stars in the most massive young cluster, NGC~346, in the SMC galaxy at the metallicity $Z=1/7 \ Z_\sun$. We find that the empirically derived mass-loss rates of O type non-supergiant stars are significantly below the theoretical expectations of \citet{2000A&A...362..295V, 2001A&A...369..574V}. The later the spectral subtype the stronger is the discrepancy between the observed and predicted mass-loss rate. Accounting for binarity, wind porosity, and large-scale wind structure in empirical spectroscopic mass-loss rate diagnostics may help to reduce the discrepancy but is unlikely to eliminate it. 

Our results show that the empirical wind luminosity relation (WLR) shows a steeper gradient than the theoretical predictions from \citet{2000A&A...362..295V, 2001A&A...369..574V} and that recent theoretical WLRs produced by \citet{Krticka2018} and \citet{2021A&A...648A..36B} for the SMC are in better agreement, but still underestimate our empirically derived gradient. We find, based on these observations and the best fit models, that the gradient of the WLR is dependent on $Z$. We find no indications of a break in the WLR around $\log{L/ L_\sun} \sim 5.4$ as suggested by  \citet{2004A&A...420.1087M, 2005A&A...441..735M}. For stars with $\log{L/ L_\sun} < 5$ only upper limits to $\dot{M}$ could be determined.

We find that the spectral type cutoff where later O types never have observable wind profiles in the UV is a lot later than previously reported. For stars with spectral type between O5 and O8 we find numerous objects with wind signatures in the UV spectra as well as those without. However, the wind signatures disappear in the UV spectra for stars later that O8--9 at the resolving power of STIS/G140L. In addittion, our multi-epoch UV observations reveal that O stars at the SMC metallicity have variable winds. In one case, SSN~18, no wind profile is observable in one epoch and a strong profile appears in another. The common nature of such features suggests that winds of massive stars at low metallicity are highly structured.

We have resolved the core of the NGC 346 cluster and obtained the spectra of the nearly complete sample of O stars in this region. The empirical HRD shows a deficiency of the O type stars located at zero-age main sequence and in the upper part of the HRD, being in agreement with previous studies on the population of massive stars in the SMC. Our results highlight once more the deficiency of massive stars in this galaxy.

The ionising energy feedback in NGC~346 is dominated by the O2 giant SSN~9 which contributes $\sim\,49$\%\ of H ionising flux, $71$\%\ of He\,{\sc i} ionising flux,  and $99$\%\ of He\,{\sc ii} ionising flux, though we note that the components of the SSN~7 binary are likely to be additional significant contributors to ionising flux based on their spectral type. We suggest that SSN~9 is largely responsible for the morphology of the H\,{\sc ii} region seen in the {\em  HST} images.

To summarise, our new in-depth study of the most massive young star cluster in the SMC galaxy, NGC~346,  enabled by the UV spectroscopic capabilities of {\em  HST}, resulted in the identification of eight new O-type stars in the cluster core, has enabled the measurement of wind properties of 19 O stars, and highlighted the key role O stars play in the ecology of young star clusters at low $Z$.

\begin{acknowledgements}This research is based on observations made with the NASA/ESA {\em  Hubble Space Telescope} obtained from the Space Telescope Science Institute, which is operated by the Association of Universities for Research in Astronomy, Inc., under NASA contract NAS 5–26555. These observations are associated with programmes GO 15112 (PI L. Oskinova), GO 15837 (PI L. Oskinova). GO 7437 (PI D. J. Lennon), GO 7667 (PI R. Gililand), GO 8629 (PI F. Bruhweiler), GO 16098 (PI J. Roman-Duval), and GO 12940 (PI P. Massey). MJR would like to thank Ian D. Howarth (UCL) for his helpful comments during the early stages of this work. This publication was has benefited from a discussion at a team meeting sponsored by the International Space Science Institute at 
Bern, Switzerland. AACS and VR
acknowledge support by the Deutsche Forschungsgemeinschaft (DFG -
German Research Foundation) in the form of an Emmy Noether Research
Group (grant number SA4064/1-1, PI Sander).
TS acknowledges funding received from the European Research Council (ERC) under the European Union’s Horizon 2020 research and innovation programme (grant agreement number 772225: MULTIPLES).  
DP acknowledges financial support by the Deutsches Zentrum  f\"ur  Luft  und  Raumfahrt  (DLR)  grant  FKZ  50  OR  2005.
YHC acknowledges the MOST grants 109-2112-M-001-040 and 
110-2112-M-001-020.
\end{acknowledgements}

%
  \bibliographystyle{aa} 
  \bibliography{bib} 


\begin{appendix}

\section{Observations}
\label{sec:obs_diary}
\begin{table}[h]
\tiny
\center
\caption{Observations list. Positions, other names, and observational material for the O stars analysed in this work.}
\begin{tabular}{c c c c c c c c c } 
\hline\hline
RA (J2000) & Dec (J2000) & Star (SSN) & Other Names & UV Observations & Optical Observations \\
\hline
00 59 04.64 & -72 10 24.17 & 7 & MPG 435, W 1, NMC 26 & HST STIS G140L & FLAMES ($3960{-}5070\,\AA$), MUSE\\
00 59 00.75 & -72 10 28.17 &  9& MPG 355, W 3, NMC 29 & HST STIS E140M/G140L & FLAMES ($3960{-}5070\,\AA$), MUSE\\
00 59 00.15 & -72 10 37.88 & 11 & MPG 342, W 4, NMC 30, OGLE SMC-ECL-3910 & HST STIS E140M/G140L  & FLAMES ($3960{-}5070\,\AA$), MUSE\\
00 58 57.39 & -72 10 33.60 & 13 & MPG 324, W 6, ELS 7, NMC 32 & HST STIS E140M/G140L  & FLAMES ($3850{-}6690\,\AA$), MUSE\\
00 59 05.98 & -72 10 33.64 & 14 & MPG 470, W 2, NMC 25, OGLE SMC108.2 37482 & HST STIS G140L & MUSE\\
00 59 01.73 & -72 10 30.64 & 15 & MPG 368, NMC 28 & HST STIS E140M & FLAMES ($3960{-}5070\,\AA$), MUSE\\
00 59 02.93 & -72 10 34.82 & 17 & MPG 396 & HST STIS G140M & FLAMES ($3960{-}5070\,\AA$), MUSE\\
00 59 06.90 & -72 10 41.49 & 18 & MPG 487, NMC 17 & HST STIS G140L & MUSE\\
00 59 06.23 & -72 10 33.44 & 22 & MPG 476 & HST STIS E140M/G140L & MUSE\\
00 59 04.02 & -72 10 51.12 & 31 & MPG 417 & HST STIS G140L & FLAMES ($3960{-}5070\,\AA$) , MUSE\\
00 59 05.89 & -72 10 50.25 & 33 & MPG 467 ELS 34, NMC 18 & HST STIS G140L & FLAMES ($3850{-}6690\,\AA$), MUSE\\
00 59 01.90 & -72 10 43.31 & 34 & MPG 370 & HST STIS G140L & FLAMES ($3960{-}5070\,\AA$), MUSE\\
00 59 07.31 & -72 10 25.80 & 36 & MPG 495, NMC 24 & HST STIS G140L & FLAMES ($3960{-}5070\,\AA$), MUSE\\
00 59 06.33 & -72 10 32.30 & 43 & MPG 481 & HST STIS G140L & MUSE\\
00 59 07.62 & -72 10 48.23 & 46 & MPG 500, NMC 16 & HST STIS G140L & FLAMES ($3960{-}5070\,\AA$), MUSE\\
00 59 05.43 & -72 10 42.40 & 57 & MPG 455 & HST STIS G140L & FLAMES ($3960{-}5070\,\AA$), MUSE\\
00 59 02.07 & -72 10 36.10 & 59 & MPG 375 & HST STIS G140L & MUSE\\
00 59 05.89 & -72 10 28.83 & 62 & MPG 468 & HST STIS G140L & MUSE\\
00 59 04.27 & -72 10 27.23 & 83 & MPG 429 & HST STIS G140L & MUSE\\
00 59 06.67 & -72 10 28.78 & 105 & MPG 486 & HST STIS G140L & MUSE\\
00 59 07.60 & -72 10 39.13 & 110 & MPG 499 & HST STIS G140L & MUSE\\
\hline 
\end{tabular}
\label{table:observation_record}
\end{table}

\section{Ionising photon rate}
\label{sec:Ionising_photons_table}

NGC~346 is an open star cluster surrounded by a lot of ionised material. For completeness we list the amount of ionising photons ($\log{Q}$) released by each individual target according to our favourite PoWR Model. The resulting ionising fluxes for $\mathrm{H}$,  \ion{He}{i} and \ion{He}{ii} are listed in Table~\ref{table:ionising_photons}. We note, that in this work only stars without a visible companion in the observed spectrum are included and that one of the major contributors, SSN~7 (one of the earliest type binaries in NGC~346), is not included here.

\begin{table}[h]
\footnotesize
\center
\caption{Ionising photon rates for each O star within our sample of NGC~346 objects.}
\begin{tabular}{c c c c} 
\hline\hline
SSN&  $\log{Q_\mathrm{H}}$ & $\log{Q_\mathrm{HeI}}$ &  $\log{Q_\mathrm{HeII}}$\\

\hline
  9 & 49.89 &     49.41 &      46.50 \\
 13 & 49.35 &     48.64 &      43.81 \\
 14 & 48.48 &     46.52 &      40.95 \\
 15 & 49.10 &     48.29 &      43.77 \\
 17 & 48.81 &     47.87 &      42.11 \\
 18 & 48.73 &     47.69 &      41.97 \\
 22 & 48.93 &     48.07 &      42.54 \\
 31 & 48.58 &     47.55 &      41.80 \\
 33 & 48.24 &     46.63 &      41.77 \\
 34 & 48.13 &     46.19 &      40.70 \\
 36 & 48.03 &     46.13 &      40.59 \\
 43 & 47.81 &     45.79 &      40.88 \\
 46 & 48.39 &     47.36 &      41.57 \\
 57 & 48.06 &     46.06 &      41.36 \\
 59 & 47.77 &     45.75 &      40.13 \\
 62 & 48.12 &     46.82 &      41.03 \\
 83 & 47.62 &     45.46 &      39.16 \\
105 & 47.40 &     45.46 &      39.54 \\
110 & 47.74 &     45.78 &      39.99 \\
\hline 
\end{tabular}
\label{table:ionising_photons}
\end{table}

\onecolumn
\clearpage

\section{Grid fit for temperature and surface gravity}
\label{sec:teff_logg_grid_fits}

The observed (MUSE) and synthetic spectra of those O-type stars for which $T_\ast$ and 
$\log\,g$ are determined by fitting a grid model.
The wings of the  He\,\textsc{ii}\,$\lambda\,4860$ line are used for measuring $\log\,g$ while the other He\,\textsc{i} and He\,\textsc{ii} lines are used to constrain  
$T_\ast$. An analysis of B stars will be presented in a subsequent paper (Rickard et al. in prep).

\begin{figure*}[h]
    \centering
    \includegraphics[width=0.77\hsize, trim={0.7cm 9.2cm 6.05cm 16.8cm},clip]{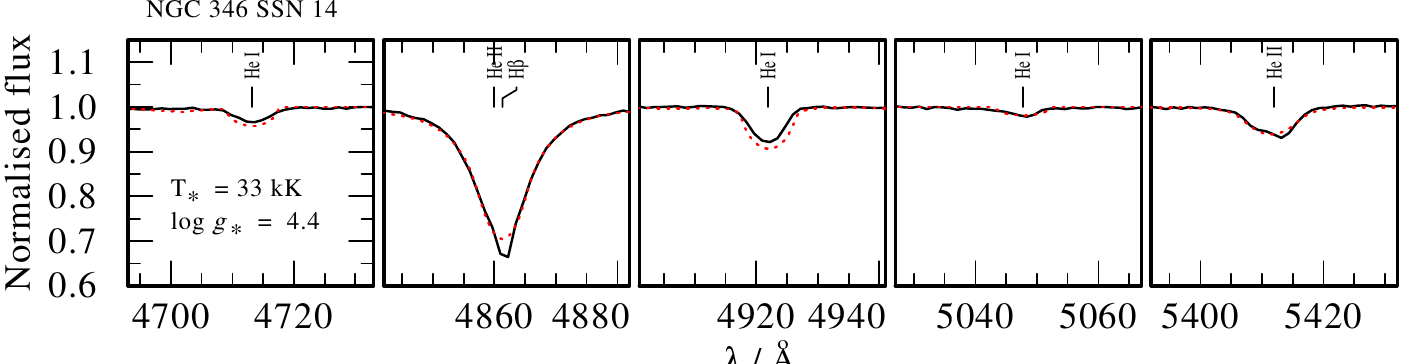}
    \includegraphics[width=0.77\hsize, trim={0.7cm 9cm 6.05cm 16.8cm},clip]{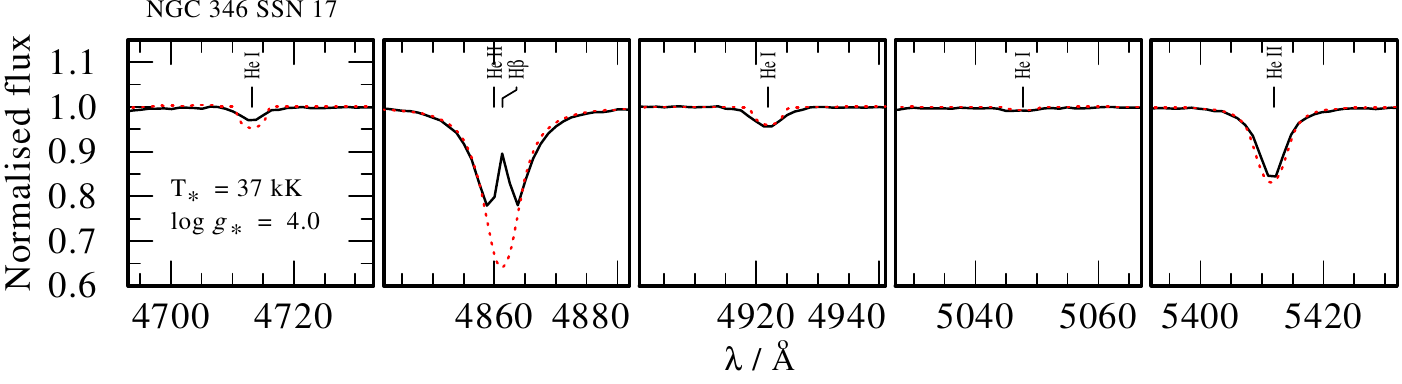}
    \includegraphics[width=0.77\hsize, trim={0.7cm 9cm 6.05cm 16.8cm},clip]{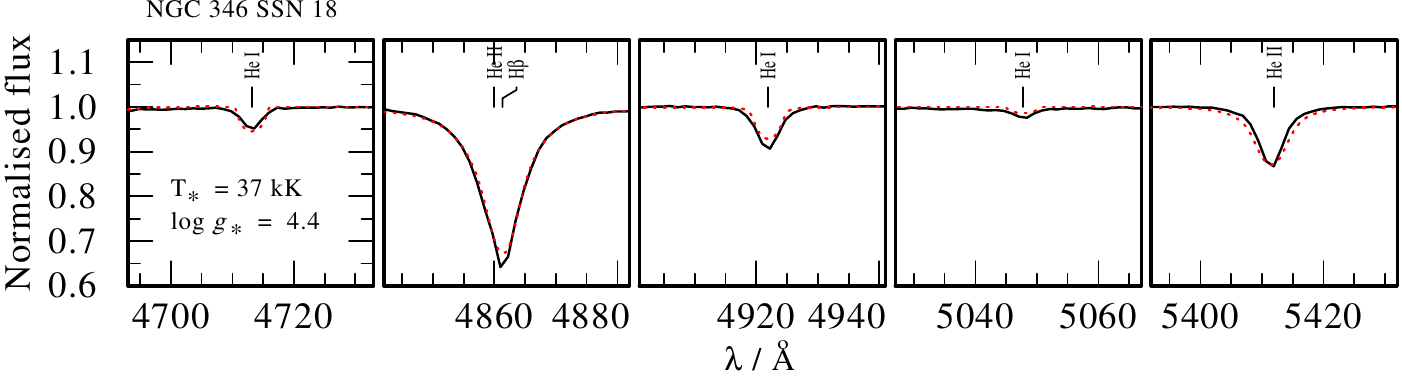}
    \includegraphics[width=0.77\hsize, trim={0.7cm 9cm 6.05cm 16.8cm},clip]{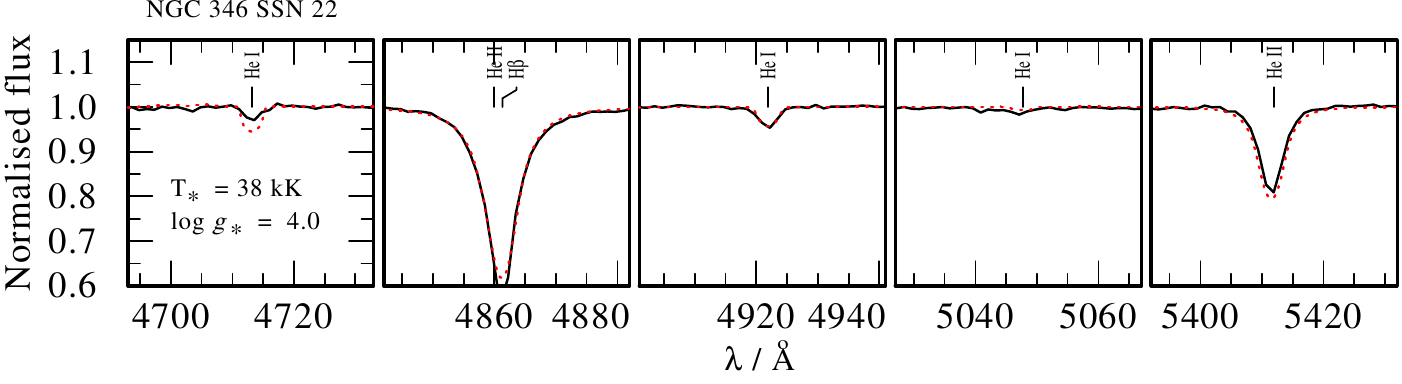}
    \includegraphics[width=0.77\hsize, trim={0.7cm 9cm 6.05cm 16.8cm},clip]{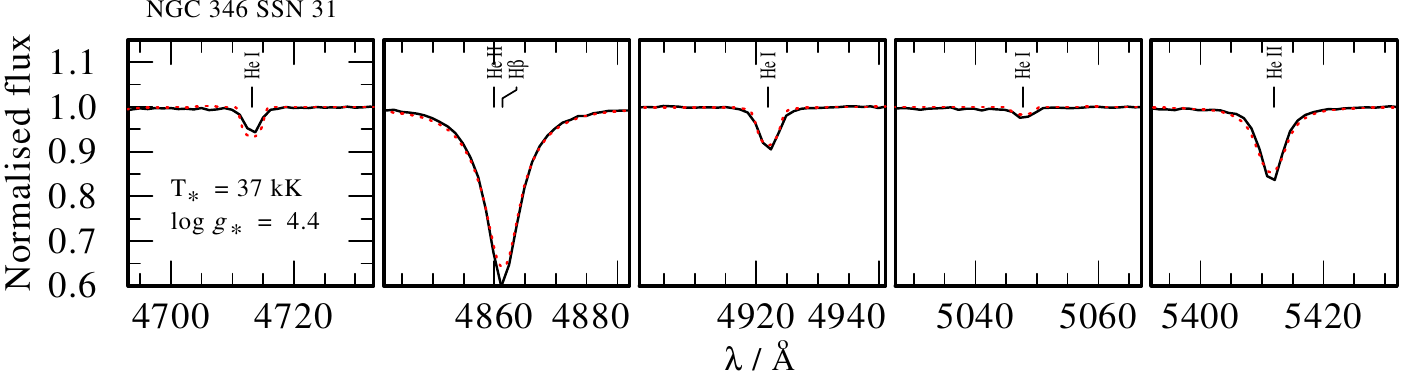}
    \caption{Synthetic Spectra Fitting to determine $T_\ast$ and $\log g$. Red line: best fitted grid model;  black line: observed MUSE spectrum.
    }
    \label{fig:MUSE_Grid_Fits}
\end{figure*}

\begin{figure*}[h]
    \centering
    \includegraphics[width=0.77\hsize, trim={0.7cm 9cm 6.05cm 16.8cm},clip]{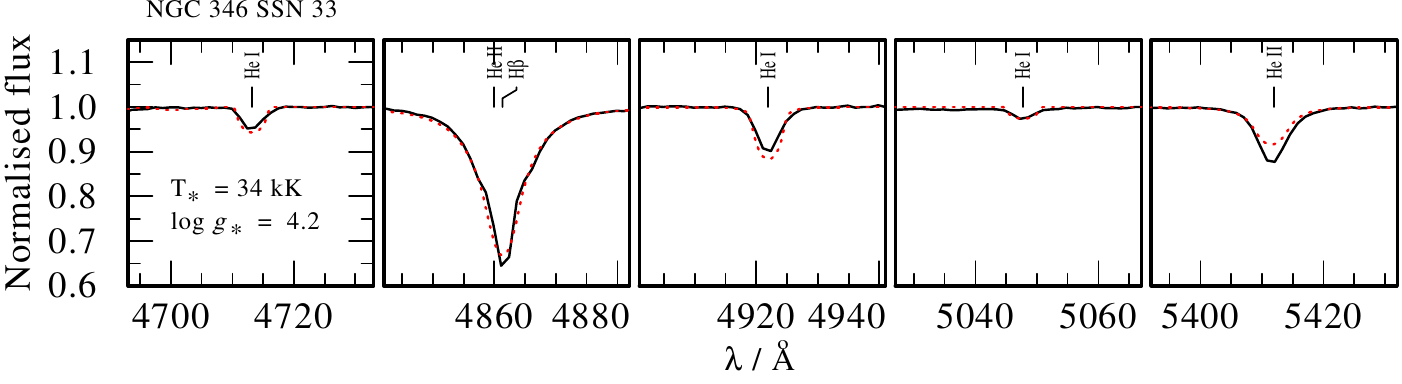}
    \includegraphics[width=0.77\hsize, trim={0.7cm 9cm 6.05cm 16.8cm},clip]{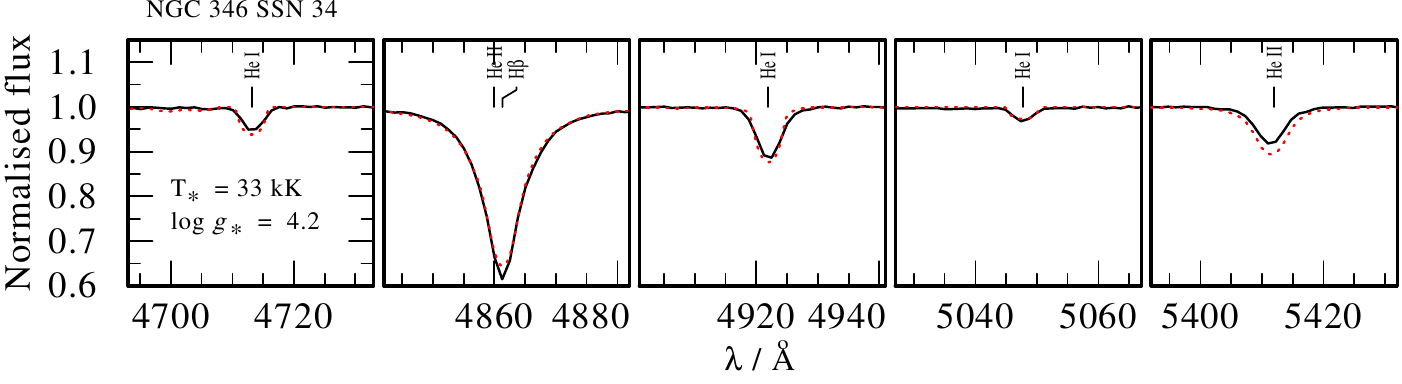}
    \includegraphics[width=0.77\hsize, trim={0.7cm 9cm 6.05cm 16.8cm},clip]{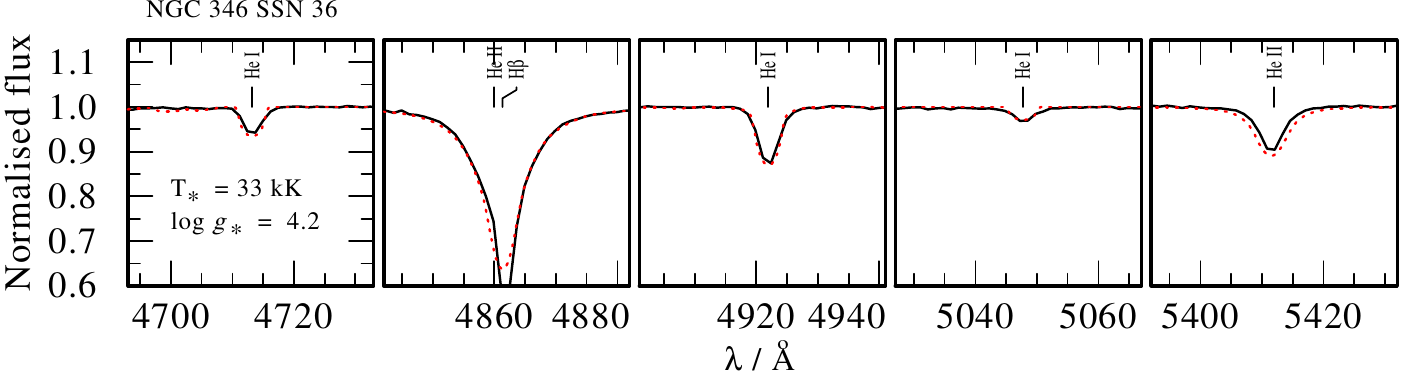}
    \includegraphics[width=0.77\hsize, trim={0.7cm 9cm 6.05cm 16.8cm},clip]{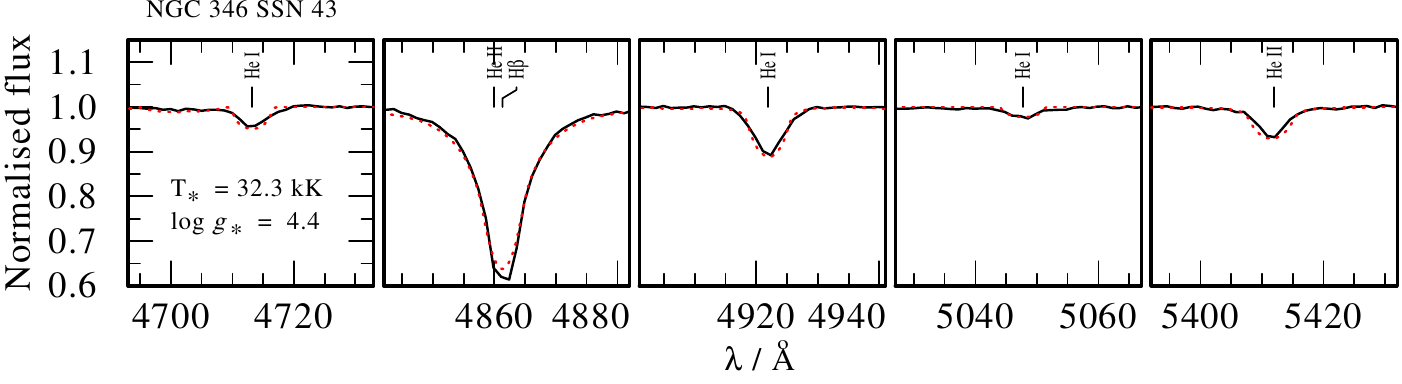}
    \includegraphics[width=0.77\hsize, trim={0.7cm 9cm 6.05cm 16.8cm},clip]{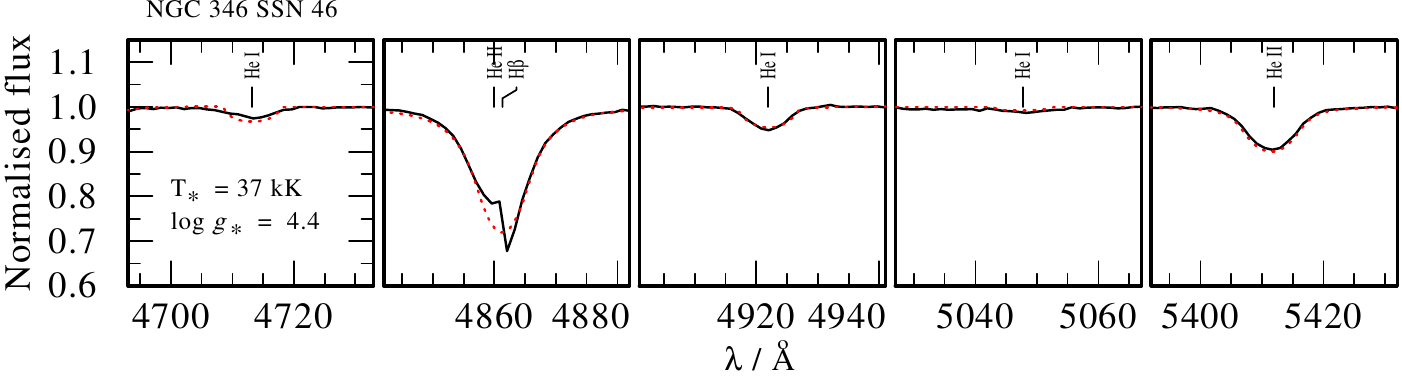}
    \caption{continued.}
\end{figure*}

\begin{figure*}[h]
    \centering
    \includegraphics[width=0.77\hsize, trim={0.7cm 9cm 6.05cm 16.8cm},clip]{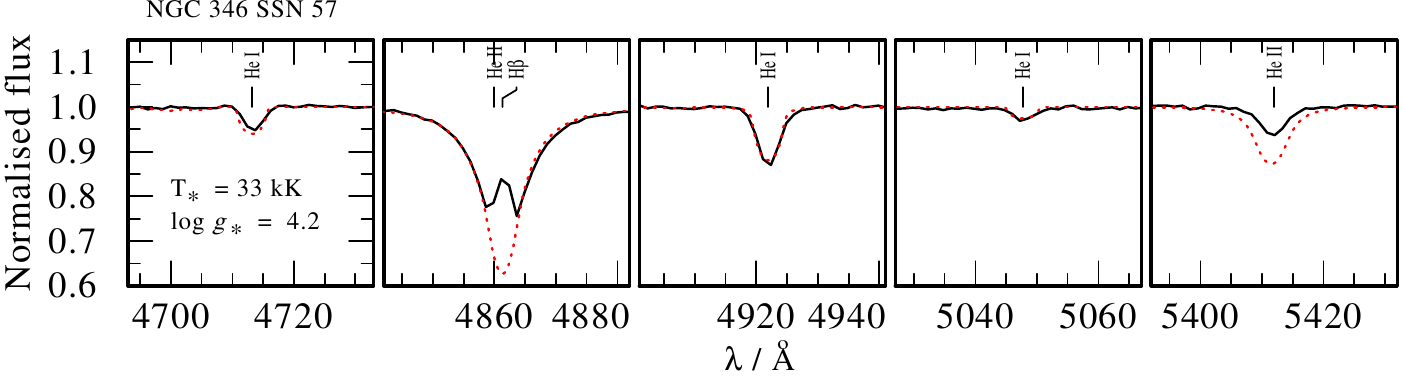}
    \includegraphics[width=0.77\hsize, trim={0.7cm 9cm 6.05cm 16.8cm},clip]{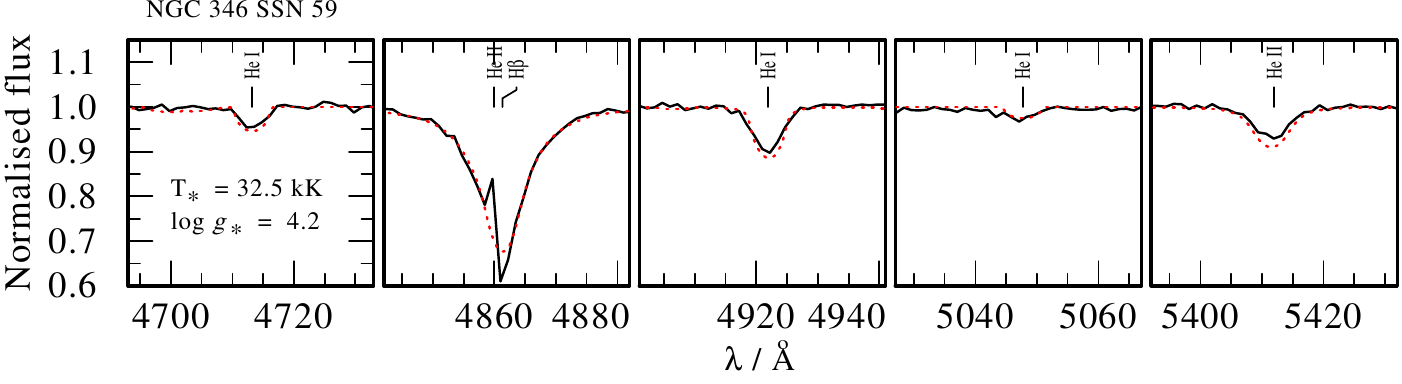}
    \includegraphics[width=0.77\hsize, trim={0.7cm 9cm 6.05cm 16.8cm},clip]{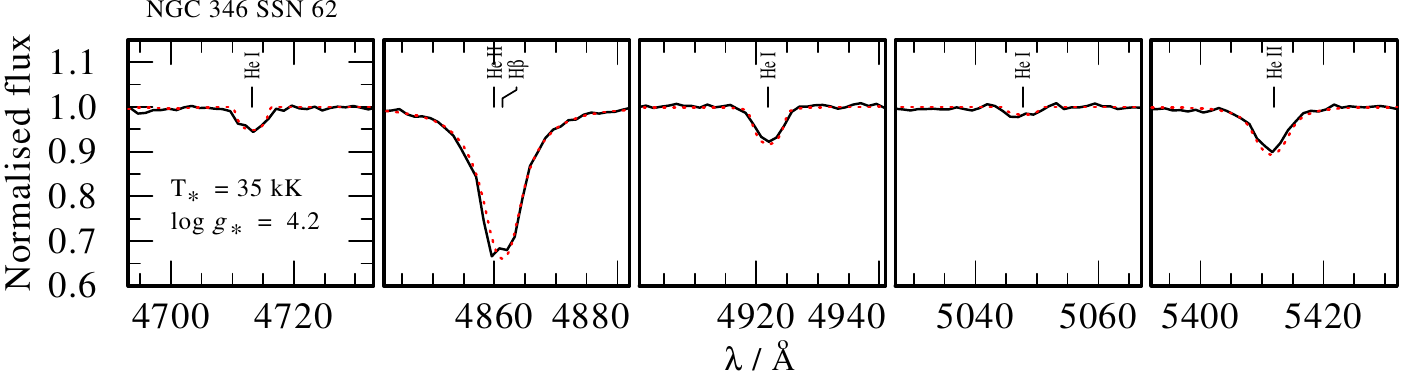}
    \includegraphics[width=0.77\hsize, trim={0.7cm 9cm 6.05cm 16.8cm},clip]{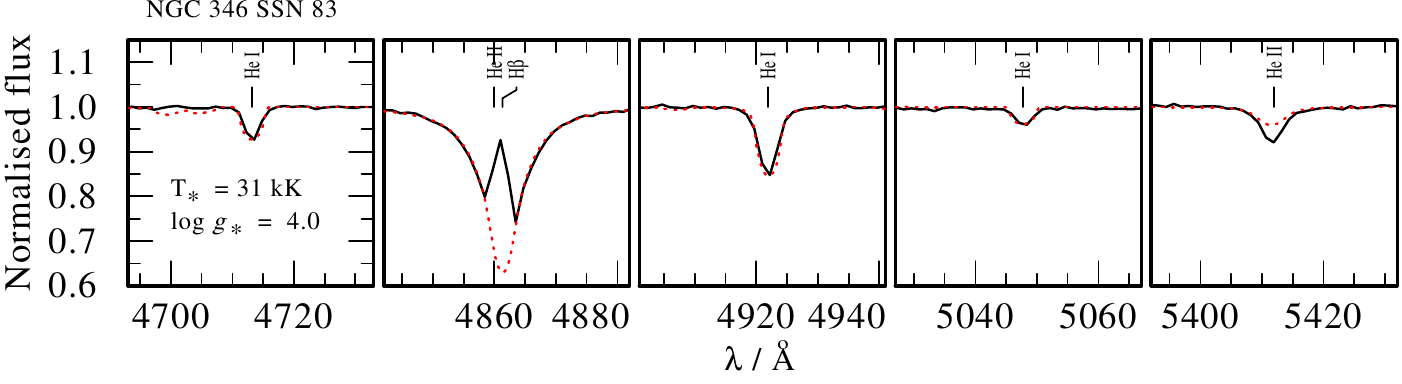}
    \includegraphics[width=0.77\hsize, trim={0.7cm 9cm 6.05cm 16.8cm},clip]{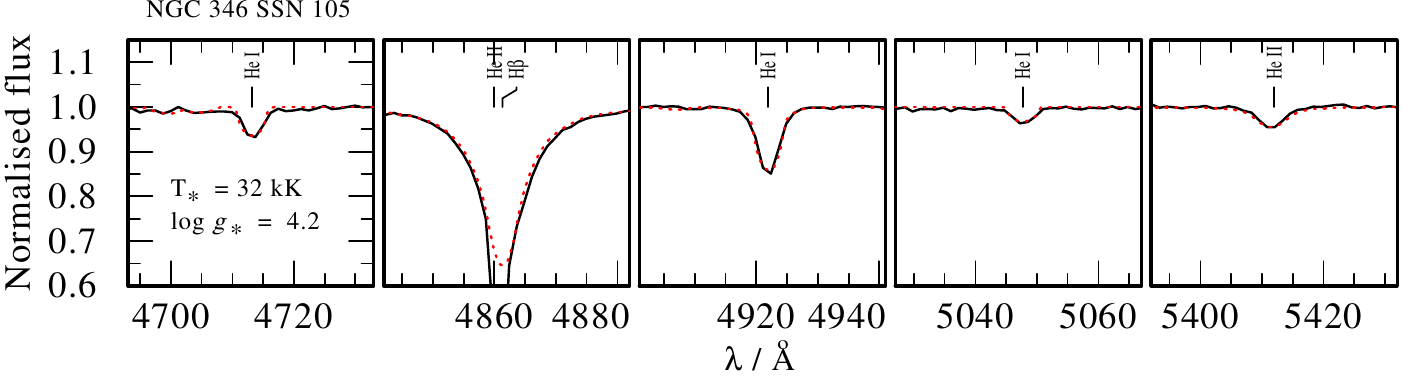}
    \includegraphics[width=0.77\hsize, trim={0.7cm 9cm 6.05cm 16.8cm},clip]{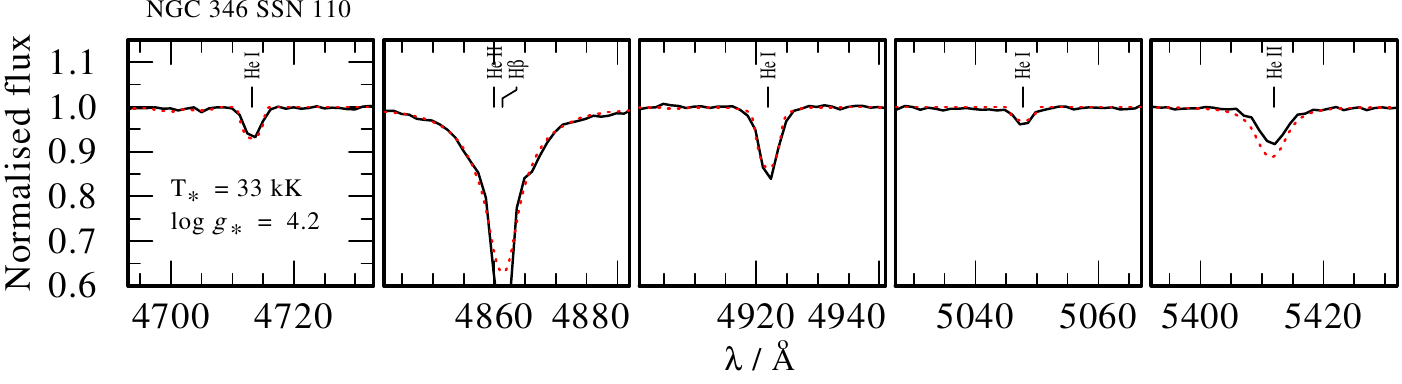}
    \caption{
    continued.}
\end{figure*}

\clearpage

\section{Spectral model fits}
\label{sec:appedix_fits}

Model continuum plots.

\begin{figure*}[h]
    \centering
    \includegraphics[width=0.9\hsize, trim={0cm 13.85cm 0cm 2.25cm},clip]{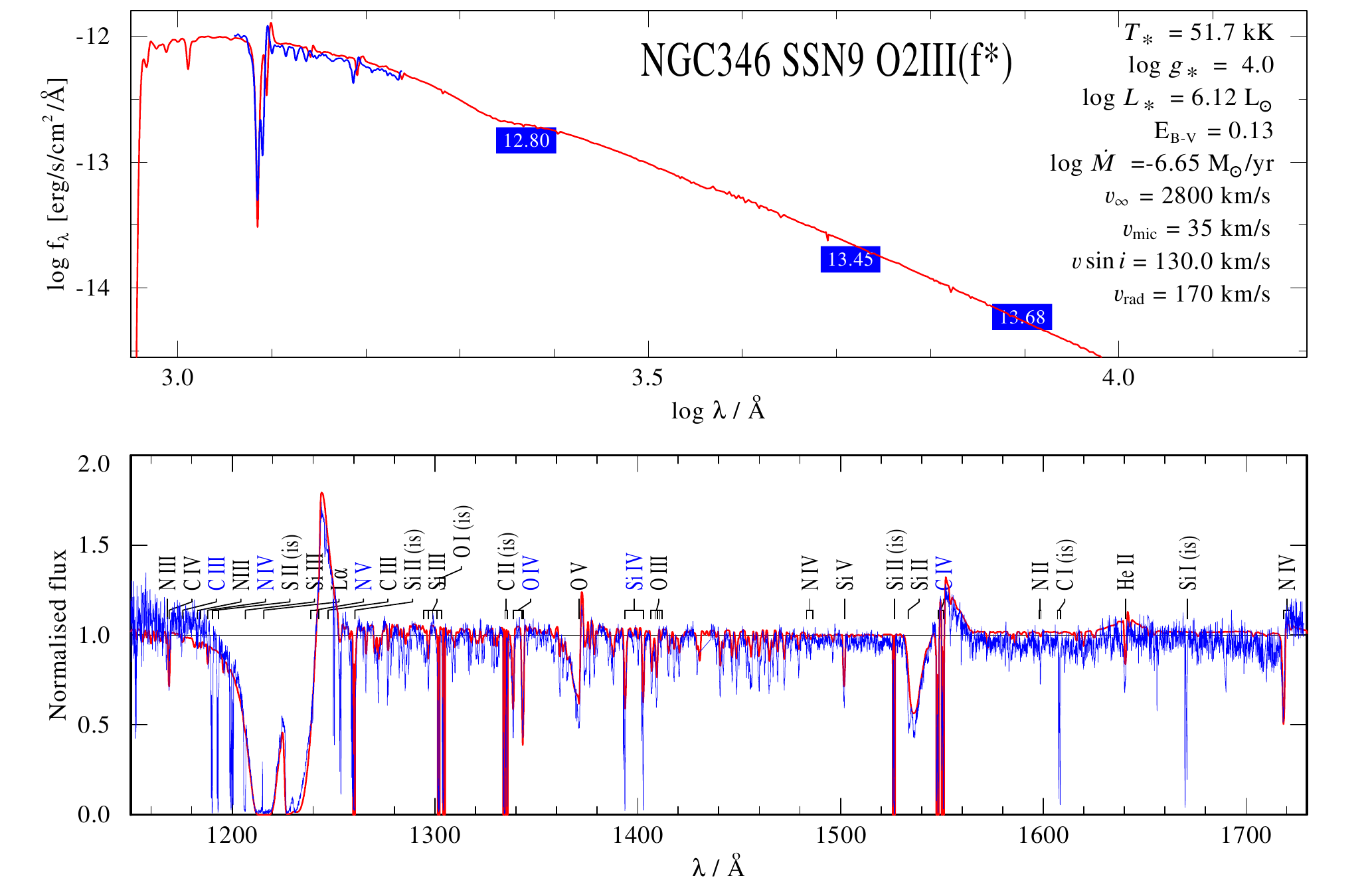}
    \caption{Spectral model fit of SSN~9. Top panel: SED fit of the model (red) to the flux calibrated observations of HST-STIS (blue). The blue squares with numbers mark the measure photometric data from \citet{2007AJ....133...44S} which are used as additional constraints for the SED fit. Middle panel: Normalised HST-STIS spectrum (blue) overplotted by our synthetic spectrum (red). Lower panels: Same as the panel above, but now zoomed on the \ion{N}{V}, \ion{O}{V} and \ion{C}{IV} UV resonance lines.}
    \label{fig:SSN_0009_p1}
\end{figure*}

\begin{figure*}
    \centering
    \includegraphics[width=0.9\hsize, trim={0cm 13.85cm 0cm 2.25cm},clip]{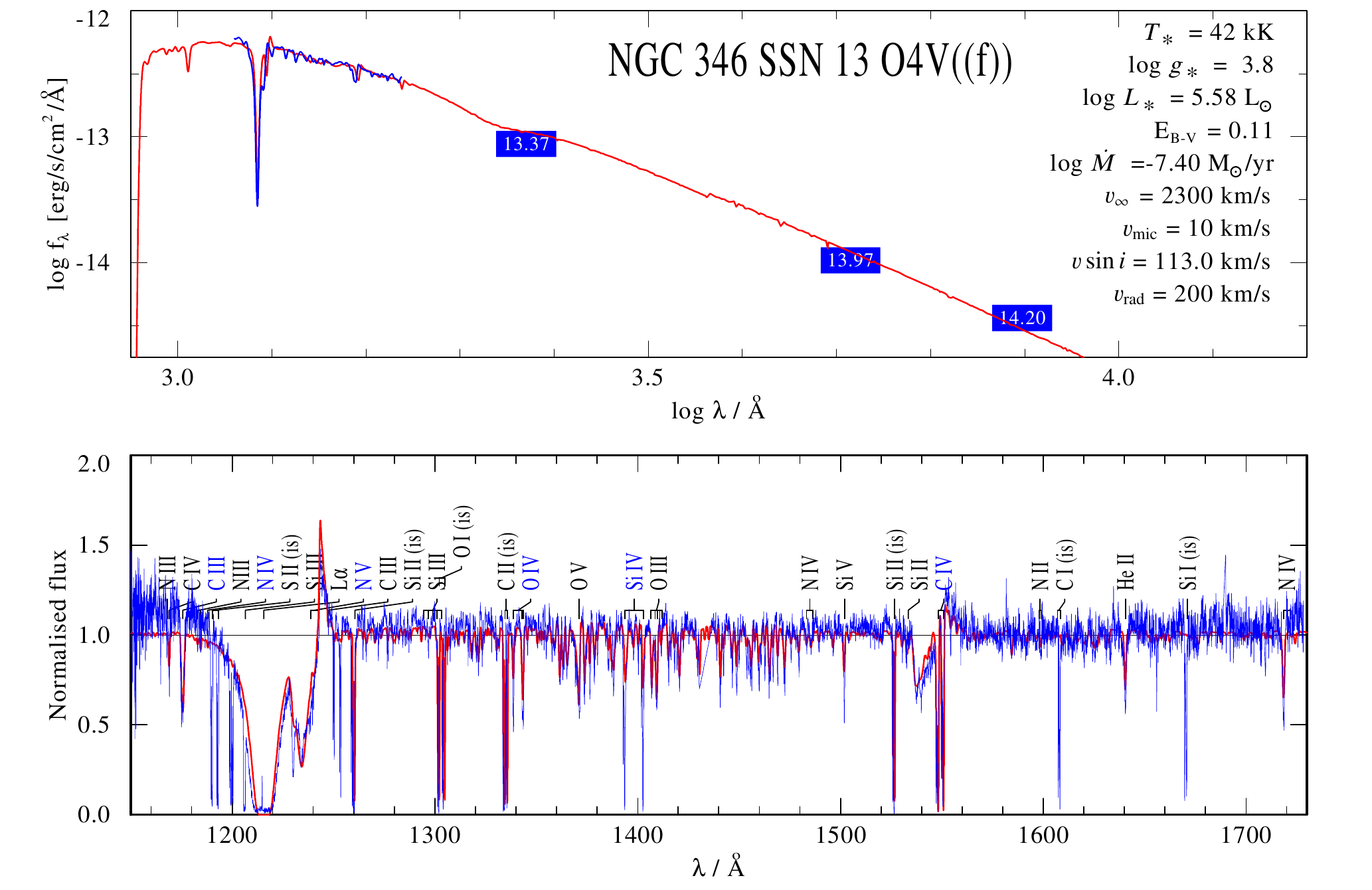}
    \caption{Same as Figure~\ref{fig:SSN_0009_p1}, but for SSN~13.}
    \label{fig:SSN_0014_p1}
\end{figure*}

\begin{figure*}
    \centering
    \includegraphics[width=0.9\hsize, trim={0cm 13.85cm 0cm 2.25cm},clip]{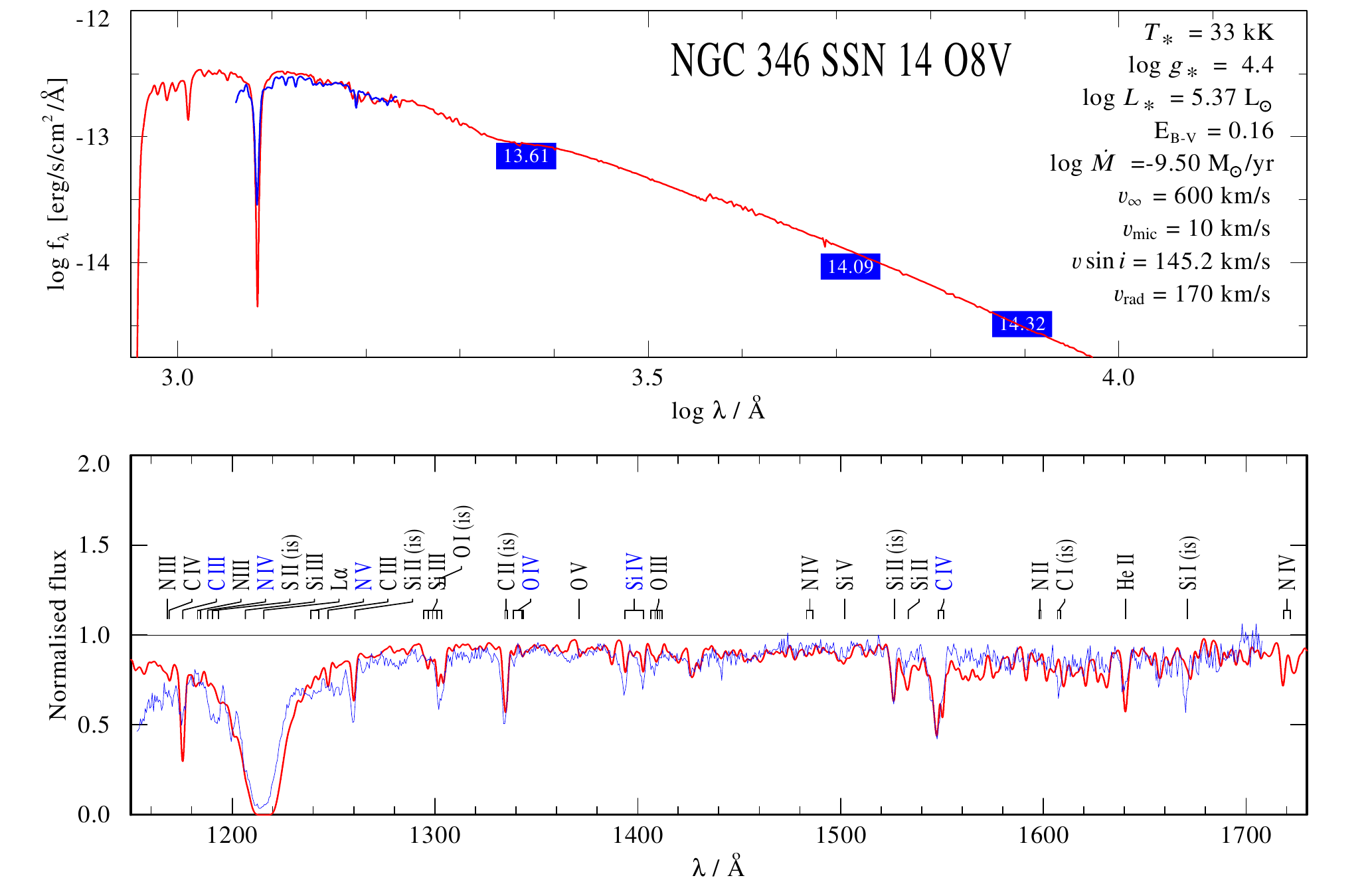}
    \caption{Same as Figure~\ref{fig:SSN_0009_p1}, but for SSN~14.
    }
    \label{fig:SSN_0014_p1}
\end{figure*}

\begin{figure*}
    \centering
    \includegraphics[width=0.9\hsize, trim={0cm 13.85cm 0cm 2.25cm},clip]{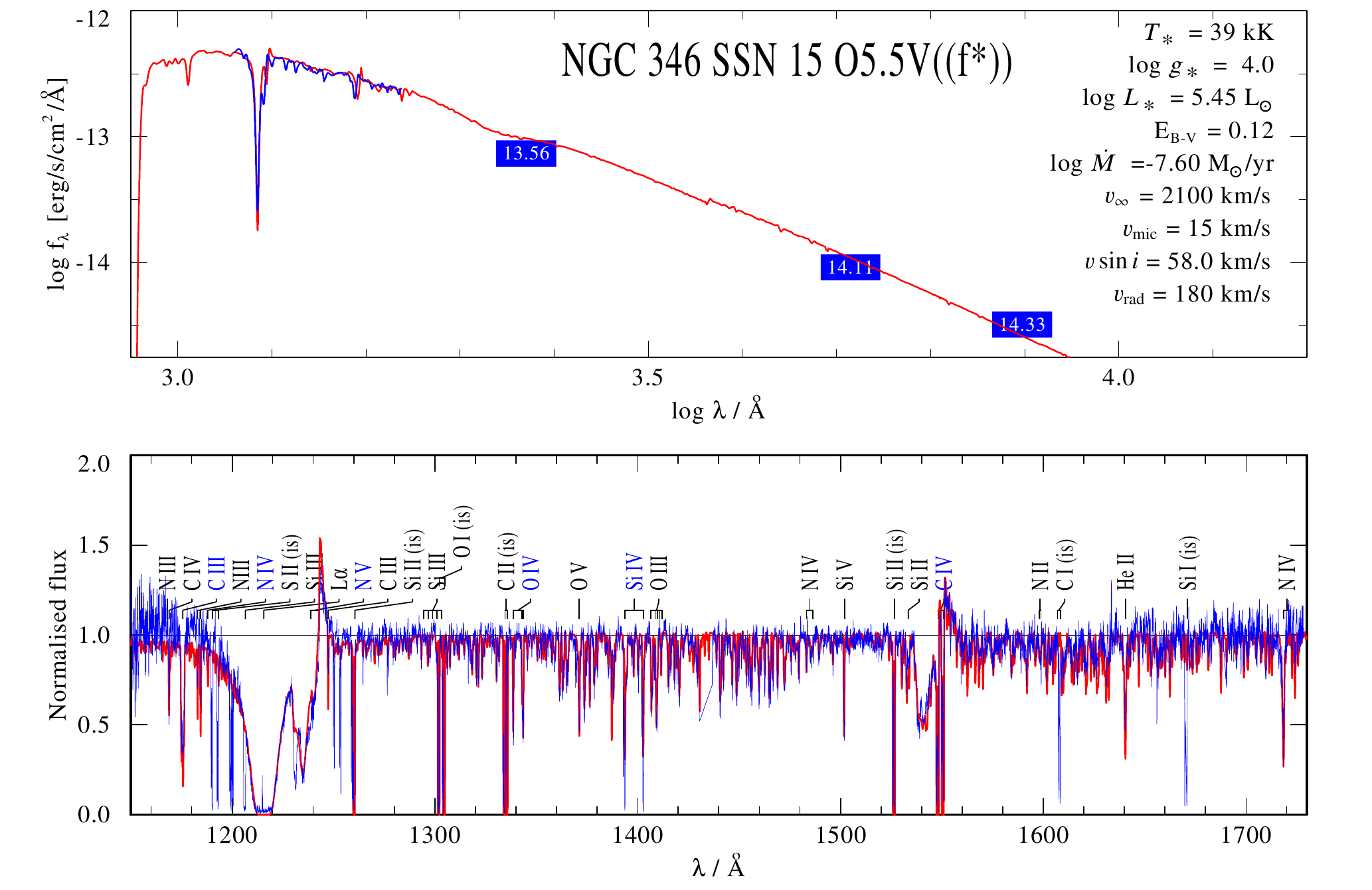}
    \caption{Same as Figure~\ref{fig:SSN_0009_p1}, but for SSN~15.}
    \label{fig:SSN_0015_p1}
\end{figure*}

\begin{figure*}
    \centering
    \includegraphics[width=0.9\hsize, trim={0cm 13.85cm 0cm 2.25cm},clip]{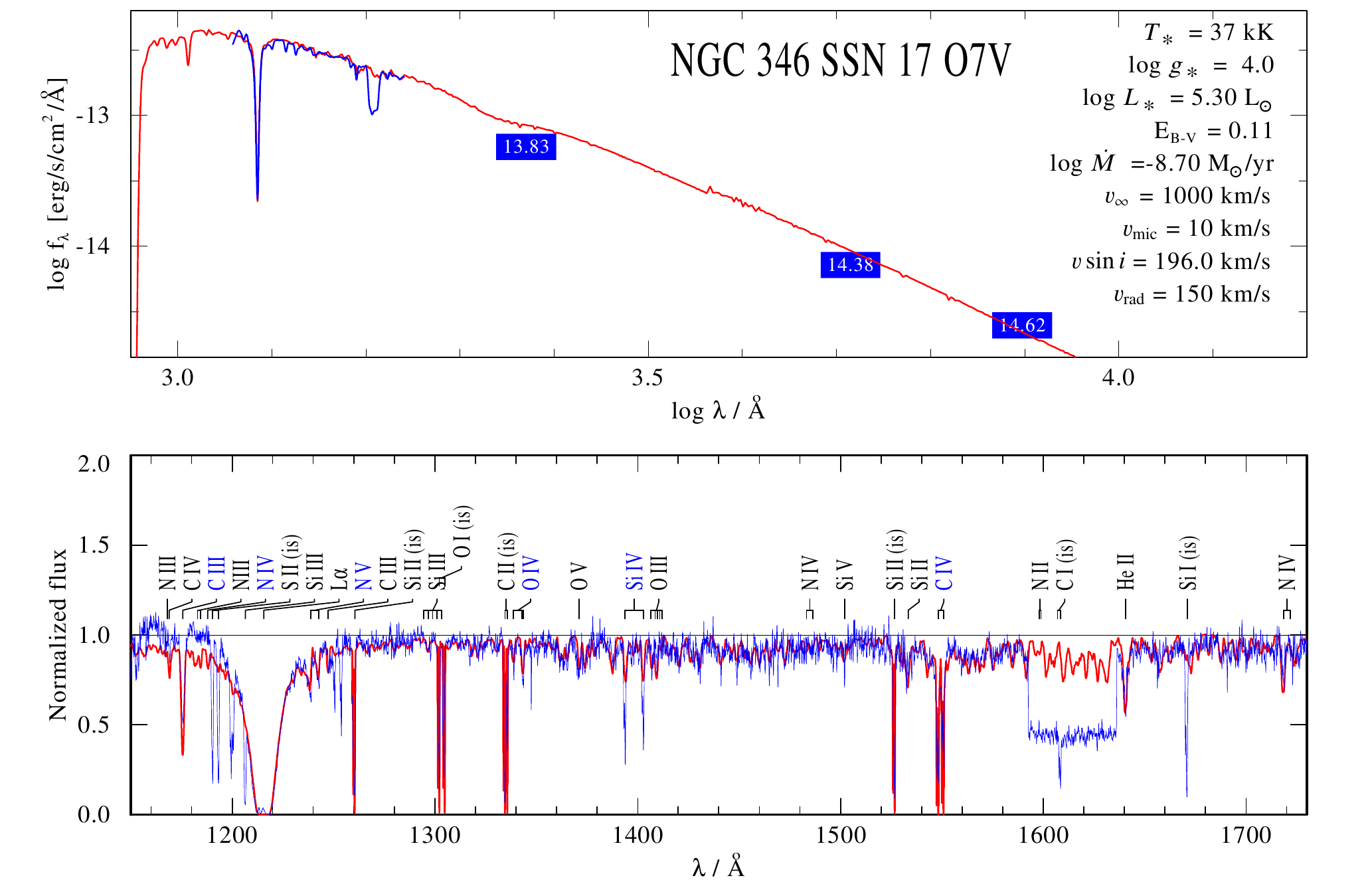}
    \caption{Same as Figure~\ref{fig:SSN_0009_p1}, but for SSN~17. HST observations contains an error between approximately $\lambda\,1590\AA$ and $\lambda\,1640\AA$ due to insufficient overlap for correct concatenation. Left shown here for reference. Does not affect fit the quality of the fit due to absence of relevant lines in this region,}
    \label{fig:SSN_0017_p1}
\end{figure*}

\begin{figure*}
    \centering
    \includegraphics[width=0.9\hsize, trim={0cm 13.85cm 0cm 2.25cm},clip]{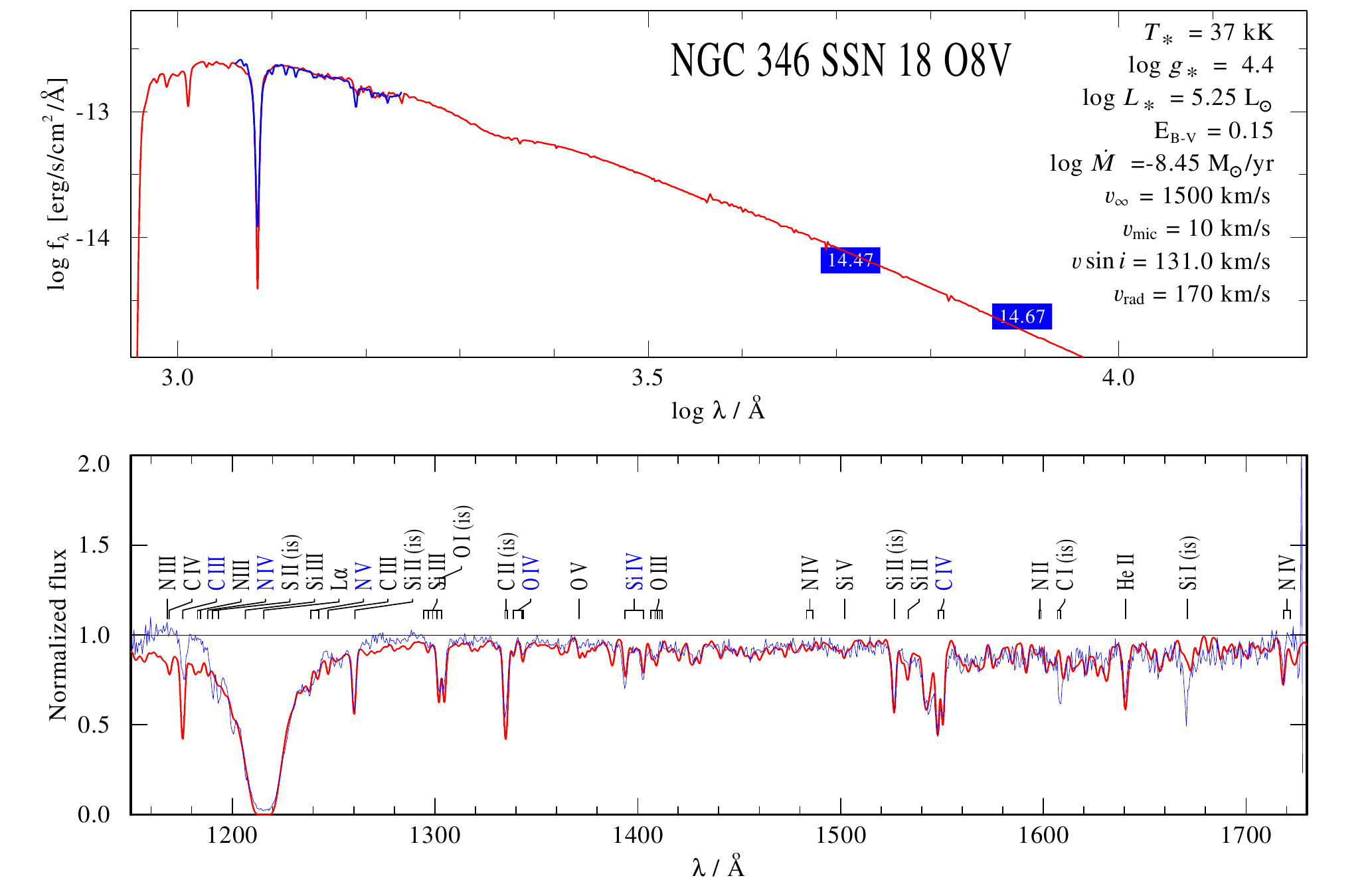}
    \caption{Same as Figure~\ref{fig:SSN_0009_p1}, but for SSN~18.}
    \label{fig:SSN_0018_p1}
\end{figure*}

\begin{figure*}
    \centering
    \includegraphics[width=0.9\hsize, trim={0cm 13.85cm 0cm 2.25cm},clip]{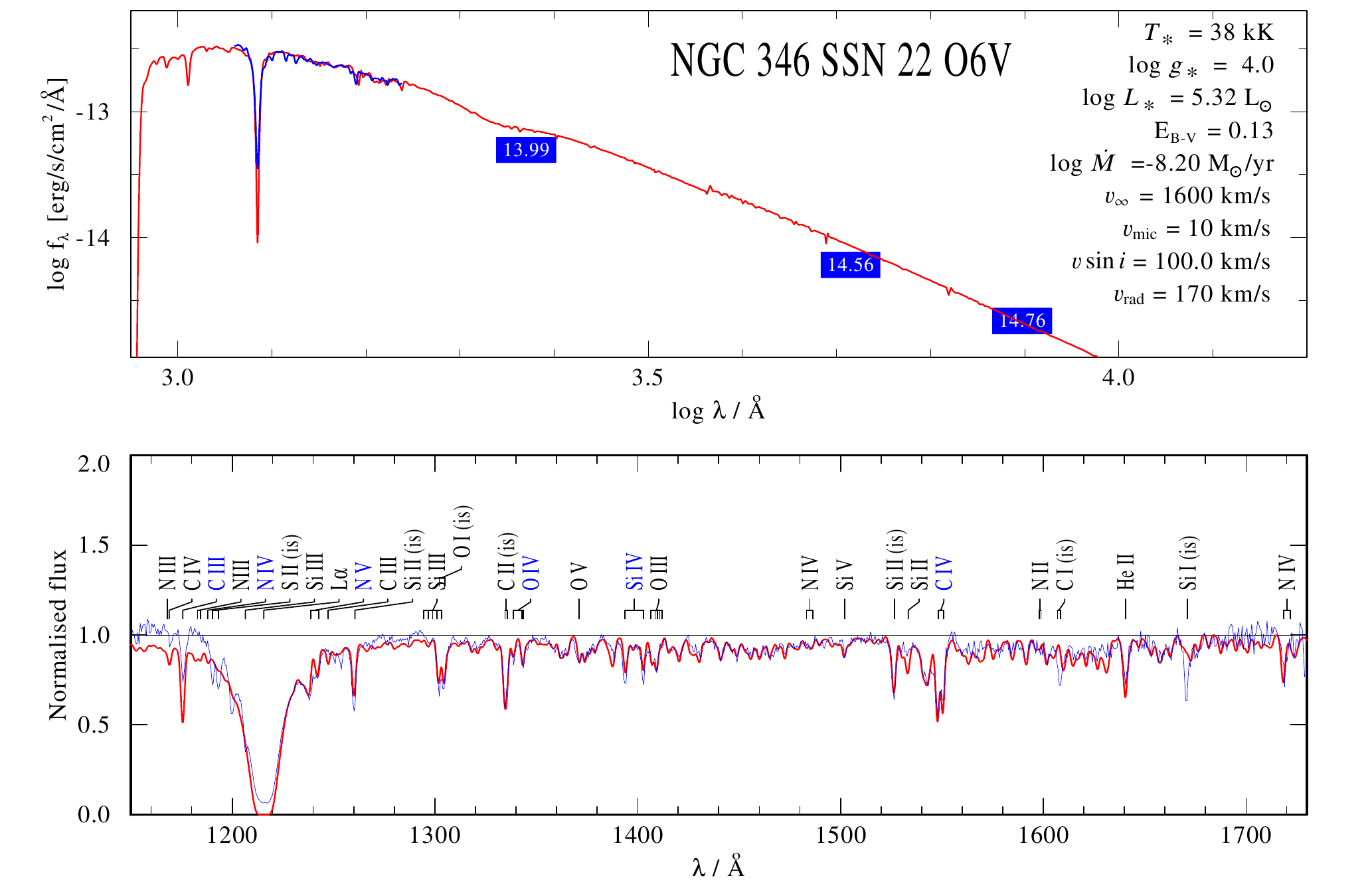}
    \caption{Same as Figure~\ref{fig:SSN_0009_p1}, but for SSN~22.}
    \label{fig:SSN_0022_p1}
\end{figure*}

\begin{figure*}
    \centering
    \includegraphics[width=0.9\hsize, trim={0cm 13.85cm 0cm 2.25cm},clip]{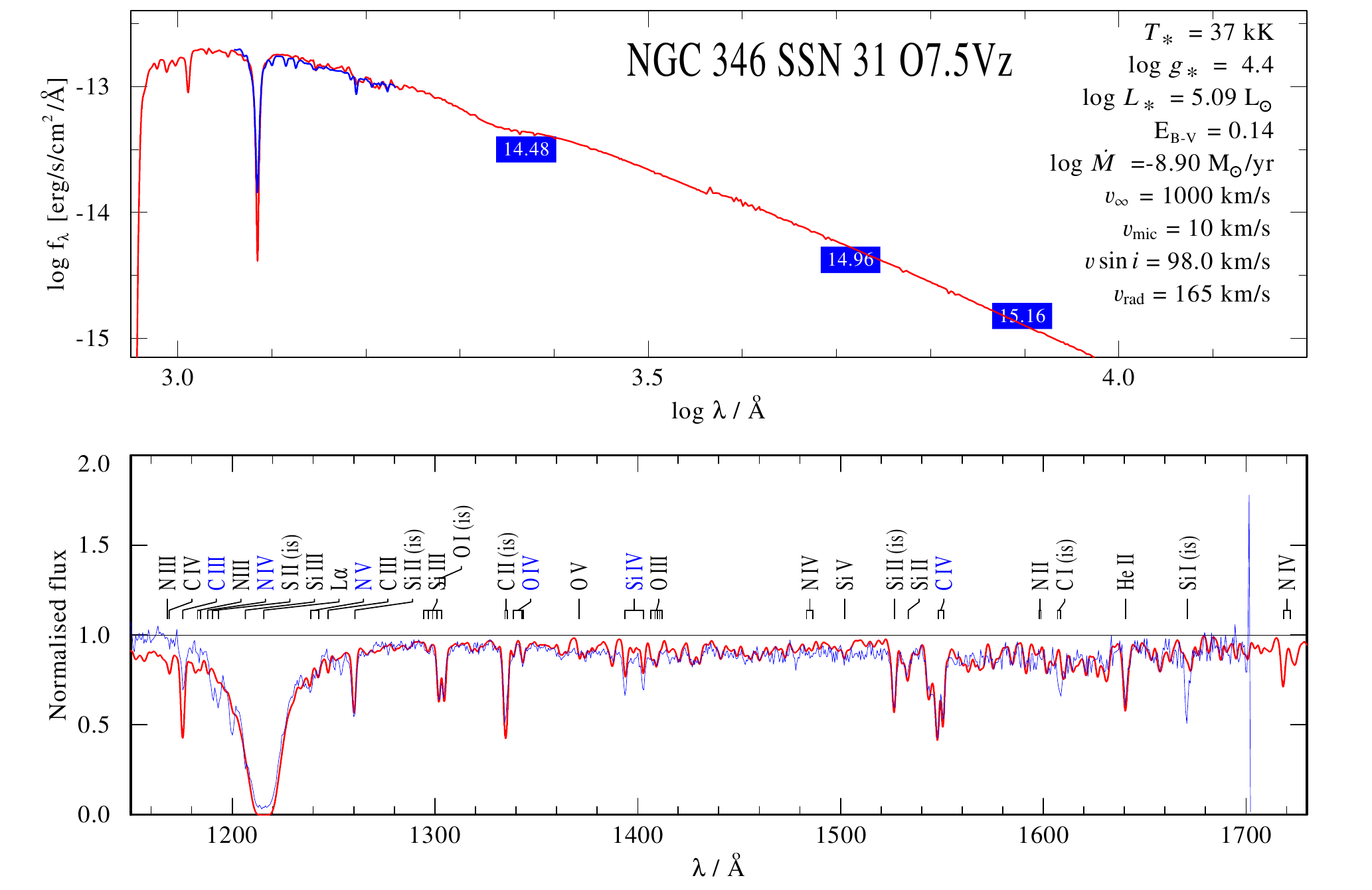}
    \caption{Same as Figure~\ref{fig:SSN_0009_p1}, but for SSN~31.}
    \label{fig:SSN_0031_p1}
\end{figure*}

\begin{figure*}
    \centering
    \includegraphics[width=0.9\hsize, trim={0cm 13.85cm 0cm 2.25cm},clip]{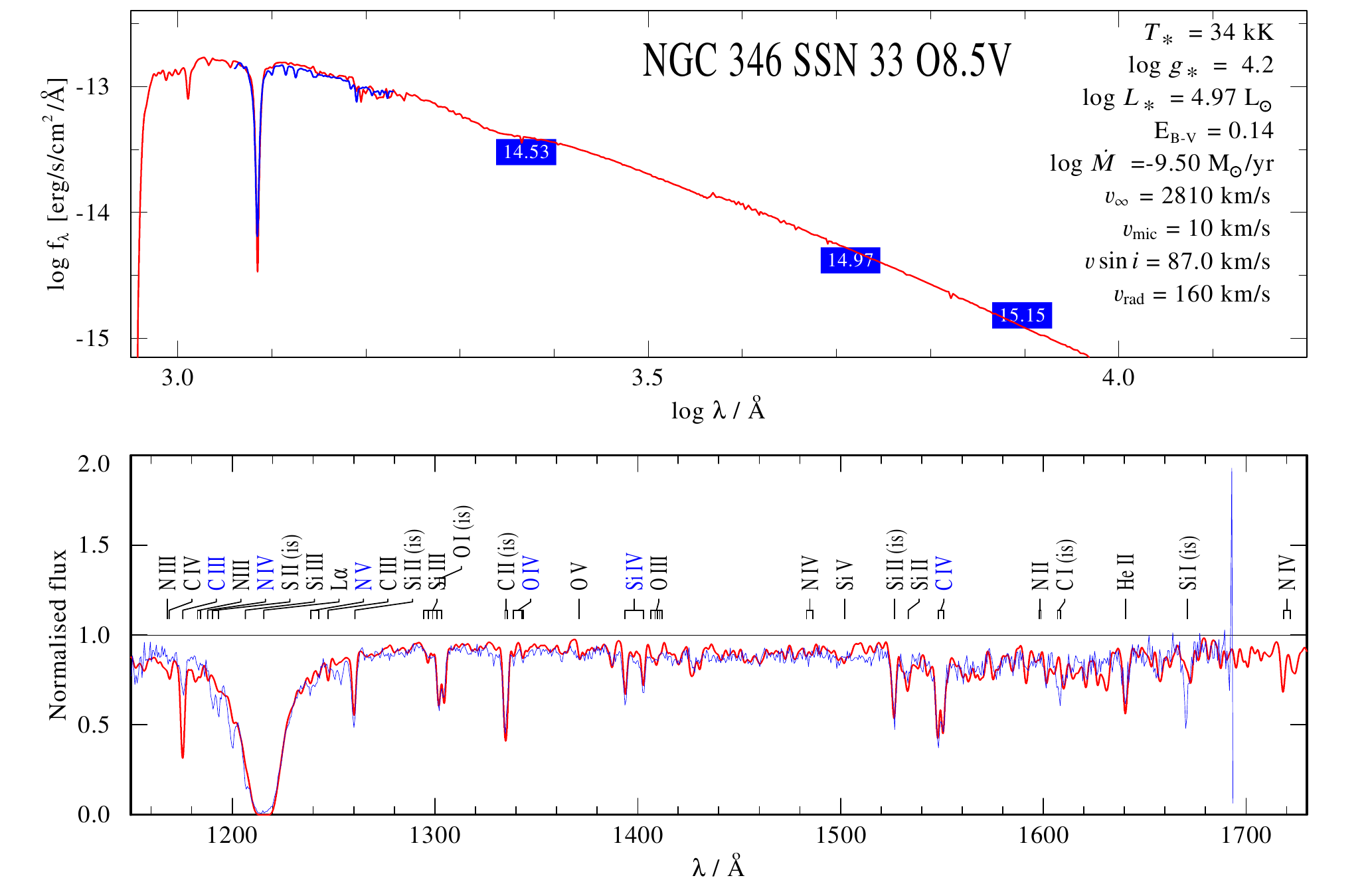}
    \caption{Same as Figure~\ref{fig:SSN_0009_p1}, but for SSN~33.}
    \label{fig:SSN_0033_p1}
\end{figure*}

\begin{figure*}
    \centering
    \includegraphics[width=0.9\hsize, trim={0cm 13.85cm 0cm 2.25cm},clip]{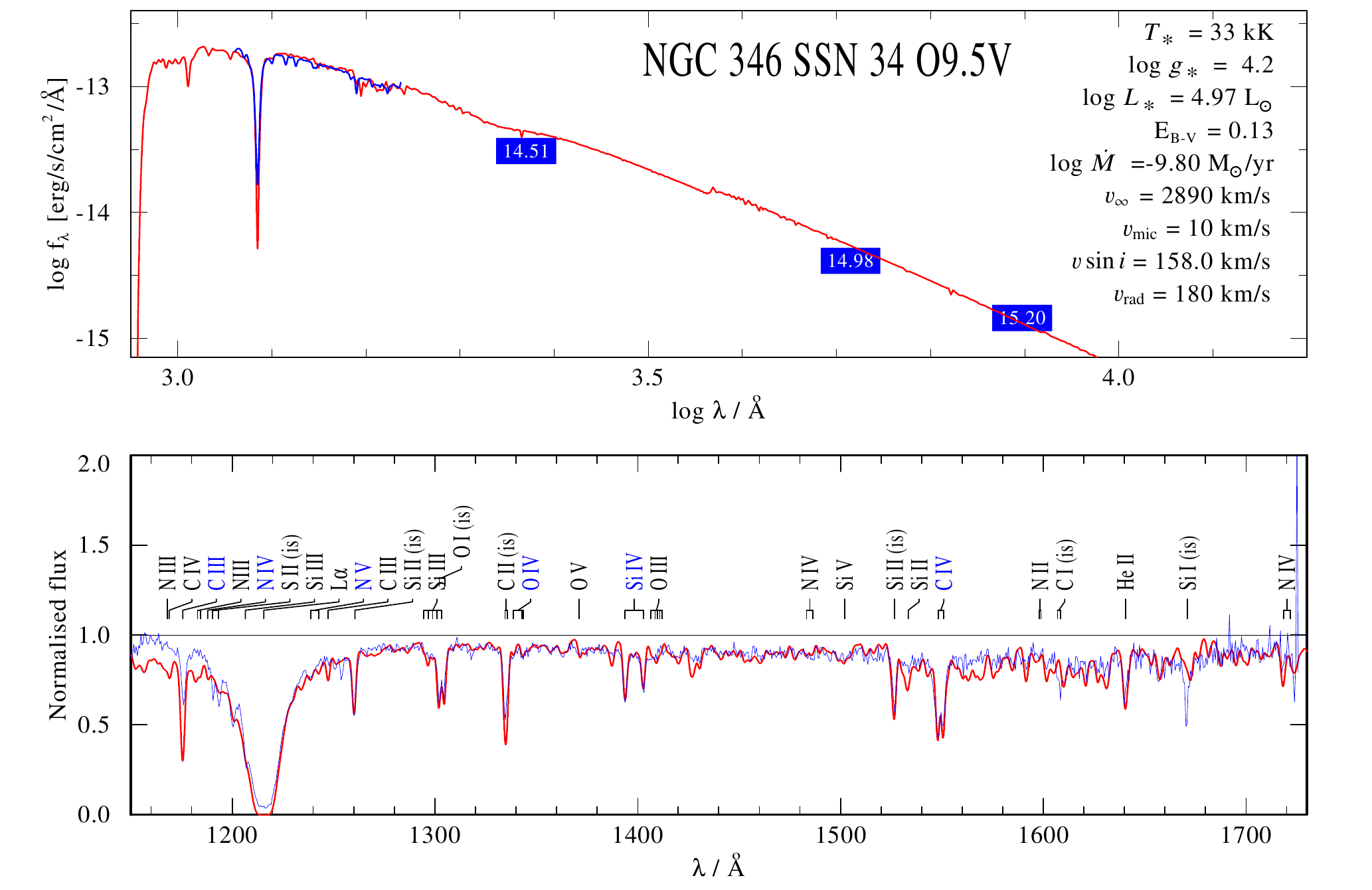}
    \caption{Same as Figure~\ref{fig:SSN_0009_p1}, but for SSN~34.}
    \label{fig:SSN_0034_p1}
\end{figure*}

\begin{figure*}
    \centering
    \includegraphics[width=0.9\hsize, trim={0cm 13.85cm 0cm 2.25cm},clip]{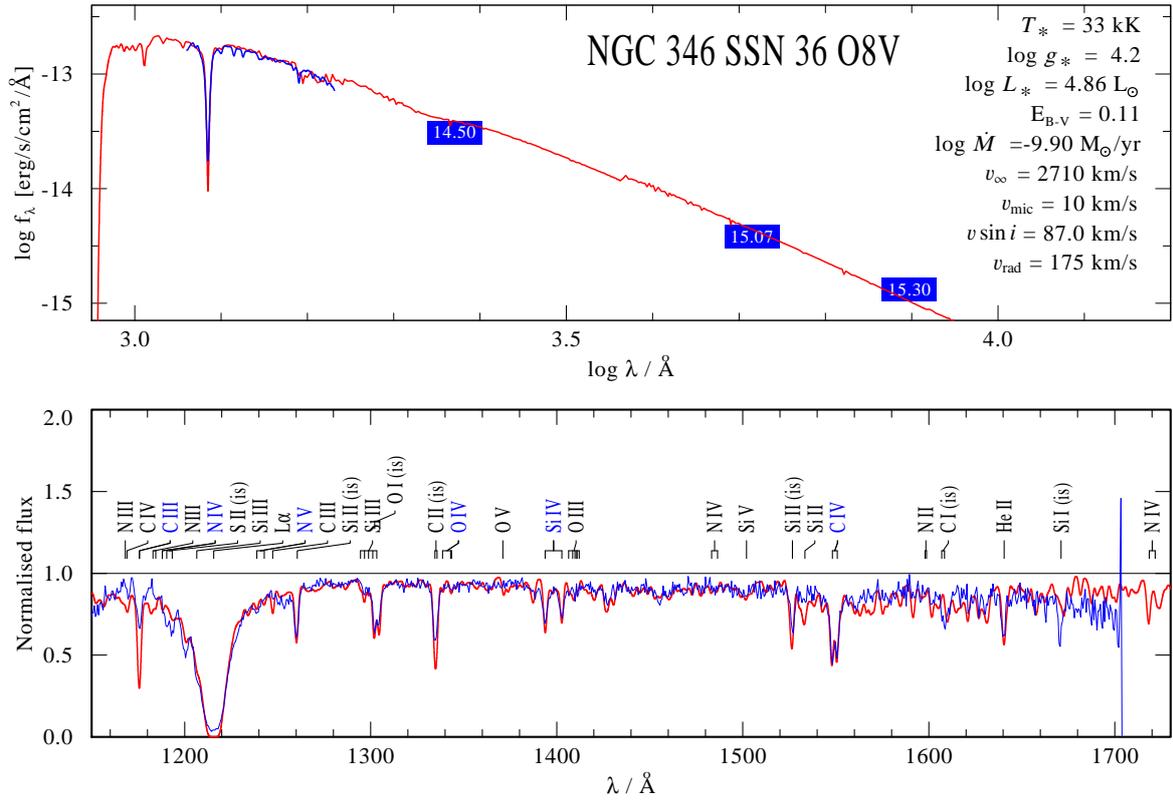}
    \caption{Same as Figure~\ref{fig:SSN_0009_p1}, but for SSN~36.}
    \label{fig:SSN_0036_p1}
\end{figure*}

\begin{figure*}
     \centering
     \includegraphics[width=0.9\hsize, trim={0cm 13.85cm 0cm 2.25cm},clip]{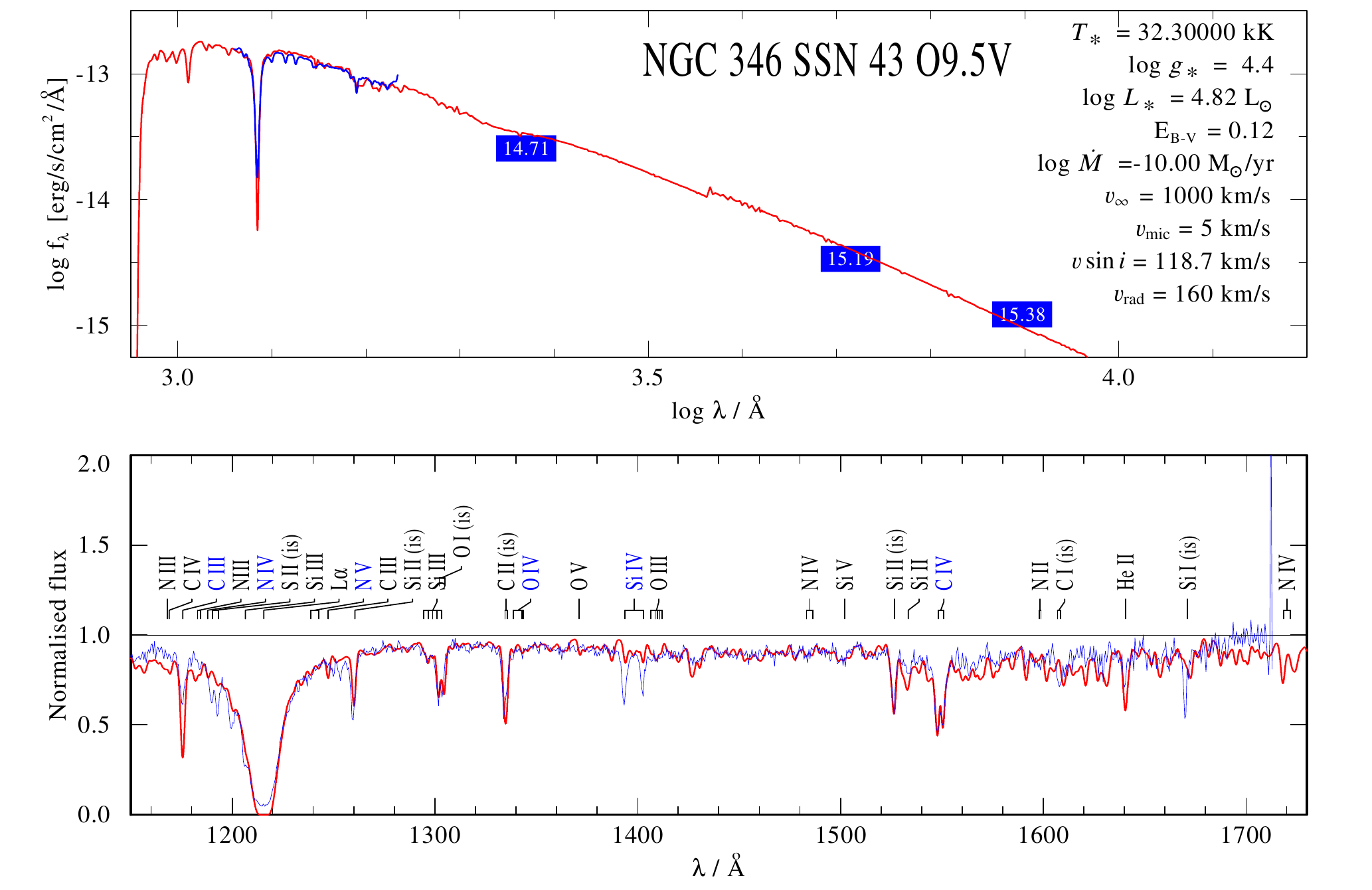}
     \caption{Same as Figure~\ref{fig:SSN_0009_p1}, but for SSN~43.}
     \label{fig:SSN_0043_p1}
\end{figure*}

\begin{figure*}
    \centering
    \includegraphics[width=0.9\hsize, trim={0cm 13.85cm 0cm 2.25cm},clip]{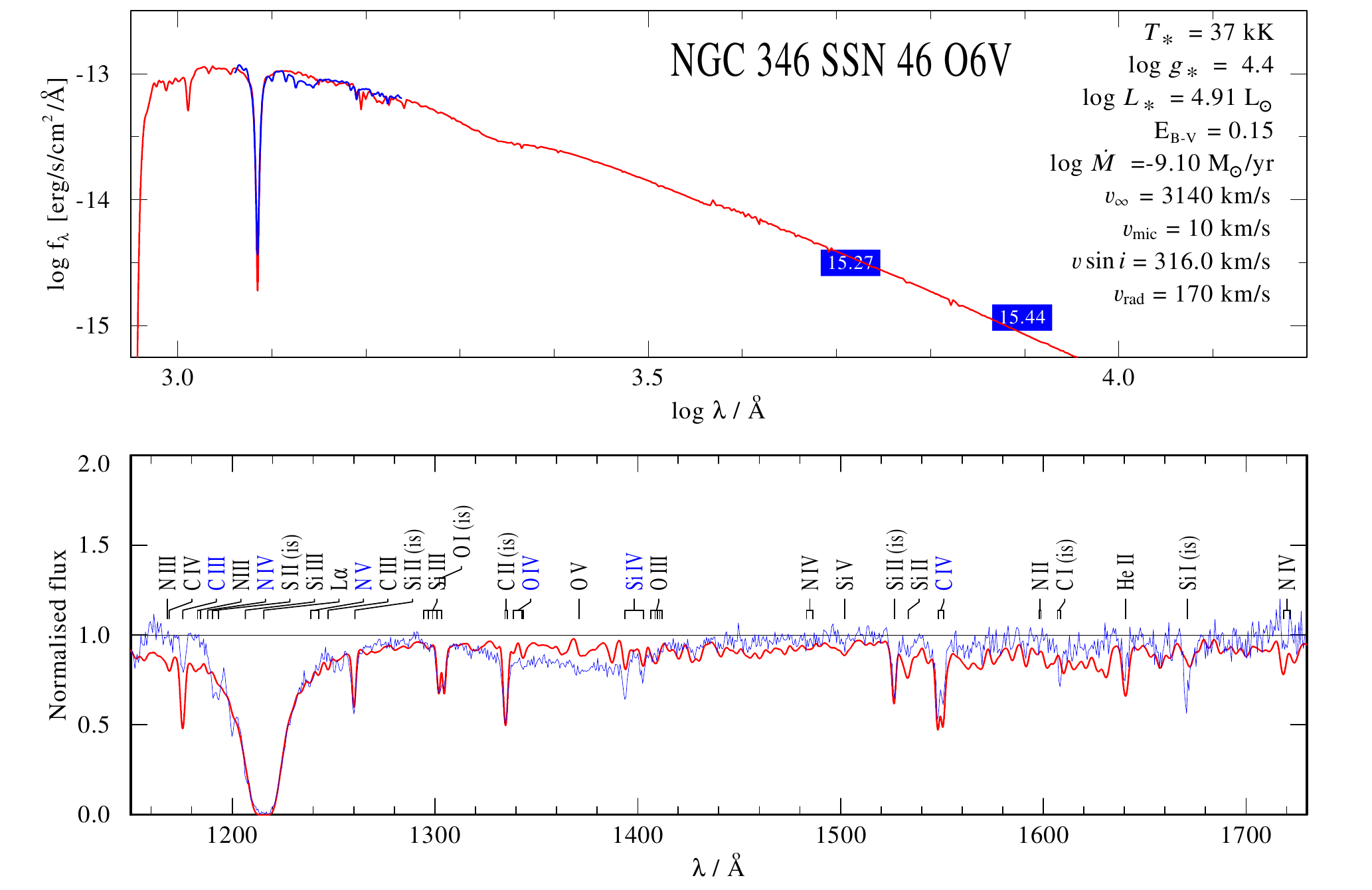}
    \caption{Same as Figure~\ref{fig:SSN_0009_p1}, but for SSN~46.}
    \label{fig:SSN_0046_p1}
\end{figure*}

\begin{figure*}
    \centering
    \includegraphics[width=0.9\hsize, trim={0cm 13.85cm 0cm 2.25cm},clip]{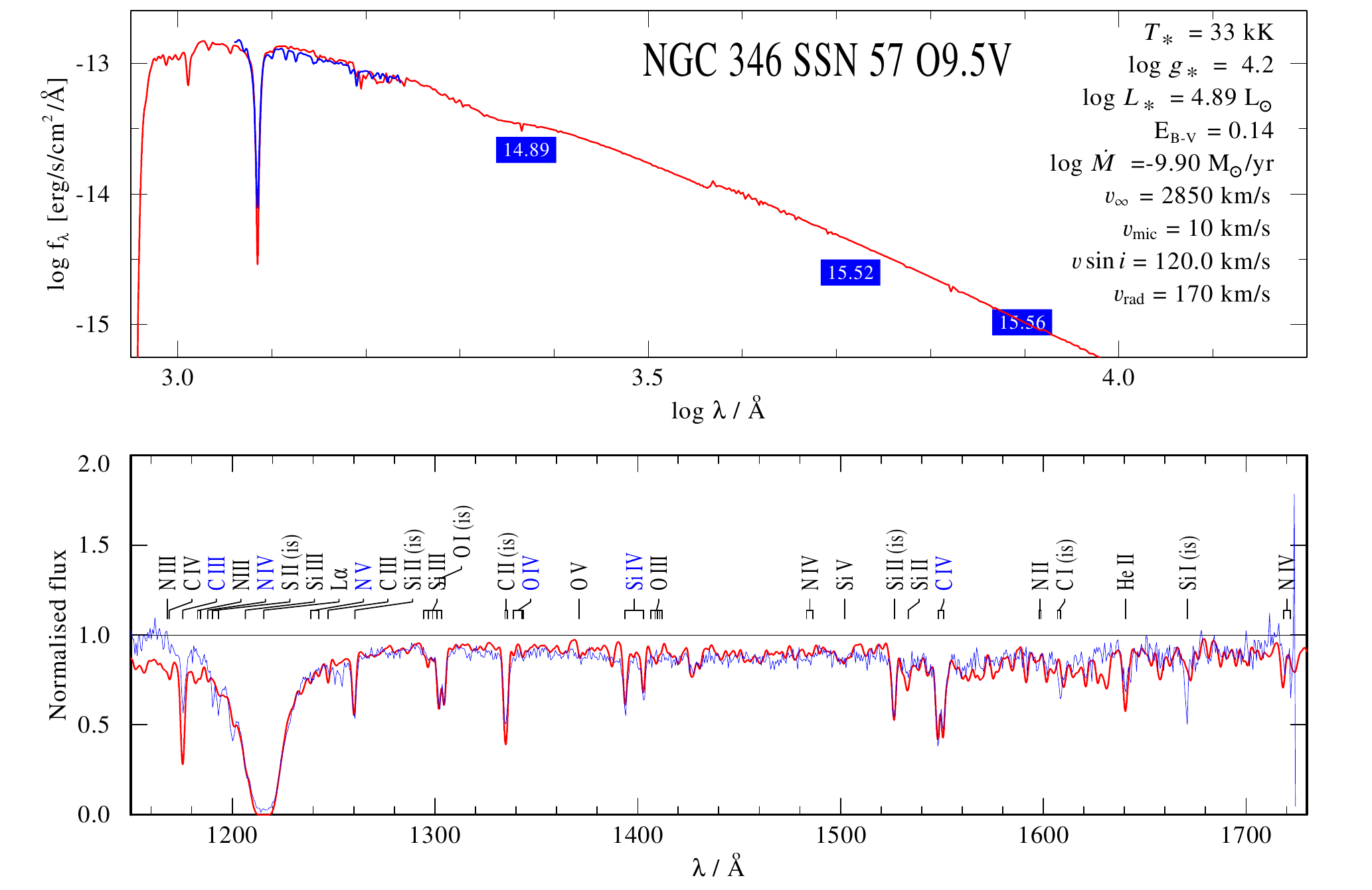}
    \caption{Same as Figure~\ref{fig:SSN_0009_p1}, but for SSN~57.}
    \label{fig:SSN_0057_p1}
\end{figure*}

\begin{figure*}
    \centering
    \includegraphics[width=0.9\hsize, trim={0cm 13.85cm 0cm 2.25cm},clip]{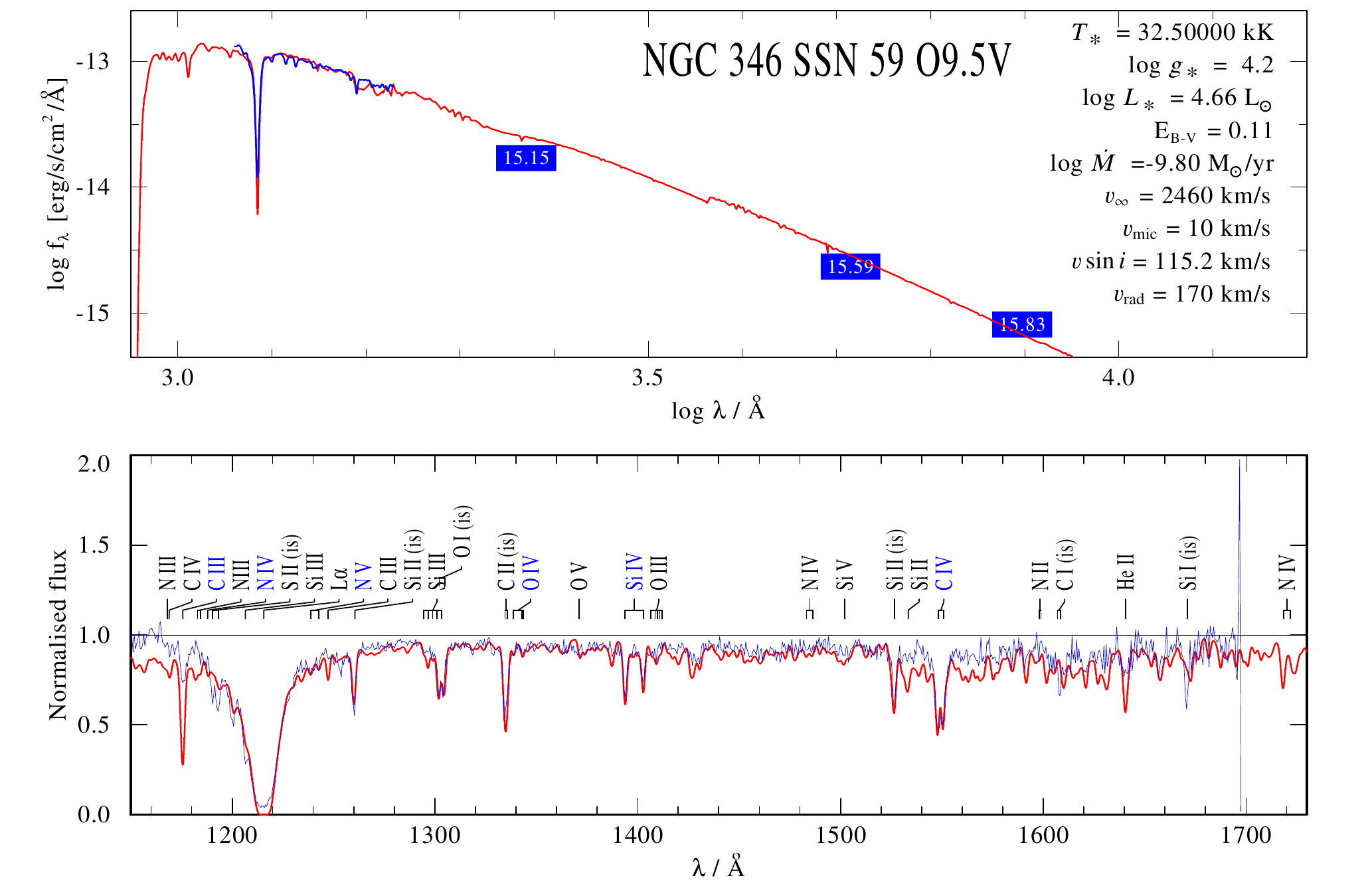}
    \caption{Same as Figure~\ref{fig:SSN_0009_p1}, but for SSN~59.}
    \label{fig:SSN_0059_p1}
\end{figure*}

\begin{figure*}
    \centering
    \includegraphics[width=0.9\hsize, trim={0cm 13.85cm 0cm 2.25cm},clip]{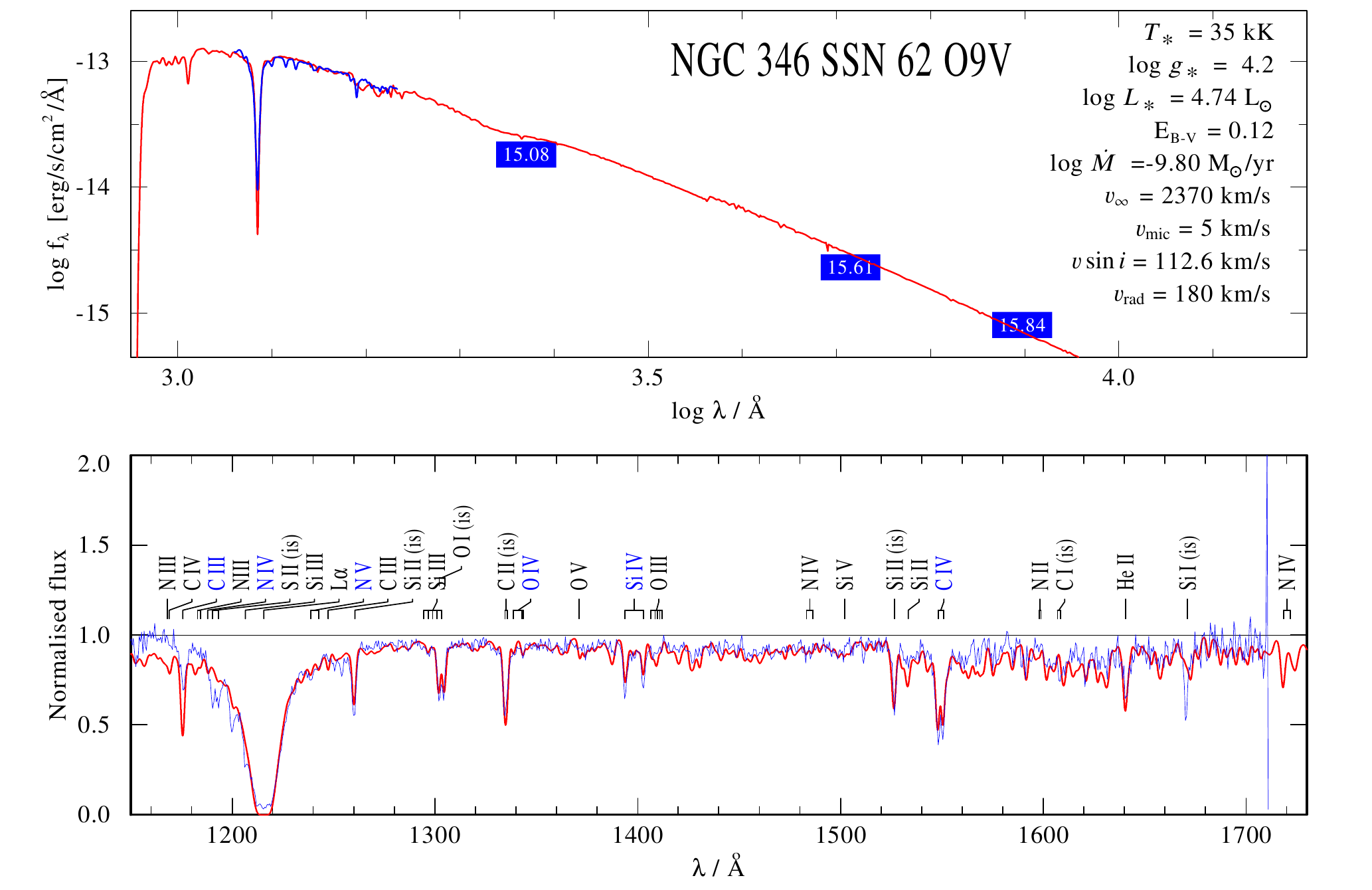}
    \caption{Same as Figure~\ref{fig:SSN_0009_p1}, but for SSN~62.}
    \label{fig:SSN_0062_p1}
\end{figure*}

\begin{figure*}
    \centering
    \includegraphics[width=0.9\hsize, trim={0cm 13.85cm 0cm 2.25cm},clip]{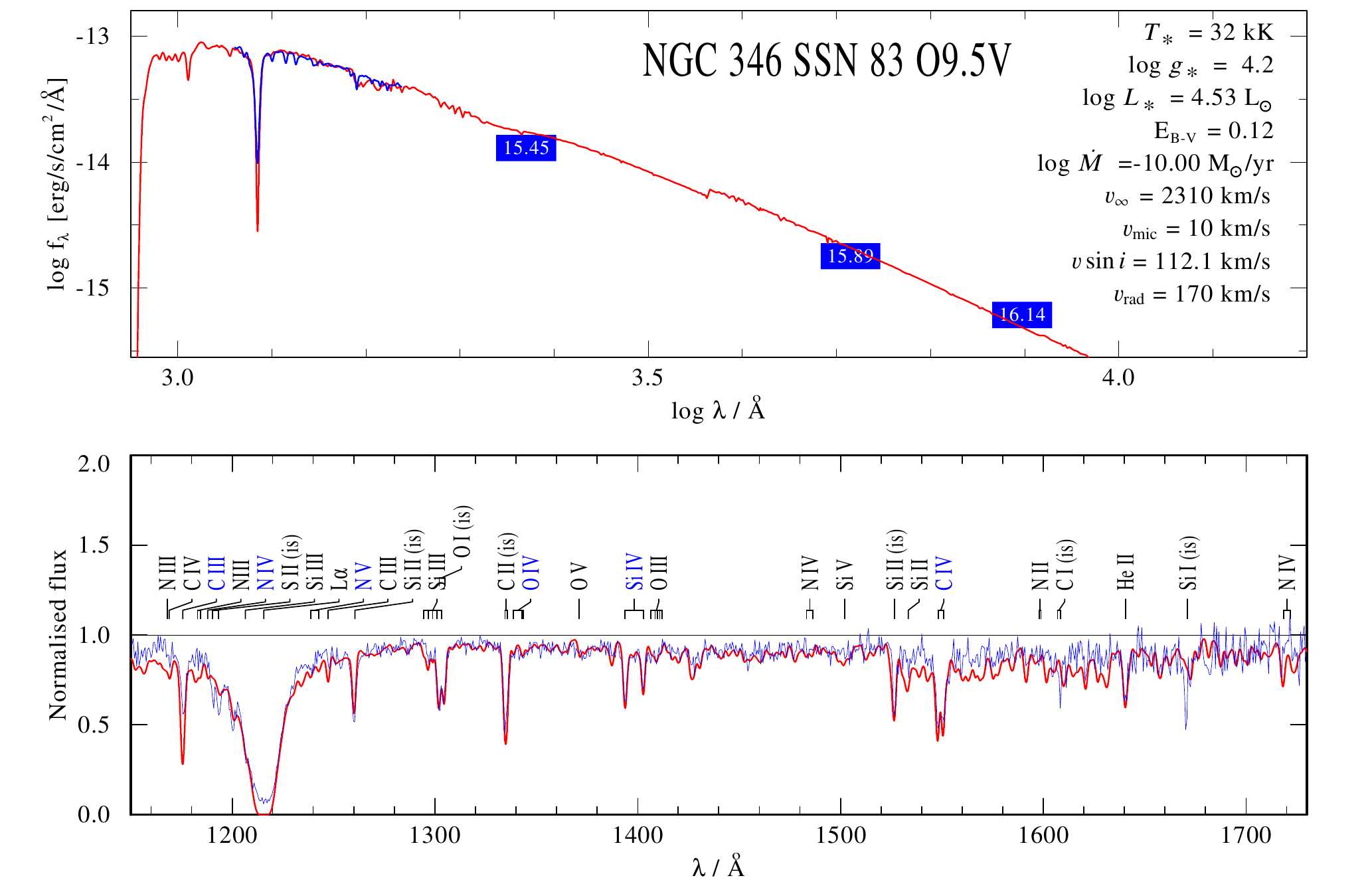}
    \caption{Same as Figure~\ref{fig:SSN_0009_p1}, but for SSN~83.}
    \label{fig:SSN_0083_p1}
\end{figure*}

\begin{figure*}
    \centering
    \includegraphics[width=0.9\hsize, trim={0cm 13.85cm 0cm 2.25cm},clip]{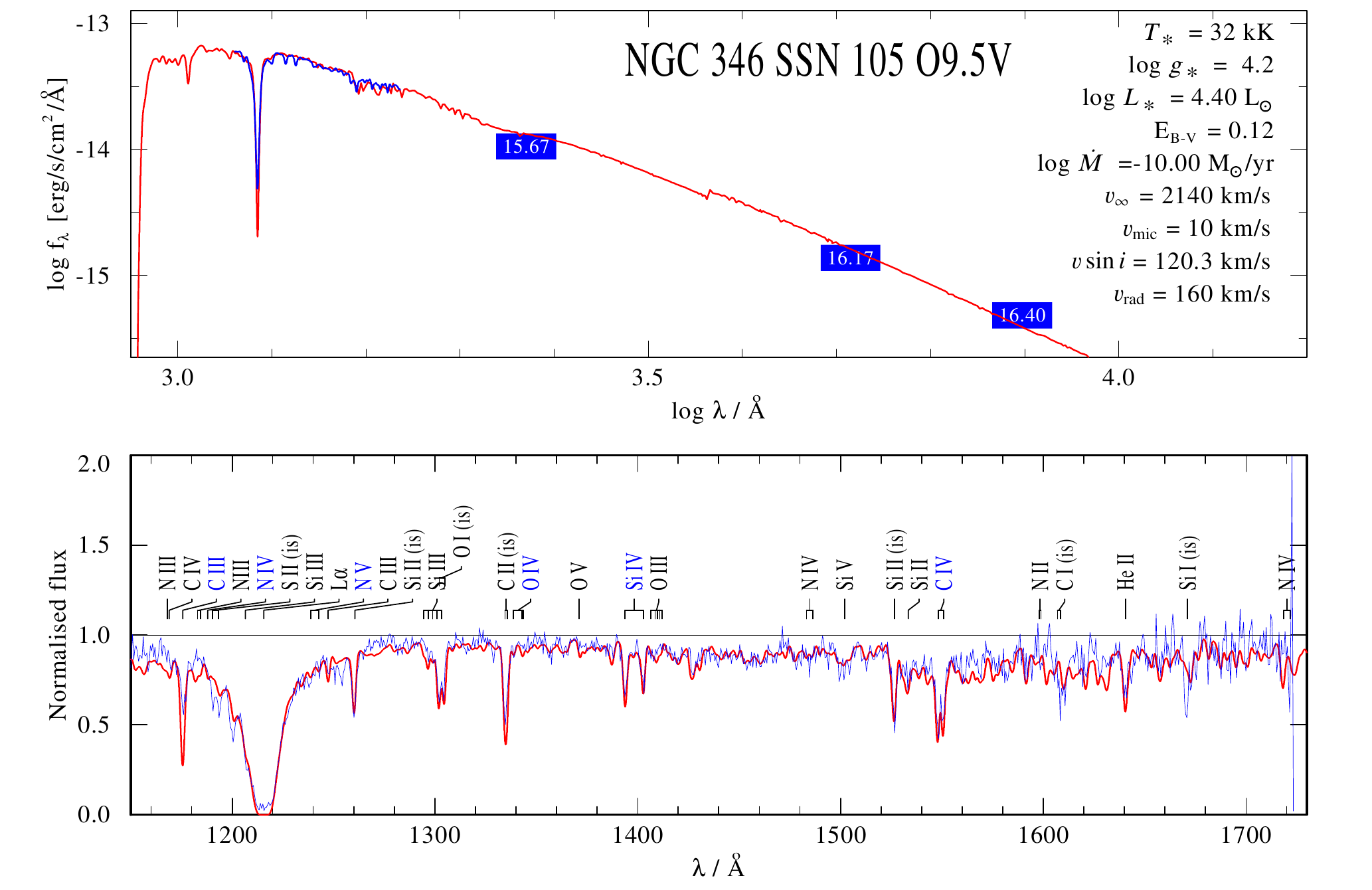}
    \caption{Same as Figure~\ref{fig:SSN_0009_p1}, but for SSN~105.}
    \label{fig:SSN_0105_p1}
\end{figure*}

\begin{figure*}
    \centering
    \includegraphics[width=0.9\hsize, trim={0cm 13.85cm 0cm 2.25cm},clip]{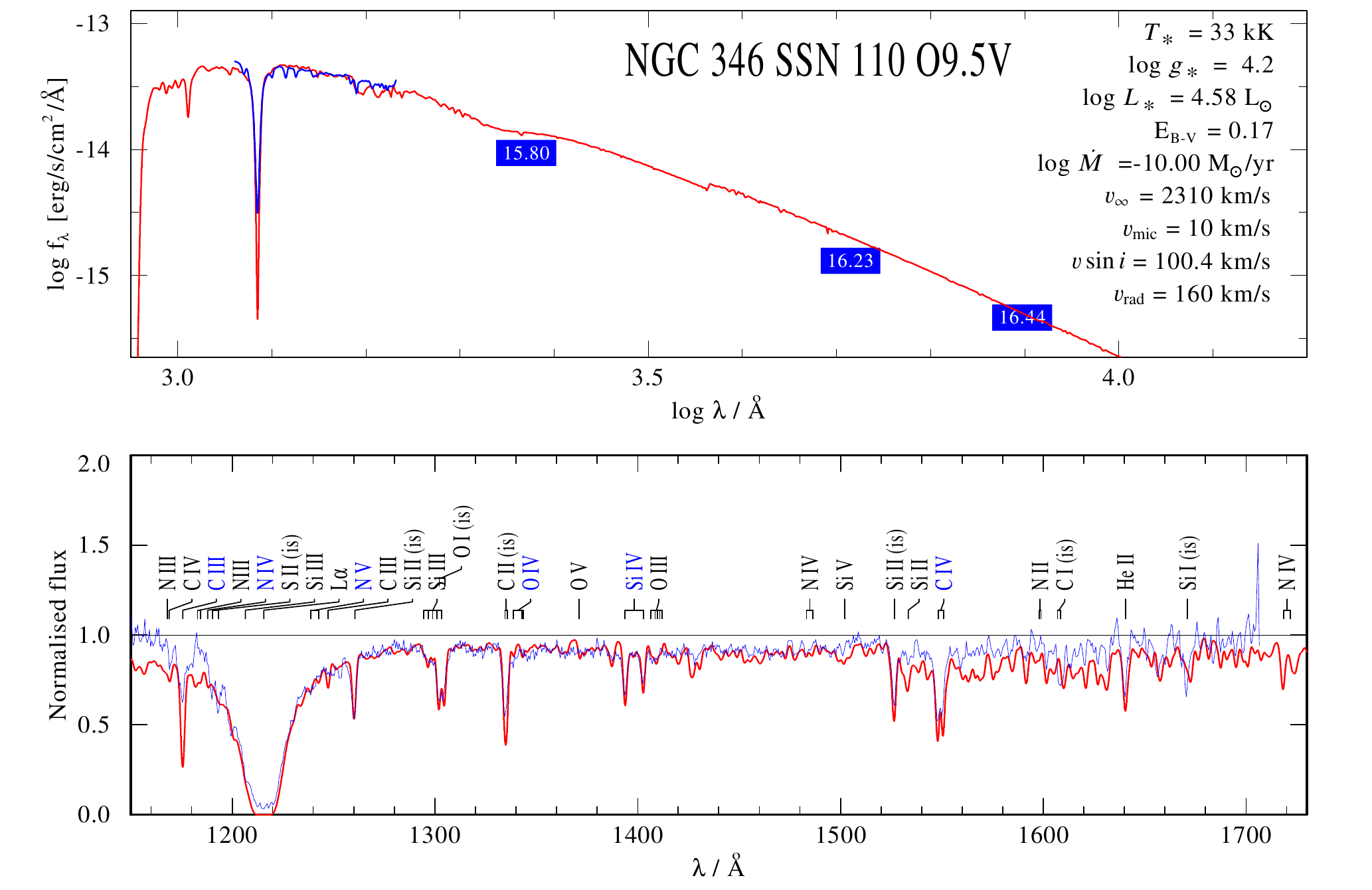}
    \caption{Same as Figure~\ref{fig:SSN_0009_p1}, but for SSN~110.}
    \label{fig:SSN_0110_p1}
\end{figure*}

\end{appendix}

\end{document}